\documentclass[twocolumn, prb, aps, superscriptaddress, longbibliography, showpacs]{revtex4-2}

\usepackage{graphicx}
\usepackage{amsmath}
\usepackage{bm}
\usepackage{hyperref}
\usepackage{siunitx}
\usepackage{amssymb}

\newcommand{\beq}{\begin{equation}}
\newcommand{\eeq}{\end{equation}}
\newcommand{\bea}{\begin{eqnarray}}
\newcommand{\eea}{\end{eqnarray}}

\newcommand{\rr}{{\bf r}}
\renewcommand{\k}{{\bf k}}

\newcommand{\q}{{\bf q}}

\newcommand{\ccfour}{{\href{https://creativecommons.org/licenses/by/4.0/}{CC BY 4.0}}}
\newcommand{\ccthree}{{\href{https://creativecommons.org/licenses/by/3.0/}{CC BY 3.0}}}

\newcommand{\STO}{SrTiO$_3$}
\begin{document}

\title{On the Remarkable Superconductivity  of FeSe and Its Close Cousins}

\author{Andreas Kreisel}
\affiliation{Institut f\" ur Theoretische Physik, Universit\" at Leipzig, D-04103 Leipzig, Germany}
\author{P. J. Hirschfeld}
\affiliation{Department of Physics, University of Florida, Gainesville, Florida 32611, USA}
\author{Brian M. Andersen}
\affiliation{Niels Bohr Institute, University of Copenhagen, Lyngbyvej 2, DK-2100 Copenhagen, Denmark}

\date{September 10, 2020}
\begin{abstract}

Emergent electronic phenomena in iron-based superconductors have been at the forefront of condensed matter physics for more than a decade. Much has been learned about the origins and intertwined roles of ordered phases, including nematicity, magnetism, and superconductivity, in this fascinating class of materials. In recent years, focus has been centered on the peculiar and highly unusual properties of FeSe and its close cousins. This family of materials has attracted considerable attention due to the discovery of unexpected superconducting gap structures, a wide range of superconducting critical  temperatures, and evidence for nontrivial band topology, including associated spin-helical surface states and vortex-induced Majorana bound states. Here, we review superconductivity in iron chalcogenide superconductors, including bulk FeSe, doped bulk FeSe, FeTe$_{1-x}$Se$_x$, intercalated FeSe materials, and monolayer FeSe and FeTe$_{1-x}$Se$_x$ on SrTiO$_3$. We focus on the superconducting properties, including a survey of the relevant  experimental studies, and~a discussion of the different proposed  theoretical pairing scenarios. In the last part of the paper, we review the growing recent evidence for nontrivial topological effects in FeSe-related materials, focusing again on interesting implications for superconductivity.
\end{abstract}

\maketitle

\renewcommand{\baselinestretch}{0.96}\normalsize
\tableofcontents
\renewcommand{\baselinestretch}{1.0}\normalsize

\vfill\eject

\section{Introduction}

{After a dozen years} 
of research into iron-based superconductivity (FeSC), a good deal has been learned about the phenomenology and microscopic origins of this fascinating phenomenon that can be generally agreed upon.  Since the bulk electron--phonon interaction is rather weak, the mechanism for pairing is almost certainly electronic, and related to the intermediate-to-strong local Coulomb interactions in these materials  \cite{HKM_ROPP,ChubukovHirschfeld2015,HirschfeldCRAS,MaitiChubukovreview}.  As the Fermi surface takes the form of small electron- and hole-like pockets centered at high symmetry points, and the bare interactions are repulsive, the most likely superconducting states change signs {between pockets.} 
As the nesting between electron and hole pockets is often particularly strong, most systems are believed to be of so-called $s_\pm$ character, where the sign change occurs between the gap amplitudes on the $\Gamma$-centered hole pockets and the $M$-centered electron pockets  \cite{Mazin2008,Kuroki2008}.
Significant gap anisotropy also exists due to the multiorbital character of Fermi surface sheets, and to the necessity of minimizing the Coulomb interaction  \cite{Maier2009A,Zhang2009}.

Strong pair scattering between the electron pockets exists as well, implying that the $d$-wave attraction is also strong,  and competes with the $s$-wave  \cite{Maiti2011}.  Transitions between competing superconducting states, as well as time-reversal symmetry breaking admixtures of the two  \cite{Thomale2012}, or of
purely $s$-wave amplitudes with different phases on different pockets  \cite{Cvetkovic2013,Maiti2013}, are therefore possible. The existence of sign-changing superconducting order has now been relatively well established in some systems via observations of the neutron resonance  \cite{Inosov2015} and  disorder properties  \cite{BalatskyRMP,AlloulRMP,EfremovPRB2011,Prozorov2014PRX,Ghigo2018,HAEM2015,Sprau2017,Du2018}.
While the exact gap structures, as well as observations of time-reversal symmetry breaking, remain controversial, the general principles outlined above are widely accepted.

Most of the consensus described above was developed in the context of the Fe-pnictide superconductors, particularly the  122 systems, but the Fe-chalcogenide materials present a completely new set of questions, and even pose challenges to the central paradigm of superconductivity established for the pnictides.  The strength of electronic correlations and spin-orbit coupling is expected to be higher in the chalcogenides, and may be responsible for the remarkable behavior  of bulk FeSe, where superconductivity condenses out of a strongly nematic state with no magnetic long-range order.  Modifying the 8 K superconductor FeSe in almost any way, including pressure, intercalation, or deposition of a monolayer film on a substrate, produces a high-$T_c$ superconductor.   The new states engendered by these modifications are thought to be related to one another, and indeed some have remarkably similar Fermi surface structures, notably lacking hole bands at the Fermi level.  Why such systems, in violation of the central paradigm established apparently quite generally for Fe-pnictides, should have the highest critical temperatures of the FeSC family, is the central current question of iron-based superconductivity research.

Finally, a wave of recent measurements and theories have offered considerable evidence for topological superconductivity in the FeTe$_{1-x}$Se$_x$ system, holding out the prospect of creating and manipulating Majorana bound states for quantum computation in these materials.  Together with the rough consensus on many aspects of Fe-pnictide superconductivity, the  challenges posed by these and other discoveries in the Fe-chalcogenide family suggest that the time is ripe to review developments in this field.  Following  reviews of mostly experimental results  \cite{Mizuguchi2010II,Shibauchi2020}, a review focused on bulk FeSe  \cite{Boehmer_2017}, and recent specialized reviews of topological aspects  \cite{Hao_review,Wu_2020_XXX}, we attempt here to synthesize what has been learned about the Fe-chalcogenide superconductors, with an emphasis on the superconducting state.   Our  goal is to  elucidate which new theoretical ideas  have been stimulated by experimental discoveries, and highlight remaining open questions in the field.

The paper is structured as follows: In Section~\ref{sec:overview}, we first give an overview of FeSe itself,   with an emphasis on its unusual band structure created by strong correlations and the implications of the tiny, highly nematic pockets at the Fermi surface.  In Section~\ref{sec:properties}, we discuss what is known about bulk FeSe's magnetic properties, a knowledge of which is essential to understand the spin fluctuation pairing interaction, and discuss measurements in the superconducting state that provide information on the highly anisotropic gap.  The spin fluctuation theory of pairing is introduced, and modifications required to explain the ``orbitally selective'' pairing reported in this system are explained.  Next, in Section~\ref{sec:pressure}, we discuss the remarkable effects of pressure and chemical pressure (via S doping on the Se site) on the FeSe phase diagram.  We then consider in Section~\ref{sec:monolayer} the FeSe monolayer system on \STO (STO) substrate, with the highest $T_c$ in the FeSC family.  We discuss various ideas that have been put forward to understand the mechanism of electron doping by the substrate, and its effect on superconductivity.  In Section~\ref{sec:intercalates}, we consider  FeSe intercalated with alkali atoms, organic molecules, and LiOH, all of which are high-$T_c$ materials, and at least some of which share similar electronic properties to the monolayer on STO. Finally in the last Section \ref{sec:topo}, we review the recent theoretical proposals and experimental evidence for non-trivial band topology in some FeSCs. We also discuss reports of topological superconductivity in these materials.

\section{Overview}
\label{sec:overview}

\subsection{Iron Pnictides}
We begin by briefly reviewing the essential ingredients in the recipe for an iron-based superconductor \cite{Paglione2010,Johnston2010,Stewart2011,HKM_ROPP,Hosono2015a,ChubukovHirschfeld2015,HirschfeldCRAS,Guterding_ab_initio_2017}.  
The original Fe-based superconductors, LaFePO and F-doped  LaFeAsO, were discovered by H. Hosono  \cite{Kamihara2006,Kamihara2008}, with structures containing square lattices of Fe atoms with pnictogen As placed in out-of-plane positions above and below the Fe plane, such that there were two inequivalent As per unit cell.
They were quickly noted to resemble  other materials classes of unconventional superconductors, such as cuprates and heavy fermion systems, by exhibiting electronic correlations which
play a significant role in  emergent ordered phases such as magnetism, nematicity, and
superconductivity  \cite{Mazin2010,Scalapino2012}. Several other materials with similar iron planes were discovered in short order, including  Ba(Fe$_{1-x}$Co$_x$)$_2$As$_2$, Ba$_{1-x}$K$_x$Fe$_2$As$_2$, BaFe$_2$(As$_{1-x}$P$_x$)$_2$, and   LiFeAs.

\begin{figure*}[tb]
\centering
{\includegraphics[width=\linewidth]{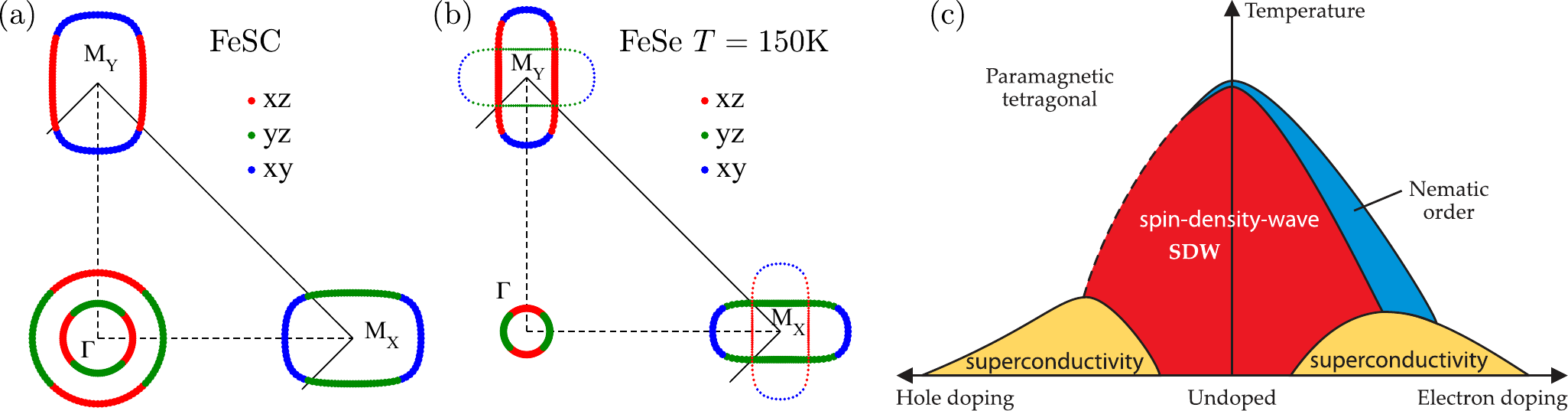}}
\caption{(\textbf{a}) Schematic Fermi surface of an iron-based superconductor (one quadrant).
 statement which creates ugly Widows and Orphans below and above the figure, even some figures do not appear on the page where these are discussed!-> This is our journal layout policy. We will adjust all figures before publication.
Colors represent majority $d$-orbital weight at each point on the Fermi surface.  Fermi pockets are depicted in the 1-Fe zone, but 2-Fe zone boundary is also shown as solid line.
(\textbf{b}) Schematic Fermi surface corresponding roughly to a high-temperature tetragonal phase of FeSe with small pockets, lifted inner sheet around $\Gamma$, and the folded electron pockets (thin dots) as expected in the 2-Fe zone.
Reprinted figure with permission from \cite{Fanfarillo2016}, copyright by the American Physical Society (2016).
(\textbf{c}) Schematic phase diagram of a Fe-pnictide superconductor, based on Ba-122.
Reproduced from \cite{ChubukovHirschfeld2015}, with the permission of the American Institute of Physics.}
\label{fig:FS_orbitals}
\end{figure*}

Like cuprates, their band structures are quite two-dimensional, but the parent compounds of the Fe-based superconductors are metallic rather than insulating.   Instead of the large Fermi surfaces seen in cuprates at optimal doping, Fe-based systems display small Fermi surface pockets centered at high symmetry points.  These pockets have almost pure Fe-$d$-character (pnictide and chalcogenide $p$-states are typically several eV from the Fermi level), but the $d$-orbital content winds around each pocket, as depicted in Figure~\ref{fig:FS_orbitals}a,b. Note that the two inequivalent As atoms implies that the correct 2-Fe Brillouin zone is one-half the size of the reference 1-Fe zone.

The $d$-spins on the Fe sites are not strongly localized in character, but to understand the low-lying magnetic states in these systems it is frequently convenient to examine the effective exchanges $J_{ij}$ between spins on sites $i,j$.
Calculations and experiments  \cite{Dai2012} both suggest that the nearest neighbor Fe exchange $J_1$ is of the same order of magnitude as the next nearest neighbor exchange $J_2$, due to the strong overlap of the pnictide $p$ orbitals with the next nearest neighbor Fe.  This unusual situation is responsible for magnetic ordering in a stripelike pattern with wave vector $(\pi,0)$ in the one-Fe zone.  In most Fe-pnictides, stripe magnetic order is dominant in the doping range near six electrons per Fe, with other magnetic orders, e.g., N\'eel order, double stripe order, and various $C_4$ symmetric phases often close by in energy~\cite{Lorenzana2008,Eremin_chub_2010,Brydon2011,Boehmer2015NC,Gastiasoro2015_DQ,Christensen_2017_multi,C4JPSJ,Christensen_2018_PRX}.  In special situations, these states are observed condense in small parts of the phase diagram, but the stripe order is generally dominant.   As the ordered magnetic state is weakened by doping, a competing superconducting dome emerges at lower temperatures (Figure~\ref{fig:FS_orbitals}c).

At higher temperatures near the edge of the magnetic phase boundary, an electronic nematic phase forms where the crystal structure is very slightly orthorhombic, but the electronic responses are found to be highly anisotropic.  Many authors have identified the nematic phase as the natural consequence of the competing spin fluctuations in  a $J_1-J_2$ spin-model \`a la Chandra--Coleman and Larkin  \cite{Chandra1990,Dai2012}, and essentially the same Ising nematic state is obtained in an itinerant picture  \cite{Fernandes2012II,Fernandes2014}.  On the other hand, one must be careful, because such a transition can in principle be driven also by orbital or lattice degrees of freedom, since the same symmetry is broken by all effects  \cite{Fernandes2014}.  For example, some authors have used the lack of long-range magnetic order to argue for a nematic transition driven by orbital fluctuations rather than spin  \cite{Yamakawa2017}.    We do not review these arguments here, but refer the interested reader to the literature.

Superconductivity in the Fe-pnictides has been
discussed extensively in the
literature  \cite{HKM_ROPP,ChubukovHirschfeld2015,HirschfeldCRAS},
generally in terms of unconventional pair states driven by repulsive  interactions
\cite{MaitiChubukovreview}.  The state believed to be realized in most such systems is
a version of the so-called $s_\pm$ state, which has the full symmetry of the
crystal but changes sign between electron and hole
pockets  \cite{Mazin2008}. Within spin fluctuation theory, discussed at greater length in
Section~\ref{subsec:theory_FeSC}, doping the system was shown to
lead to increased competition with $d$-wave pairing  \cite{Graser2009,Zhang2009,Maiti2011}.
The $s$ and $d$ states are shown schematically in {Figure}~\ref{fig:gaps}a,b.
There has also been considerable interest in the possibility of time reversal symmetry breaking pair states, e.g., $s+id$~\cite{WCLee2009,Thomale2012} or $s+is$ (the latter implying the existence of at least three gaps, with complex phases determined by interband interactions)  \cite{Stanev2010,Maiti2013}.  Strong evidence for such states in overdoped Fe pnictides has been provided recently by $\mu$SR  \cite{Grinenko2020}.

The multiband nature of the Fe-based superconductors implies that the effect of  nonmagnetic disorder, often a key probe of unconventional superconductivity,
is more subtle than
in single-band systems.  In the presence of sign changing, e.g., $d$ or $s_\pm$ states,
any significant scattering connecting  Fermi surfaces hosting gaps of opposite sign
will be pairbreaking and suppress $T_c$.  Dopants substituting for Fe correspond to a strong
localized impurity potential, and
therefore cause significant large momentum scattering; in general such impurities are pairbreaking.
On the other hand,
many chemical substituents lie outside the Fe-pnictogen plane, and therefore have a much weaker
and longer range potential in the
real space of the Fe plane. These defects are expected, and generally observed, to suppress $T_c$ at
a much slower rate, because they scatter largely within the same Fermi surface sheet.    Note that chemical substitution, in addition to providing an impurity potential,
can also affect  the electronic structure
by doping with carriers or by chemical pressure.  It is therefore important  to  compare
$T_c$ suppression with an independent measure of disorder,
such as the residual resistivity.  This problem
was discussed by
Wang et al.  \cite{YWang2013}, who proposed that almost any rate of $T_c$ suppression
was possible in an $s_\pm$ system, depending
on the ratio of intra- to interband scattering.  This conclusion is already present   in
the important early work of   Golubov and Mazin  \cite{GolubovMazin1997}.

Some unusual pairing states can be driven by disorder in multiband systems.  Efremov~et al.~\cite{EfremovPRB2011}
showed that in a system with asymmetric $s_\pm$ pairing, i.e., two gaps different in size, a transition
$s_\pm\rightarrow s_{++}$ can take place simply because disorder averages the two gaps; whether this occurs before
$T_c$ vanishes depends on the characteristics of the interaction matrix in band space.  Theoretical aspects of this transition, of which there has been at
least one convincing observation   \cite{Ghigo2018}, have been reviewed
by Korshunov et al.  \cite{Korshunov2016}.  In addition, it has been pointed out that it is sometimes energetically
favorable to pass from an $s_\pm$ state to a $s_{++}$ state via an intermediate, disorder-driven
$s+is$ state  \cite{Stanev2014,Babaev2017,Garaud2018}.

\subsection{How FeSe Is Different from Pnictides}
\label{subsec:different}

In the last few years, the field of   Fe-based superconductivity has been driven largely by  studies of bulk FeSe and its close cousins, including FeTe$_{1-x}$Se$_x$, doped or ``dosed'' FeSe, thin layers of FeSe or FeTe$_{1-x}$Se$_x$ on SrTiO$_3$ (STO) substrates, and a number of intercalated FeSe compounds. Reasons for the  attention focused on this class of systems include improved sample control
\cite{Boehmer2013} and a series of surprising discoveries that remain topics of considerable current controversy; (1) peculiar nematic effects including highly anisotropic electronic properties, in the absence of long-range magnetic order; (2) unusual low-energy electronic structure compared to other FeSCs, (3) tunable superconducting critical transition temperatures $T_c$, and (4) evidence for topologically non-trivial bands and associated topological superconductivity.

\begin{figure*}[tb]
\centering
{\includegraphics[width=\textwidth]{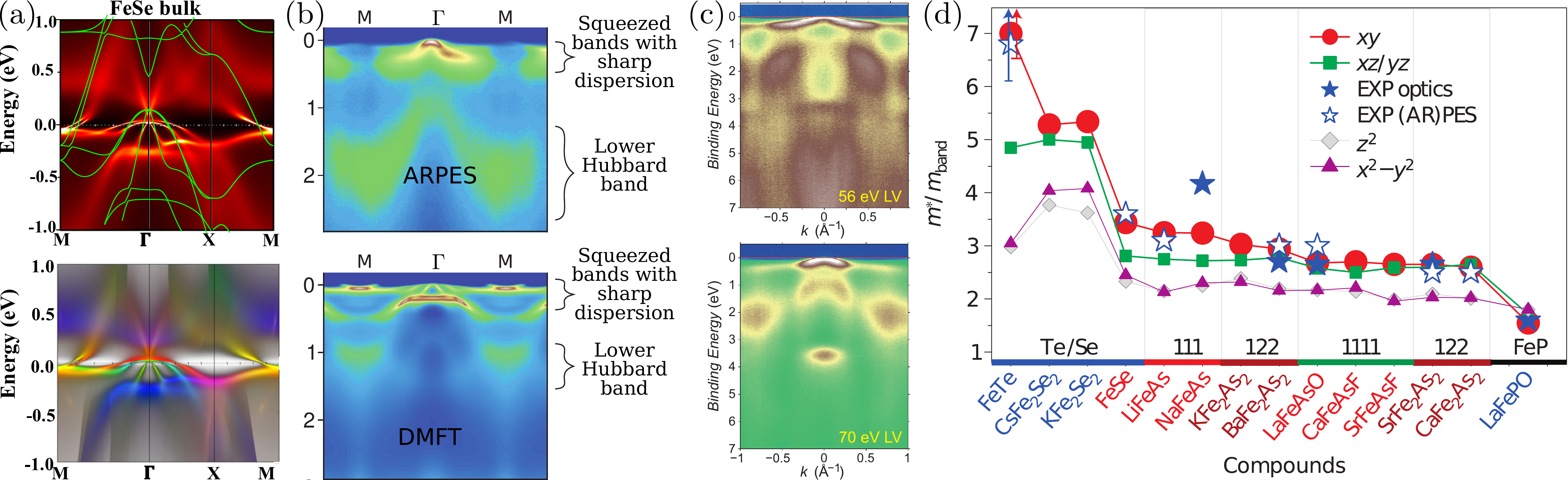}}
\caption{Effects of correlations in FeSe and related compounds.
(\textbf{a}) Spectral function for FeSe computed within a DFT + DMFT approach together with the bare band structure computed by DFT only (green lines) evidencing strong renormalization of the band structure (upper panel). Lower panel: Orbital resolved spectral function color coded. Reprinted figure with permission from   \cite{Haule16}, copyright by the American Physical Society (2017) (blue: $d_{z^2}$ and $d_{x^2-y^2}$, green: $d_{xz}$ $d_{yz}$, red: $d_{xy}$). (\textbf{b}) Details of the spectral function as measured in ARPES (top) and DMFT results (bottom) for FeSe showing the renormalized bands and the appearance of Hubbard bands as consequence of correlations~\cite{Evtushinsky2016}.
(\textbf{c})~ARPES data on FeSe in the M-$\Gamma$-M direction at $T=10$ K measured
at different incident photon energies and plotted to high binding energies, where broad features are found. Reproduced from~\cite{Watson_hubbard2017}. \ccfour.
(\textbf{d}) Relative mass enhancements for a number of different compounds showing the significance of correlations in FeSC and the trend of increasing correlations for the chalcogenide materials and FeTe (reprinted with permission from Springer  \cite{Yin2011}, copyright (2011)); see also reference~\cite{deMedici2014}.}
\label{fig:DMFT}
\end{figure*}

In this work we
concentrate on the Fe chalcogenides, with the focus on the FeSe system and its cousin
materials created by replacing Se by sulfur or tellurium, electron-doping by intercalation, preparation of thin films, and application of pressure.   FeSe with excess Fe was discovered to be an 8~K superconductor in 2008  \cite{Hsu2008}, but a cold vapor deposition technique was required to reliably make high-quality, stoichiometric, crystals  \cite{Boehmer2013}.
A summary of interesting properties of this compound  has been provided in earlier reviews \cite{Yoshikazu2011,Boehmer_2017,Coldea_2017_review}, reviews
on monolayer FeSe can be found in references~\mbox{\cite{Sadovskii2016,Wang2017,Huang2017,Liu_2020}.}
As for FeTe, it turns out to be the most stable compound of the 11 chalcogenides in its pristine form~\cite{Mizuguchi2010II},  which may, together with the presence of interstitial Fe,  be responsible for the difficulty of making homogeneous samples doped with Se away from the FeTe point.
FeTe  exhibits a double stripe magnetic structure,  appears to be more strongly correlated than the other FeSC (see Figure \ref{fig:DMFT}d), and is expected to have larger spin-orbit coupling (SOC).

As opposed to most other FeSCs, FeSe develops no static magnetic order at ambient pressure.  At high temperature, the crystal is tetragonal, but makes a transition to an orthorhombic structure below 90 K, and undergoes no further ordering until the superconducting transition at 8--9 K. The entire phase immediately below the structural transition is referred to as the nematic phase, displaying very strongly anisotropic responses to external fields although the change in the lattice constant at the transition is only about 0.1\%.
The reasons for the absence of static magnetism and the microscopic origin and nature of the dominant nematic order  are the subject of considerable debate, which we do not attempt to discuss or resolve here.

Angle-resolved photoemission spectroscopy (ARPES), scanning tunneling microscopy (STM) and quantum oscillations (QO) show that FeSe bulk crystals exhibit
tiny hole and electron pockets at the Fermi surface quite different from other FeSCs and very different from the results of standard first-principles calculations. We discuss the spectroscopic  data in Section \ref{subsec:bandstructure} below. The exact description of the low-energy electronic structure and  Fermi surface is under intense debate at the time of writing.
However, a few qualitative aspects are clear.  First of all, the correlation effects appear to be quite strong relative to the pnictides, a conclusion reflected in a number of observables.  The effective masses in the different orbital channels, extracted by comparing a variety of observables~\cite{Yin2011,deMedici2014}, are found to be substantially larger for Fe-chalcogenides than for Fe-pnictides---see Figure~\ref{fig:DMFT}---a~trend captured quite well by dynamical mean field theory (DMFT) in general  and studied for FeSe, e.g., in references \cite{Aichhorn2010,Lanata2013,Haule16,Evtushinsky2016,Watson_hubbard2017}.  Secondly, the mass renormalizations appear to be more strongly orbitally differentiated in these materials, with the largest renormalizations occurring in the $d_{xy}$ case  \cite{Yi2017}. Finally, the size and shape of the pockets is strongly renormalized, in a manner not captured by DMFT.  The pocket shrinkage relative to DFT is observed in virtually all Fe-based superconductors, but is particularly severe in FeSe.  Several authors have recently pointed out that renormalizations of this type can be obtained only with a nonlocal treatment of the self-energy  \cite{Ortenzi2009,Zantout2019,Bhattacharyya2020}.

One aspect of the band structure of bulk FeSe which is particularly remarkable is the fact that the Fermi energies of hole and electron pockets (band extrema) in the low temperature phase near $T_c$ are quite small, of order 5--10 meV (see Section \ref{subsec:bandstructure} below).  By contrast, typical Fermi energies in pnictides are $\sim$${\cal O}$(50 meV)   This implies that the $\cal O$(2 meV) superconducting gaps observed spectroscopically in Fe-chalcogenides are nearly as large as the Fermi energies, an observation that has led to the search for effects characteristic of the BCS-BEC crossover regime.
While the most straightforward consequences of BEC behavior, a characteristically broadened specific heat transition and a pseudogap, are not
observed,
there are significant anomalies in transport and NMR that have lent credence to the suggestion~\cite{Kasahara2016}.
It is also true that the properties of a multiband system in the BCS-BEC crossover regime are not well studied; while there are many predictions for single band systems, and a number of two-band calculations, few appear to be appropriate for FeSC with both hole and electron bands present simultaneously (see, however, reference~\cite{Chubukov2016A}).  This is of course a crucial distinction,  since the chemical potential will be pinned or nearly so in a compensated system, suppressing canonical BEC crossover effects.  The discussions surrounding this fascinating possibility were reviewed recently in reference \cite{Shibauchi2020}.

In Section~\ref{subsec:gap_expt}, we review spectroscopic data from measurements capable of determining the superconducting gap structure.  Since superconductivity condenses out of a $C_2$ symmetric nematic normal state, it is not surprising that the gap function determined in experiment reflects this symmetry breaking.
The degree of anisotropy, however, is very surprising; the momentum structure of the superconducting gap is extremely distorted relative to $C_4$-symmetry despite the tiny orthorhombicity of the underlying crystal structure. This interesting property of the superconducting gap has given rise to a variety  of different theoretical suggestions for the origin of the gap structure. We regard it as a healthy development, largely driven by FeSCs, that theoretical models are competing to best describe such measured gap ``details'', as opposed merely to overall symmetry properties.  We now sketch some of these theoretical approaches.

\subsection{Theoretical Approaches to Pairing}
\label{subsec:theory_FeSC}
A model of the electronic structure for the FeSCs often employed for theoretical calculations is a multiband tight binding model with the kinetic energy
term   \cite{Graser2009,Eschrig09,Fernandes_models_review_2017}
\begin{equation}
H_0=\sum_{ij\sigma \ell \ell'} t^{\ell \ell'}_{ij} c_{i\ell\sigma}^\dagger c_{j\ell'\sigma},
\label{eq_tb}
\end{equation}
where $c_{i\ell\sigma}^\dagger$  creates an electron in Wannier orbital $\ell$ with spin $\sigma$.
Note that $\ell$ is an orbital index with $\ell\in(1, \ldots, 5)$ corresponding to the states which have dominating character of the five Fe $3d$ orbitals $(d_{xy},d_{x^2-y^2},d_{xz},d_{yz},d_{3z^2-r^2})$. Extensions of such models to include the $p$ orbitals of the pnictogen or chalcogen atoms are sometimes used, although usually not needed to describe the low energy properties.
To the kinetic energy is added a  Hubbard--Kanamori (general on-site) interaction,
\begin{eqnarray}
H_{\text{int}} &=& {U}\sum_{i,\ell}n_{i\ell\uparrow}n_{i\ell\downarrow}+{U}'\sum_{i,\ell'<\ell}n_{i\ell}n_{i\ell'}
+\nonumber\\	& + &
{J}\sum_{i,\ell'<\ell}\sum_{\sigma,\sigma'}c_{i\ell\sigma}^{\dagger}c_{i\ell'\sigma'}^{\dagger}c_{i\ell\sigma'}c_{i\ell'\sigma}
+\nonumber\\ & + &
{J}'\sum_{i,\ell'\neq\ell}c_{i\ell\uparrow}^{\dagger}c_{i\ell\downarrow}^{\dagger}c_{i\ell'\downarrow}c_{i\ell'\uparrow} , \label{H_int}
\end{eqnarray}
where  $U$ is the usual Hubbard interaction between opposite spins, $J$ is the Hund's rule exchange, $U'$ is the bare interorbital interaction, $J'$ is a pair hopping term,  and $n_{i\ell\sigma}=c_{i\ell\sigma}^\dagger c_{i\ell\sigma}$ ($n_{i\ell}=n_{i\ell\uparrow}+n_{i\ell\downarrow}$) denotes the (total) density operator.  The parameters ${U}$, ${U}'$, ${J}$, ${J}'$ are related in the case of spin rotational invariance by $U'=U-2J$, and $J=J'$, i.e., the two quantities $U$ and $J/U$ fix the interactions   \cite{Kuroki2008,a_kemper_10}.

For a given choice of the parameters, one must search for a superconducting instability.  Since $U,U',J,J'$ are all repulsive, mean field theory does not initially appear to be a useful approach, since bare interactions are repulsive.  We discuss below in Section~\ref{sec:SOC_driven} some interesting recent results which show that this is not necessarily true in the multiorbital pairing case, particularly if spin-orbit coupling is strong.  Nevertheless, the most reliable way to find an attractive pair channel is to study the effective interaction vertex in the Cooper channel generated by the exchange of particle-hole pairs.

One popular approximation to this effective vertex goes under the name of random phase approximation (RPA), and dates back to the ideas of Schrieffer  \cite{BerkSchrieffer}.  The pairing vertex is proportional to the generalized particle-hole
susceptibility in the paramagnetic state  \cite{Graser2009}
\begin{align}
\chi_{\ell_1 \ell_2 \ell_3 \ell_4}^0 (q) & = - \sum_{k,\mu,\nu } M_{\ell_1 \ell_2 \ell_3 \ell_4}^{\mu\nu} (\k,\q)  G^{\mu} (k+q) G^{\nu} (k),   \label{eqn_supersuscept}
\end{align}
where we have adopted the shorthand notation $k\equiv (\k,\omega_n)$ for the momentum and frequency. The weight factors $M$ are given by
\begin{align}
M_{\ell_1 \ell_2 \ell_3 \ell_4}^{\mu\nu} (\k,\q)& =& a_\nu^{\ell_4} (\k) a_\nu^{\ell_2,*} (\k) a_\mu^{\ell_1} (\k+\q) a_\mu^{\ell_3,*} (\k+\q),\nonumber 
\end{align}
where the $a_\nu^\ell(\k)$ are the matrix elements of the unitary transformation that diagonalize the kinetic energy.
The Green's function describing band $\mu$ is given by
\begin{align}
G^\mu(\k,\omega_n)=[{i \omega_n -  E_\mu(\k)}]^{-1}.
\end{align}

Calculating the interacting susceptibility within RPA, where bubble diagrams are included, one gets
\begin{align}
\label{eqn:RPA}
\chi_{1\,\ell_1\ell_2\ell_3\ell_4}^{\rm RPA} (\q,\omega) &= \left\{\chi^0 (\q,\omega) \left[1 -\bar U^s \chi^0 (\q,\omega) \right]^{-1} \right\}_{\ell_1\ell_2\ell_3\ell_4}
\end{align}
and
\begin{widetext}
\begin{eqnarray}\label{eq_fullGammanodress}
{\Gamma}_{\ell_1\ell_2\ell_3\ell_4} (\k,\k')\! &=&\!\frac 12 \left[3\bar U^s \chi_1^{\rm RPA} (\k-\k') \bar U^s 
+   \bar U^s - \bar U^c \chi_0^{\rm RPA} (\k-\k') \bar U^c +  \bar U^c \right]_{\ell_1\ell_2\ell_3\ell_4} \\
{\Gamma}_{\nu\mu} (\k,\k')  &=& \mathrm{Re}\sum_{\ell_1\ell_2\ell_3\ell_4} a_{\nu}^{\ell_1,*}(\k) a_{\nu}^{\ell_4,*}(-\k)
{\Gamma}_{\ell_1\ell_2\ell_3\ell_4} (\k,\k') \;  a_{\mu}^{\ell_2}(\k') a_{\mu}^{\ell_3}(-\k')\,
\end{eqnarray}
\end{widetext}
are the pairing vertices in the orbital and band basis, respectively. Here $\bar U^c$ is the analog of the interaction in the charge channel and $\chi_0^{\rm RPA}$ is the corresponding charge susceptibility
in the RPA approximation   \cite{Graser2009}.
The susceptibility is then approximated by the static
susceptibility, i.e., at zero frequency, and the appearance of a superconducting instability can be sought by solving the linearized gap equation
\begin{equation}\label{eqn:gapeqn}
-\frac{1}{V_G}  \sum_\mu\int_{\text{FS}_\mu}dS'\; \Gamma_{\nu\mu}(\k,\k') \frac{ g_i(\k')}{|v_{\text{F}\mu}(\k')|}=\lambda_i g_{i}(\k)\,
\end{equation}
for the eigenvalues $\lambda_i$ and the eigenvectors $g_i(\k)$,  where FS$_\mu$ and $v_{F \mu}$ are the Fermi pocket and Fermi velocity corresponding to band $\mu$, respectively. 

\begin{figure*}[tb]
\centering
{\includegraphics[width=\textwidth]{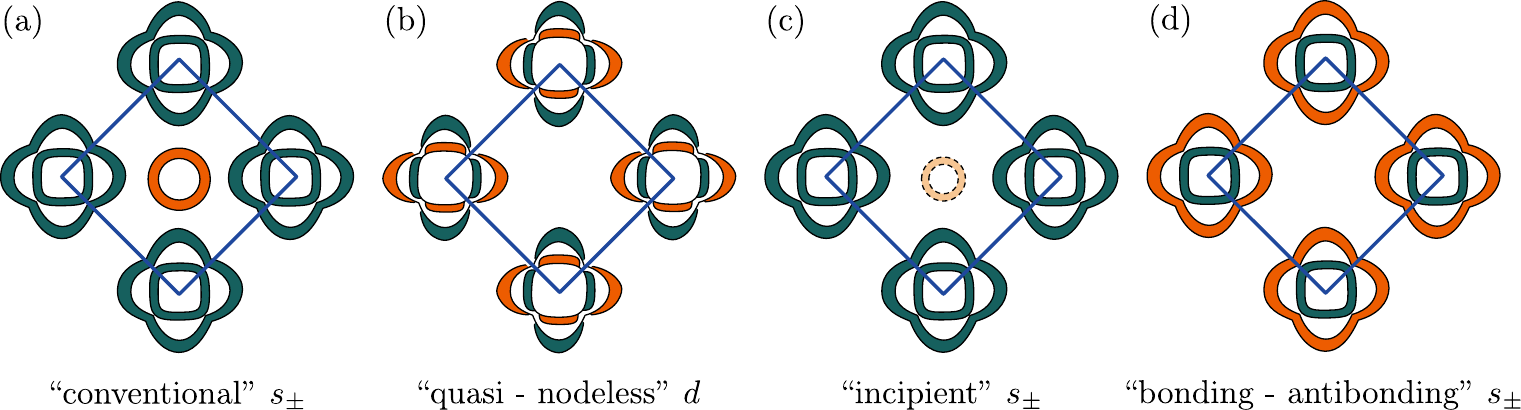}}
\caption{Schematic pictures of candidate
superconducting order parameters depicted in 2-Fe Brillouin zone for a tetragonal system.
Gaps depicted by the thickness of the green ($\Delta>0$) and orange ($\Delta<0$) lines on simple Fermi surface pockets. (\textbf{a}) Conventional $s_\pm$ state is driven by strong pair scattering between inner hole pocket(s) and electron pockets.  Subsequent states depicted do not have hole pockets at Fermi level. (\textbf{b}) $d$-wave is driven by  scattering between electron pockets, (\textbf{c}) ``incipient $s_\pm $ is driven by resonantly enhanced scattering processes connecting incipient hole band states with the electron states at the Fermi level (Section~\ref{subsubsec:incipient}), and (\textbf{d}) ``bonding--antibonding'' $s_\pm$ is driven by scattering between electron pockets supplemented by strong hybridization.}
\label{fig:gaps}
\end{figure*}

The mathematics of the pairing interaction in the multiorbital system are straightforward but not completely transparent.  Nevertheless, it is relatively easy to anticipate what kinds of gap structures may be favored in a given situation by examining the structure of the generalized susceptibility.  Scattering processes $\ell,\k\rightarrow \ell',\k'$ are favored if they nest an electron with a hole pocket or, more generally, connect Fermi surface segments with opposite Fermi velocity.  Interorbital scattering processes are suppressed relative to intraorbital ones  \cite{Zhang2009,a_kemper_10}.  For a Fermi surface like that of  Figure~\ref{fig:FS_orbitals}a, which shows a standard pnictide like Fermi surface at $k_z=0$, obvious scattering processes include that at $(\pi,0)$ from the $d_{yz}$ section of the inner $\Gamma$-centered hole pocket to the $d_{yz}$ section of the electron pocket at $M_X$, which drives the $s_\pm$ interaction leading to the first state shown in Figure~\ref{fig:gaps}a.  This is, roughly speaking, the situation in BaFe$_2$As$_2$, and explains the fact that the inner hole pocket and electron pocket gaps are the largest.

In FeSe, however, stronger interactions renormalize the inner hole band downward, leaving only a very small vestige of the outer hole pocket, and shrink the electron pockets correspondingly, so the $s_\pm$ process is less favored.  When electron doping effects are included, the hole pockets can disappear completely from the Fermi level, but leave a residual interaction with the incipient outer hole band (see Section~\ref{subsubsec:incipient}; state also depicted in Figure~\ref{fig:gaps}c). Competing with the e-h  process is the $d_{xy}\rightarrow d_{xy}$ process between the electron pockets, a scattering vector parallel to $(\pi,\pi)$ but smaller in magnitude, which drives both the $d$-wave and, including hybridization between the two electron pockets, the bonding--antibonding $s_\pm$ with gap sign change between the electron pockets also shown in Figure~\ref{fig:gaps}d.

\section{Bulk FeSe\label{sec:properties}}
\subsection{Electronic Structure of FeSe}
\label{subsec:bandstructure}

From a general perspective, the electronic structure of FeSe is very similar to
other FeSCs in the sense that the states at the Fermi level are mostly of Fe-d character, where states of $d_{xy}$ and $d_{xz/yz}$ symmetry dominate at the Fermi level.
This picture was established initially  theoretically within DFT, and also been verified experimentally. However, the electronic structure of FeSe is more complex than anticipated, and because band energy scales are very small, it evolves with temperature even more than typical FeSCs  \cite{Brouet2013,Kushnirenko_2017}.
One important issue that has received too little attention relates to the $d_{xy}$ band that sometimes results in a $\Gamma$-centered hole pocket in FeSC, e.g., in LiFeAs.  According to ARPES  it appears that a band of $d_{xy}$ character does  not cross the Fermi level, while ab-initio calculations predict the existence of such a Fermi surface sheet  \cite{Eschrig09}, i.e., realistic models for the electronic structure cannot be derived from those ab-initio calculations.

Indirect measurements of the electronic structure as a function of temperature are magnetotransport investigations where a sharp increase of the resistance below $T_\mathrm{s}$  \cite{Huynh2014,Knoener2015} was found. This was subsequently  interpreted as changes in the carrier density and the mobility in the orthorhombic phase  \cite{Huynh2014,Sun2016II,Sun2016III,Terashima2016II,Watson2015II,Ovchenkov2017} which in this view point towards drastic changes of the low energy properties and the Fermi surface.

To account for such changes in the electronic structure upon entering the orthorhombic phase, it has been proposed that an orbital order term with a sign change of the orbital splittings from $(0,0)$ to $(\pi,0)$,
\begin{align}
H_{OO}&=\Delta_b \sum_{\mathbf k}(\cos k_x -\cos k_y)\bigl[ n_{xz,\mathbf k}+n_{yz, {\mathbf k}}\bigr]
\notag\\ &
+\Delta_s\sum_{\mathbf k}\bigl[n_{xz,{\mathbf k}}-n_{yz,{\mathbf k}}\bigr]\,,
\label{eq_oo}
\end{align}
should be present in the kinetic energy.
Here $\Delta_b$ and $\Delta_s$ are the values of a bond order and site order term  \cite{Scherer_2017, Jiang_16,Sprau2017}. In addition, an orbital ordering term in the $d_{xy}$ orbital is allowed by symmetry and has been included in some more recent works  \cite{fernandes14,Christensen2019arXiv190804889C}.
From a LDA+U perspective an
off-diagonal orbital order term that lowers the overall symmetry has been found to be the ground state, a result that would allow a band hybridization such that the Y-pocket is lifted  \cite{Long2019arXiv190703562L}.

The electronic structure has been measured by a number of ARPES investigations which have been reviewed briefly above  \cite{LiuJPCM2015,Pustovit2016,Coldea_2017_review}.
From a theoretical perspective, the spectral function
\begin{equation}
A(\k,\omega)=-\frac 1 \pi \mathop\text{Im}  G^R(\k,\omega),
\end{equation}
is measured in these experiments to a good approximation. The spectral function is turn related to the
retarded Green's function
\begin{equation}
G^R(\k,i\omega_n)=\frac{1}{i\omega_n-E_{\k}-\Sigma(\k,i\omega_n)}\label{eq_int_G}\,,
\end{equation}
where  $E_{\k}$ is bare kinetic energy, as, for example, described by the eigenvalues of Equation~(\ref{eq_tb}), and electronic correlations (or other scattering processes) are parametrized by the self energy $\Sigma(\k,i\omega_n)$.

\begin{figure*}[tb]

\centering
{\includegraphics[width=\textwidth]{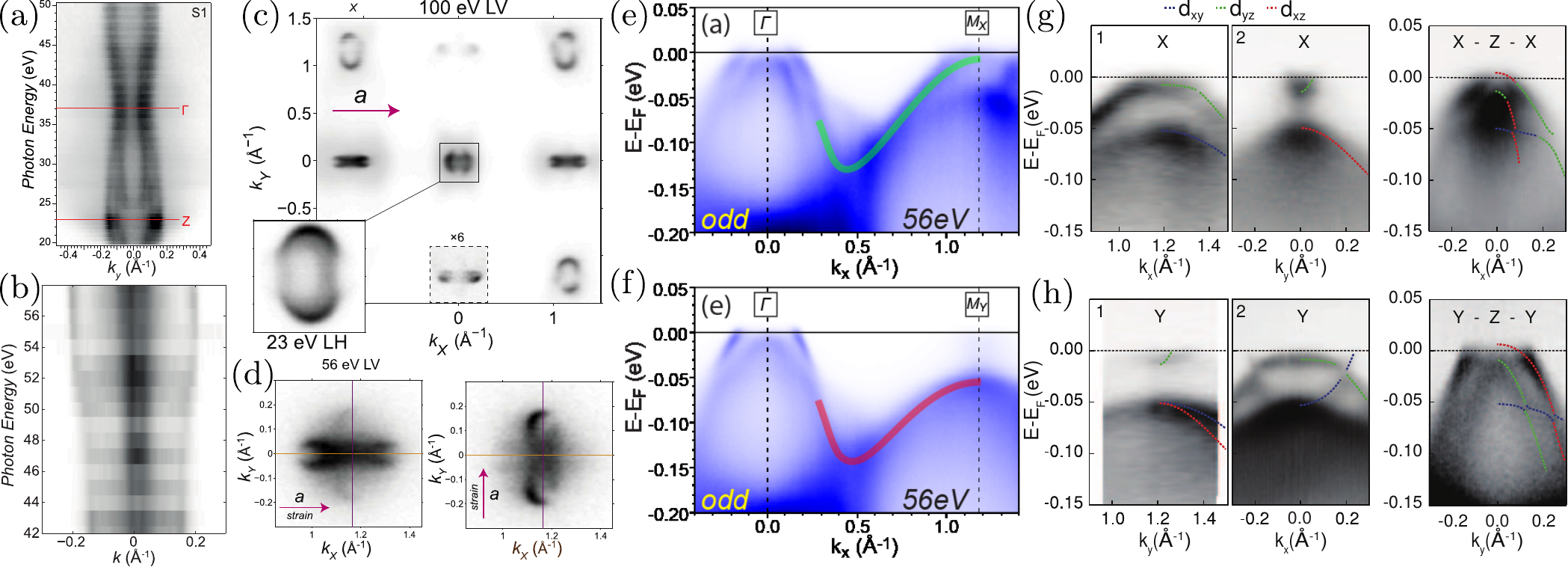}}
\caption{ARPES investigations of the electronic structure in FeSe.
(\textbf{a}) Scanning the $k_z$ dispersion of the quasi-two-dimensional
hole band of FeSe which crosses the Fermi level at 10 K by measurement of the photon-energy dependence of the
momentum distribution curves (MDC) in ARPES.
Reprinted figure with permission from   \cite{Watson2015}, copyright by the American Physical Society (2015).
(\textbf{b}) Photon-energy dependence of the MDC through the M point at 10 K, which corresponds to the $k_z$ dispersion of the electron pocket. The A point where pockets are largest corresponds to photon energy of 56 eV.
Reproduced from   \cite{Watson2016}. \ccthree.
(\textbf{c}) Fermi surface map of an ``accidentally detwinned sample'' taken at 100 eV in LV polarisation. Note that although the
selection rules alternate between the first and second $\Gamma$ points, the elongation of the hole pocket is along the b direction in both
locations. This is clarified in the inset which shows a detailed map of the hole Fermi surface.
(\textbf{d}) (left) Fermi surface map around the A point as obtained with 56 eV LV (vertical) polarisation and (right) equivalent
measurements, but with the sample rotated by 90$^\circ$.
Reproduced from   \cite{Watson2017II}.  \ccthree.
(\textbf{e}) Measured dispersions on detwinned FeSe along $\Gamma$-$M_X$ at photon energy of 56 eV (close to $k_z = \pi$) with odd polarization.
(\textbf{f}) Same measurement, but along $\Gamma$-$M_Y$ to identify the dispersion of the corresponding bands marked with red and green lines.
Reproduced from   \cite{Yi2019arXiv190304557Y}. \ccfour.
(\textbf{g}) Dispersions and orbital characters of bands in FeSe via polarization dependent ARPES (s-polarized 56 eV light). High symmetry cuts along the $k_x$- and $k_y$-directions near the X point and along the X-Z-X direction near the zone center showing the holelike band dispersions.
(\textbf{h}) Similar measurements but with the sample rotated by 90~degrees (light polarization along a-direction).
Reproduced from   \cite{Huh2019arXiv190308360H}. \ccfour.}
\label{fig:ARPES_FeSe}
\end{figure*}

We give here a summary of the findings, which are based on the
identification of the
peak of the spectral function, together with a polarization analysis.
At high temperatures in the tetragonal phase, ARPES measurements find one holelike cylinder at the $\Gamma$ point  of $d_{xz}/d_{yz}$ character (see Figure~\ref{fig:ARPES_FeSe}a). Additionally, there is one holelike band with the same orbital  character which  does not cross the Fermi level.  This band is split from the other holelike band  due to a spin-orbit coupling of 10--15 meV and therefore pushed below the Fermi surface  \cite{BorisenkoSO,Watson2016II}.  At the X and Y points in the 1-Fe Brillouin zone,
two electronlike cylinders of $d_{yz}/d_{xy}$, respectively $d_{xz}/d_{xy}$, character are present, as expected from DFT. Even in the tetragonal phase,
the sizes of the Fermi surface pockets are significantly smaller than predicted from DFT  \cite{Rhodes2017,Kushnirenko_2017}, and additionally, the predicted holelike band of $d_{xy}$ character exhibits a significant band shift  downward in energy, such that it does not cross the Fermi level at all. The need for models of the electronic structure consistent with ARPES data has lead to the proposal of  ``engineered''
models for the electronic structure where the hoppings in Equation~(\ref{eq_tb}) have been adjusted to match experimental findings of the spectral positions of the observed bands  \cite{Mukherjee2015,Kreisel2017,Sprau2017}.

Upon entering the nematic phase, the $\Gamma$ pocket is elongated and a modification of relative weights of $d_{xz}/d_{yz}$ character occurs.  The splitting to the other holelike band has been estimated as   $\sim$10~meV~\cite{Suzuki2015} (15 meV in references \cite{Watson2016,Fedorov2016}), which must be interpreted as the
combined effect of SOC and a splitting due to orbital ordering.
The latter effect can be modelled by suitable choice of the orbital order terms in Equation~(\ref{eq_oo}). One electronlike sheet at the X-point becomes peanut-shape like, as seen by experiment, while the $Y$-pocket remains quite elongated along $y$.  Experimentally, the shape, orbital character, and even the existence of the $Y$-Fermi pockets, are controversial.

Early ARPES experiments on FeSe were done on twinned samples, making it difficult to separate nematicity from the effect of averaging over twin domains.  Subsequent measurements were performed on detwinned crystals and were able to resolve the electronic structure on one orthorhombic domain; see Figure~\ref{fig:ARPES_FeSe}c,d.
Those measurements could only observe one of the the two crossed ``peanut-shaped'' electron pockets at the X point   \cite{Suzuki2015,Shimojima2014,Watson2017II,Rhodes2018}; similar conclusions were reported recently using nano ARPES  \cite{rhodes2020} within individual nematic domains. Various explanations for  this dramatic consequence of nematicity were offered, including selection rules specific to ARPES  \cite{Watson2017II}, strong correlation effects rendering some of the electronic states less coherent  \cite{Sprau2017}, unusual band shifts and hybridization of bands   \cite{Huh2019arXiv190308360H,Yi2019arXiv190304557Y}; some data of these experiments are shown in Figure~\ref{fig:ARPES_FeSe}e--g. Finally, evidence for an additional band splitting was reported, which could be due to magnetism or SOC on the surface  \cite{li2019evidence}. These issues may seem rather arcane to the newcomer to the field, but they are  essential to the goal of understanding how correlations  affect the electronic structure that is essential for deducing the pairing~interaction.

\begin{figure*}[tb]

\centering
{\includegraphics[width=\textwidth]{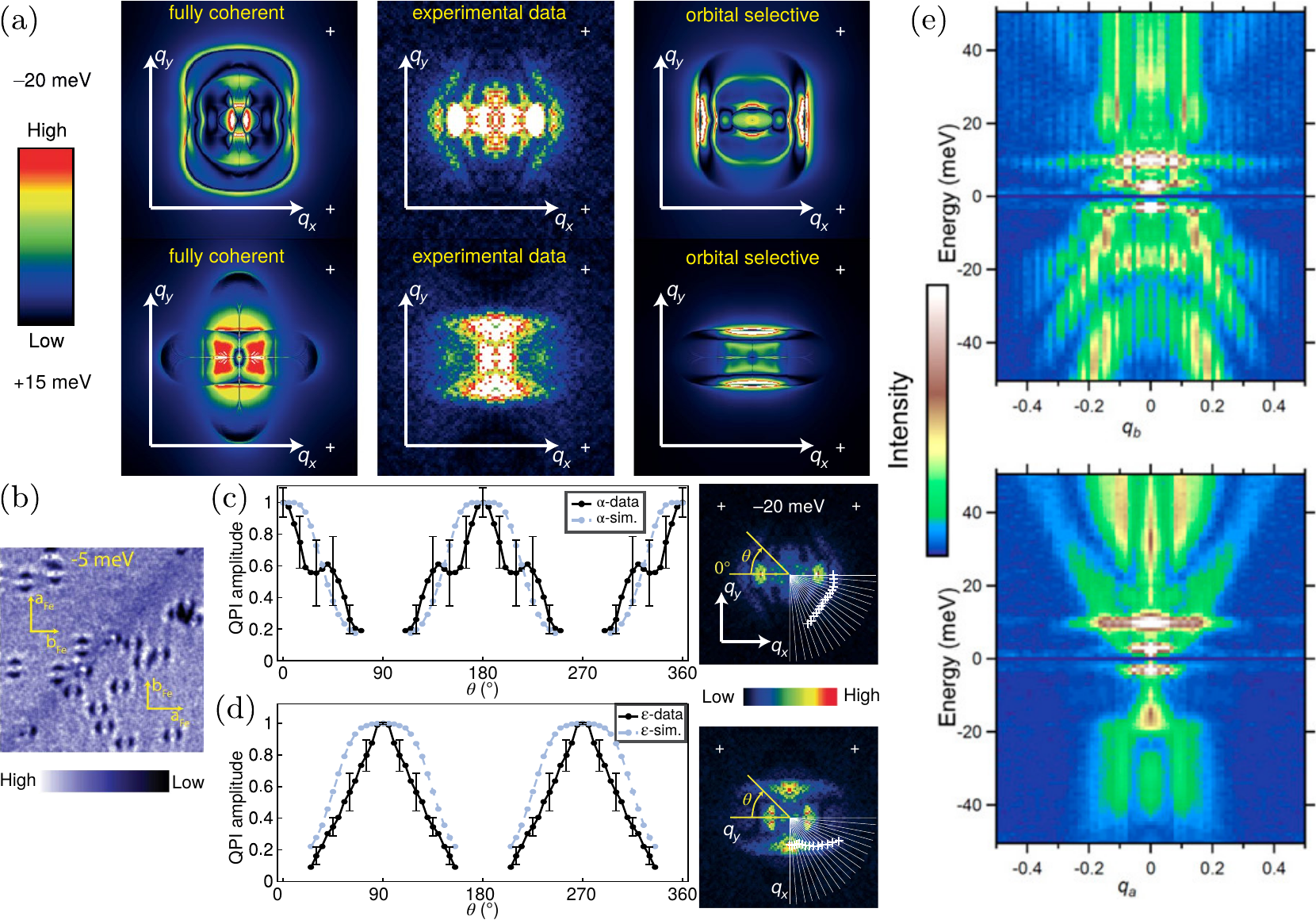}}
\caption{Quasiparticle interference investigations on the electronic structure of FeSe.
(\textbf{a}) (left) Predicted QPI
signatures of intraband scattering in a fully coherent electronic structure at two selected energies of $-$20 meV and +15 meV, (center) measured QPI in FeSe at the same energies and (right) predicted signatures assuming an orbital selective quasiparticle (OSQP) picture  \cite{Kostin2018}.
(\textbf{b})  QPI pattern in real space as standing density waves aligned horizontally or vertically show that the direction of the dominant waves rotates across the twin boundary  \cite{Kostin2018}.
(\textbf{c}) QPI intensity from the holelike band $\alpha$ and extraction of the QPI amplitude as a function of angle  around the pocket together with a simulation of the amplitude assuming OSQP.
(\textbf{d}) Similar analysis for the electronlike band $\varepsilon$. Reprinted with permission from Springer   \cite{Kostin2018}, copyright (2018).
(\textbf{e}) Band dispersions obtained from QPI pattern at zero magnetic field. The pair of sharp intensity peaks at $E=\pm 2\,\mathrm{meV}$ is due to the opening of the superconducting gap. The distinct dispersion properties along the orthogonal $q_b$  (top) and $q_a$  (bottom) directions are clearly visible  \cite{Kasahara2014}.}
\label{fig:QPI}
\end{figure*}

Recently, evidence for the second pocket at the Y point has been reported in the literature as well~\cite{Kostin2018,Kushnirenko2018} suggesting that a detailed understanding of this issue might also be related to shifts of the $d_{xz}$ and $d_{yz}$ orbital states in the orthorhombic state. Depending on the assignments of peaks in the measured spectral function to bands at the X-point, a large splitting of \mbox{50~meV~\cite{Shimojima2014,Nakayama2014,Watson2015,Huh2019arXiv190308360H}} or much smaller splitting of 10 meV  \cite{Watson2016,Watson2017II} has been deduced, while in both scenarios the sign of the splitting is reversed between the $\Gamma$ point and the X point  \cite{Suzuki2015}---similar to findings in Ba122~\cite{Pfau2019}.
The four branches of oscillation frequencies observed in quantum oscillations of the resistivity~\mbox{\cite{Terashima2014,Watson2015II,Audouard2015EPL_Hc2}} correspond to the extremely small areas of the Fermi surface sheets covering only few percent of the Brillouin zone.
Estimates for the Sommerfeld coefficient using the areas and effective masses from these investigations are in agreement with specific heat data  \cite{Terashima2014,Lin2011,Hardy2019}, {presented in Figure \ref{fig:thermodynamic_FeSe}}g. 
By assigning certain oscillation frequencies to two cylinders (from hybridized electronlike bands in the 2-Fe zone) this is consistent with the presence of the Y-pocket, while an assignment of the frequencies to maximal and minimal areas of one corrugated cylinder would agree with the absence of the Y-pocket.  
STM measurements {of quasiparticle interference (QPI)} \cite{Sprau2017,Kostin2018} (see Figure~\ref{fig:thermodynamic_FeSe}g), which
are able to measure within a single domain, agree on the size and shape of the $\Gamma$-centered Fermi pocket and the X-pocket with the deductions of the ARPES measurements.  As~expected from the layered structure of FeSe, only~a weak $k_z$ dispersion is found, rendering the pockets as weakly corrugated cylinders  \cite{Watson2015,Zhang2015_PRB} (see Figure~\ref{fig:QPI}).

Since the FeSe system exhibits a large temperature range where the nematic state is stable, it can be used to
test a number of theoretical scenarios describing nematicity in FeSC.
The nematic state in FeSe exhibits the lower (orthorhombic)
symmetry via an Ising  nematic type order parameter  \cite{Fernandes2014}.   Since conventional  softening of the lattice orthorhombicity
via a static linear coupling  was excluded as the sole driving force for the structural transition early on  \cite{Gallais2013},  spin, orbital/electronic charge instabilities of the electronic structure remain as candidates for driving the transition.  These have been studied  extensively by Raman scattering measurements  \cite{Gallais2013,Gallais2015III,Wu_Blumberg2016,Thorsmolle2016,Gallais2016} and  time resolved ARPES  \cite{Shimojima2019,fanfarillo2019photoinduced}.   A frequently encountered argument in favor of orbital/charge fluctuations begins by noting that while in FeSe  lattice distortion, elastic softening and elasto-
resistivity measurements associated with the structural
transition at $T_s$ are comparable to other FeSC  \cite{Fernandes_models_review_2017}, nuclear magnetic resonance (NMR) and
inelastic neutron scattering measurements do not detect
sizable low energy spin
fluctuations above $T_s$ (as, e.g., in Ba122)  \cite{Imai2009,Baek2016}.   However, we argue below in Section~\ref{subsec:magnetic} that there is a simple explanation for the apparent ``lack'' of high temperature fluctuations, so the spin nematic explanation cannot be ruled out on this basis.

The interacting electron gas with multiple orbital degrees of freedom can be unstable
against unequal occupation of the $d_{xz}$ and the $d_{yz}$ orbital. Taking into account
nearest neighbor Coulomb interactions of strength $V$,
\begin{equation}
\label{eq:nncoul}
H_{V} = V \sum_{\langle i, j \rangle ,\ell,\ell'} n_{i \ell} n_{j \ell'}\,,
\end{equation}
it turns out that the general changes in low temperature electronic structure
can be well described by a mean field approach already  \cite{Scherer_2017,Jiang_16}, giving rise to the  orbital order terms described in Equation~(\ref{eq_oo}). An alternative
explanation for differing signs of orbital   order on $\Gamma$ and $X,Y$ has been revealed by a renormalization group analysis, where a solution of this type was shown to be driven by the $d$-wave orbital channel  \cite{Xing2018}.

Another theoretical approach  \cite{Fanfarillo2016} involves starting from anisotropic spin fluctuations in the nematic phase, parametrizing them by a bosonic spin-fluctuation propagator of the form
\begin{equation}
\label{bapp}
B_{X/Y}(\omega)=\frac{1}{\pi}\frac{\omega\omega_{0}}{(\omega^{X/Y}_{sf}(T))^2+\Omega^2}\,.
\end{equation}
Here $\omega_0$ is a constant while $\omega^{X/Y}_{sf}(T)=\omega_0(1+T/T_{\theta})$ is the characteristic energy scale of spin modes at the $X$ and $Y$ point of the Brillouin zone.  Such an approach, using  Equation~(\ref{bapp}) to calculate the interband self-energy \`a la Ortenzi et al.  \cite{Ortenzi2009}, is also capable of qualitatively capturing the evolution of the electronic structure upon entering the nematic state, while at the same time describes the basic properties of the Fermi surface shrinking  despite neglect of the full momentum dependence of the spin-fluctuation propagator  \cite{Fanfarillo2016}.
In the same framework, the shifts of bands relative to DFT results can be explained qualitatively; these together with  strongly anisotropic spin fluctuations
might be crucial to understand the superconducting order parameter  \cite{Benfatto2018,Kreisel2017}.

In addition to band shifts and orbital order, another important effect of the one-particle self-energy is to create  decoherence, i.e., a reduction in quasiparticle spectral weight.
The effective mass $m^*$ is strongly renormalized,
\begin{equation}
\frac{m}{m^*}=  \left.\frac{1+\frac{\partial }{\partial E_\k} \mathop{\mathrm{Re}}\Sigma(\k,\omega)}
{1-\frac{\partial }{\partial \omega} \mathop{\mathrm{Re}}\Sigma(\k,\omega)}\right |_{E_\k=0,\, \omega=0},
\label{eq_def_m_star}
\end{equation}
and the quasiparticle weight on the Fermi surface (FS)
\begin{equation}
Z_{\k}=\Biggl(1-\frac{\partial\mathop{\mathrm{Re}} \Sigma(\k,\omega)}{\partial\omega}\Bigr|_{\mathrm {FS}}\Biggr)^{-1},
\label{eq_def_Z}
\end{equation} deviates from unity,
although the latter is considerably more difficult to { measure}. It is well known that
these effects depend significantly on  orbital channel, and that  $d_{xy}$ orbital states are generally the most strongly correlated  \cite{deMedici2014,Medici_review, Yin2011,Haule16,Yi2017}; see Figure~\ref{fig:DMFT}d.  In the nematic phase, the renormalizations of the $d_{xz}$ and $d_{yz}$ orbitals will generally be different.  Whether the ARPES data imply   a strongly differentiated
coherence between the $d_{xz}$ and the $d_{yz}$ orbital  \cite{Kreisel2017}, e.g., to understand the absence of the Y-pocket,
is currently  controversial.
Experimental investigations have proposed  as an alternative explanation strong shifts of the bands close to the Y point together with an orbital hybridization such that
the Y-pocket is diminished or a gap opens  \cite{Huh2019arXiv190308360H,Yi2019arXiv190304557Y}.   Within a tight-binding approach, it is, however, difficult to construct a hybridization term that will lift the Y-pocket entirely away from the Fermi surface, while preserving all symmetries of the crystal.

A combination of decoherence of $d_{xy}$ states  (without significant $d_{xz,yz}$ decoherence), nematic order, spin-orbit coupling and/or surface hybridization has been proposed to account for the  observed configuration of the bands at the $X$ point  \cite{Christensen2019arXiv190804889C}. The second pocket (in the one Fe zone appearing at the $Y$ point) is in this scenario of dominant $d_{xy}$ character everywhere in the nematic state (instead of $d_{xz}$) and difficult to observe spectroscopically due to its decoherence.

Finally, recent experimental measurements using neutron scattering and X-ray diffraction, together with an analysis of the pair distribution function found that short range orthorhombic distortions even at high temperatures can account for
the experimental findings  \cite{Koch_2019,Frandsen_2019}. Possible theoretical scenarios for this observation might be locally induced nematicity by impurities  \cite{Steffensen_2019} or local displacements of the atomic positions from its ideal tetragonal symmetry positions which are found to be stabilized within a DFT calculation using large elementary cells of overall tetragonal symmetry. The latter scenario was also found account for unusual band shifts at the $\Gamma$ point  \cite{Wang2019arXiv191102670W},  in particular the suppression of the $xy$ band.  However, in this formulation it is not  a priori clear what makes the electronic structure of FeSe special.

\subsection{Magnetic Properties\label{subsec:magnetic}}

\subsubsection{Long Range Order}

\label{subsubsec:lro}

Unlike most Fe-pnictide parent compounds, bulk FeSe  does not order magnetically at ambient pressure,  but magnetic order occurs with the application of relatively small pressures (Section~\ref{sec:pressure}), suggesting that ordered magnetism is ``nearby'', and that strong magnetic correlations play a significant role.  The statement that no long-range order exists is based primarily on neutron diffraction experiments 
\cite{McQueen2009,Wang2015}, which observe no magnetic Bragg peaks.  However, more exotic magnetic states with quadrupolar magnetic order, resulting from competition of various dipolar states, have been proposed  \cite{Wang_Z_16,Hu_2016arXiv160601235H}.  Such order would only be visible indirectly in a typical diffraction experiment.

\begin{figure*}[tb]

\centering
\includegraphics[width=\linewidth]{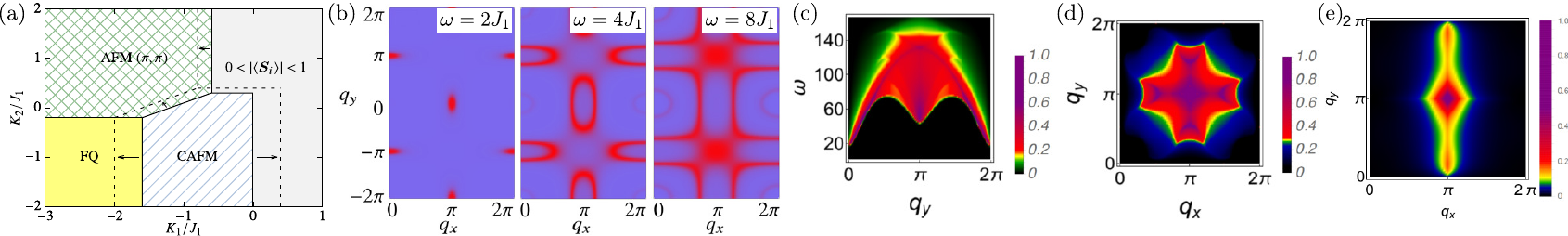}
\caption{Spin fluctuations in FeSe from localized models (\textbf{a}) Variational mean-field phase diagram for a model of localized $S=1$ spins with bilinear ($J_i$) and biquadratic ($K_i$) Heisenberg interactions. Phases are ferroquadrupolar order (FQ), antiferromagnetic N\'eel order (AFM), and columnar antiferromagnetic order (CAFM).
Breaking of the $C_4$ shifts the phase boundaries (dashed line)  \cite{Wang_Z_16}. (\textbf{b}) Expected dynamical structure factor in the FQ phase at different energies.
Reprinted figure with permission from   \cite{Wang_Z_16}, copyright by the American Physical Society (2016).
(\textbf{c}--\textbf{e}) Magnetic structure factor from a Schwinger boson mean field theory calculation in the nematic spin liquid phase plotted along a momentum path (\textbf{c}), in the Brillouin zone for fixed energy of (\textbf{d}) 50 meV, summed over two nematic domains and (\textbf{e}) in a single domain. Reprinted figure with permission from   \cite{She2018}, copyright by the American Physical Society (2018).}
\label{fig:localized_theory}
\end{figure*}

To explain the background of such proposals, we note that magnetism in FeSCs has frequently been discussed in terms of a Heisenberg model with localized spins, which indeed can describe the spin-wave modes in the  observed ordered phases.  To account for the lack of dipolar order and aspects of the low-energy spin modes in FeSe,  the Hamiltonian must contain bilinear and biquadratic couplings of spin operators $\mathbf{S}_i$,
\begin{align}\label{Eq:HamTot}
H &= \frac{1}{2}\sum_{i,\delta_n}
\left\{J_n \mathbf{S}_i\cdot\mathbf{S}_j + K_n (\mathbf{S}_i\cdot\mathbf{S}_j)^2\right\}.
\end{align}
Here $j=i+\delta_n$, and $\delta_n$ connects site $i$ and its $n$-th
nearest neighbor sites.
Frustration among competing magnetic states was proposed to explain the absence of magnetic order by Glasbrenner et al.  \cite{Glasbrenner2015}, who compared the energies of various stripe and N\'eel states within DFT, and showed that they were within a few meV of each other for FeSe, whereas in Ba122 and other pnictides, the simple $(\pi,0)$ stripe state was  lower in energy than competing states by a large margin.  They then discussed the competition among these states within a localized spin model with biquadratic exchange, and showed that estimates from ab-initio approaches of the coefficients $J,K$ in Equation~(\ref{Eq:HamTot}) put FeSe near a multicritical point in the magnetic phase diagram where several stripelike states were nearly degenerate.  It was argued that under these circumstances, quantum fluctuations would prevent ordering. Intriguingly, Glasbrenner~et~al. also noted that all such states were consistent with the observed nematic order, suggesting that the robust nematic order observed in FeSe was also a consequence of these magnetic fluctuations.
Other groups have sought explanations  starting from the same spin model, but argued
that further
frustration in the biquadratic couplings including $K_n$ up to $n=3$ can explain the absence of dipolar magnetism in FeSe  \cite{Wang_Z_16} and stabilize quadrupolar order; see phase diagram in Figure~\ref{fig:localized_theory}a. To our knowledge, there is no definitive evidence for such order in experiment.
The suppression of the biquadratic couplings upon application of pressure on FeSe should make the magnetic order reappear, as proposed recently  \cite{Hu_2016arXiv160601235H}.
Similar ideas were put forward independently in a description of FeSe as a paradigmatic quantum paramagnet  \cite{WangLee2015NatPhys_FeSe-Para-Nematicity}.    Finally,  this type of argument was also advanced in the context of the monolayer FeSe when calculating Boltzmann weighted spectra for different spiral magnetic configurations of similar energy  \cite{Shishidou2018}.

\subsubsection{Spin Fluctuations in Normal State}
\label{subsubsec:sf_normal}

\begin{figure*}[tb]

\centering
\includegraphics[width=\textwidth]{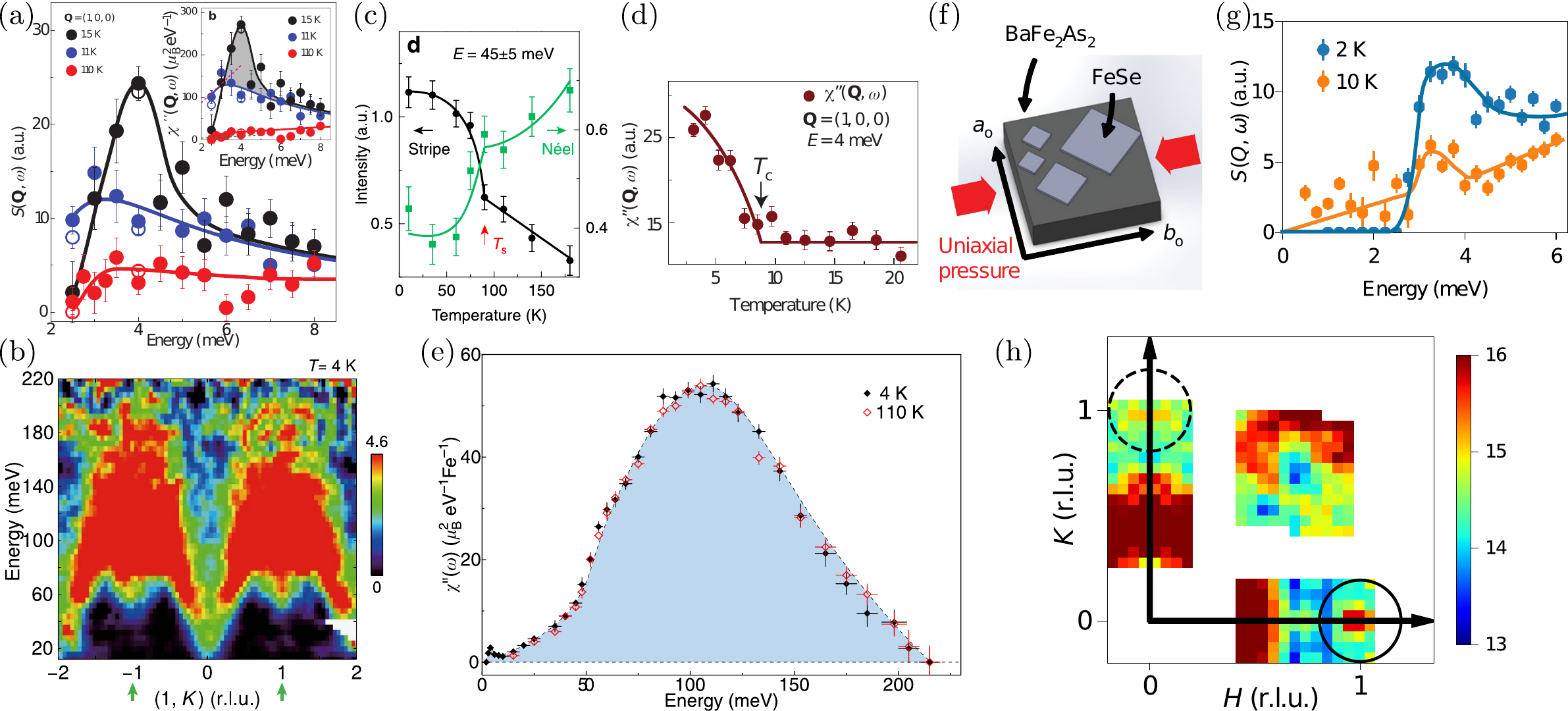}
\caption{Neutron scattering investigations on FeSe.
(\textbf{a}) Energy dependence of spin fluctuations for FeSe in the
superconducting and normal state. The dynamic spin correlation function $S(\mathbf{Q},\omega)$
exhibits an enhancement at low temperatures and a resonance feature. (inset) Bose-population
factor corrected and normalized data with the resonance spectral weight as the shaded area.
Reprinted with permission from Springer   \cite{Wang2015}, copyright (2016).
(\textbf{b}) Dispersions of the stripe and Ne\'eel spin fluctuations in FeSe at 4 K visualized as energy-momentum
slice of the spin fluctuations in units of mbar sr$^{-1}$meV$^{-1}$f.u.$^{-1}$~\cite{Wang2015GS}.
(\textbf{c}) Temperature dependence of the intensities of the spin fluctuations showing a weight transfer from N\'eel to stripe upon cooling.
Reproduced from   \cite{Wang2015GS}. \ccfour.
(\textbf{d})~The temperature dependence of the dynamic spin correlation shows a kink at $T_c$  \cite{Wang2015}.
(\textbf{e}) Sum of the stripe and N\'eel spin fluctuations $\propto \sum_\q \chi ``(\mathbf{Q},\omega)$, showing the resonance mode at small energies, but few differences of the overall local susceptibility.
(\textbf{f}) Neutron scattering on detwinned FeSe: Sample arrangement with FeSe crystals glued on large single crystals of  BaFe$_2$As$_2$ which are put under a uniaxial pressure  \cite{Chen2019}.
(\textbf{g}) Scattering intensity $S(\mathbf{Q}, \omega)$ integrated around (1, 0) above ($T = 10$ K) and below ($T = 2$ K) $T_c$
on twinned FeSe.
(\textbf{h}) Normal-state spin fluctuations in detwinned FeSe on a selected energy of $E=8.5$\,meV exhibiting strong intensity at $(1,0)$ (black circle),  and negligible intensity at (0,1) [units of $\pi/a$].
Reprinted with permission from Springer   \cite{Chen2019}, copyright (2019).}
\label{fig:neutron_FeSe}
\end{figure*}

Hints to the  microscopic origins of magnetic  correlations in FeSe  can be found in the complex temperature and momentum dependence of  magnetic fluctuations, and its imprints on the nematic state and superconductivity; see Section~\ref{subsubsec:sf_SC} below.
Experimentally, these correlations have been studied using NMR and inelastic neutron scattering experiments 
\cite{Kotegawa2012,Boehmer2015,Vaknin_16,Wang2015GS,Wang2015,Baek2015,Baek20}, with the latter  summarized in Figure~\ref{fig:neutron_FeSe}.
In the spin nematic scenario, these fluctuations are argued to drive the nematic order, eventually~leading
to a divergence of the nematic susceptibility.  Alternatively,  nematicity is proposed to arise through
orbital or charge fluctuations  \cite{Baek2015,Gallais2013}.
Recall the NMR data on FeSe~\cite{Imai2009,Baek2015} are rather different from the Fe-pnictides, where a strong upturn in the spin-lattice relaxation time $1/(T_1T)$ beginning well above $T_s$ is taken to signal the onset of strong spin fluctuations at high temperature.  In~FeSe, this upturn is visible only below $T_s$ and just
above $T_c$, suggesting rather weak spin fluctuations in the vicinity of the nematic transition, and leading to suggestions of the primacy of orbital fluctuations.

On the other hand, the spin fluctuations of bulk FeSe have been measured in detail using inelastic neutron scattering on powder samples  \cite{Rahn2015} and
(twinned) crystals  \cite{Wang2015,Wang2015GS,Ma_2017},
revealing a complex dependence on temperature and momentum transfer.
Magnetic fluctuations of stripe-type and N\'eel-type were detected  \cite{Wang2015GS}, revealing a transfer of spectral weight at  energies $\lesssim 60$ meV away from N\'eel-type fluctuations as temperature decreased and the system entered the nematic phase; see Figure~\ref{fig:neutron_FeSe}c. A large fluctuating moment of \mbox{$\sim 5.1$ $\mu_\mathrm{B}^2$/Fe}, corresponding to an effective spin of $S\sim0.74$ was estimated  \cite{Wang2015GS},  which is almost unchanged from high temperatures $T>T_s$ to very low temperatures, as evidenced by the local susceptibility presented in Figure~\ref{fig:neutron_FeSe}e.
The overall bandwidth is found to be smaller than in 122-type FeSC systems, and a sizeable low energy spectral weight grows below $T_\mathrm{s}$  \cite{Wang2015} which agrees with findings from NMR  \cite{Imai2009,Baek2015,Vaknin_16} and is in line with the proposal of competition between stripe-type and N\'eel-type magnetic ordering vectors as suggested also by Raman spectroscopy  \cite{Baum2019}.
The presence of spin fluctuations at low energies indicates the proximity of the system to a magnetically ordered state which can be realized
by tuning the system with pressure.

In the context of the inelastic neutron data, the early NMR results on FeSe  \cite{Imai2009,Baek2015} that suggested weak spin fluctuations (Section~\ref{sec:overview})  present above $T_s$ and seemed to point to an orbital fluctuation-driven nematic transition should be re-examined.    It is important to remember that the spin-lattice relaxation is local, i.e., $1/(T_1T)\propto {\rm Im}\, \sum_{\q} \chi ``(\q,\omega)/\omega$; i.e., spin fluctuations at all wavelengths contribute. Inelastic~neutron scattering measurements show quite different temperature dependence for N\'eel $(\pi,\pi)$ and stripe $(\pi,0)$ spin fluctuations, such that at low energies $\omega$, N\'eel fluctuations dominate at high $T$, whereas stripe fluctuations dominate at low $T$, as shown in  Figure~\ref{fig:neutron_FeSe}c.  Both are summed in $1/(T_1 T)$ and since one is increasing and one decreasing with $T$ can  give a relatively flat total $T$ dependence~\cite{Mukherjee2015} as observed in experiment.    The conclusion is that the stripelike fluctuations are indeed present, and~are additionally strongly enhanced above $T_s$, just as in, e.g., Ba122; the difference with respect to the Fe-pnictides is the existence in FeSe of the strong fluctuations at other wavevectors at higher temperatures near the nematic~transition.

Since most inelastic neutron experiments have been performed on twinned crystals, it has been difficult to test whether or not the nematic state creates a significant anisotropy of spin fluctuations.
In order to discriminate in a neutron scattering experiment between fluctuations with momentum transfer $(\pi,0)$ and $(0,\pi)$, one
needs to (a) place a reasonably large amount of the sample material into the beam and (b) simultaneously detwin the crystals. This has been achieved by gluing many (small) crystals of FeSe on large crystals of Ba-122 and mechanically detwinning the Ba-122 (see Figure~\ref{fig:neutron_FeSe}f) such that
the detwinning ratio of the FeSe crystals could also be observed using elastic neutron scattering. Then, by measuring the nominal
signal from momentum transfer of  $(\pi,0)$ and $(0,\pi)$, one can correct the data and extract the spin fluctuations with momentum transfer of $(0,\pi)$, finding   no measurable signal at low energies  \cite{Chen2019} (Figure~\ref{fig:neutron_FeSe}g,h).
This approach is restricted to the energy window where the Ba-122 has a gapped spin fluctuation spectrum.

\begin{figure*}[tb]

\centering
{\includegraphics[width=\linewidth]{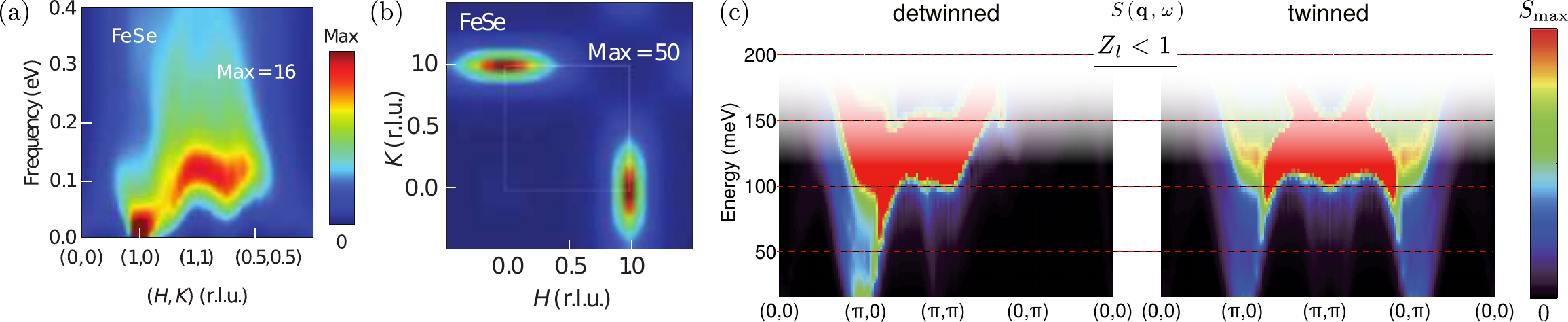}}
\caption{Spin fluctuations in FeSe.
(\textbf{a}) Dynamic spin structure factor $S(q,\omega)$ of FeSe in the tetragonal phase calculated using DFT + DMFT, plotted along high symmetry directions, and (\textbf{b}) plotted plane $(H,K)$ at $k_z=\pi$.
Reprinted with permission from Springer   \cite{Yin2014}, copyright (2014).
(\textbf{c}) Structure factor from a tight binding model using reduced coherence of some orbital channels plotted along high symmetry directions. For the twinned result, an average over two orthorhombic domains was performed.
Reprinted figure with permission from   \cite{Kreisel2018}, copyright by the American Physical Society (2018).}
\label{fig:itinerant_theory}
\end{figure*}

Thus any theory of FeSe must account for the extremely anisotropic spin fluctuation in the nematic normal state of FeSe. Any calculation of the spin-fluctuation spectrum from standard first principles methods then
has the drawback that the low temperature nematic phase is not correctly captured and any non-magnetic tetragonal calculation yields a $C_4$ symmetric spin fluctuation spectrum, as shown in Figure~\ref{fig:itinerant_theory}a,b where the corresponding dynamic structure factor shows low energy weight at $(\pi,0)$, but also at $(0,\pi)$. More phenomenological approaches indeed yield anisotropic spin fluctuations, for example, by construction  of bosonic propagators of anisotropic spin modes  \cite{Fanfarillo2016} which can be justified from a derivation of an effective action including quartic terms in the nematic order 
\cite{Fanfarillo_mismatch2018}, or by assuming
strongly orbitally selective quasiparticles with reduced coherence  \cite{Kreisel2018} (Figure~\ref{fig:itinerant_theory}c, resembling the measured spectrum on a twinned crystal, as shown in Figure~\ref{fig:neutron_FeSe}b); reduced coherence in the $d_{xy}$ orbital has also been found to be necessary to explain nematic fluctuations in Fe$_{1+y}$Te$_{1-x}$Se$_x$~\cite{jiang2020nematic}.
In~reference \cite{Fanfarillo_mismatch2018}, it was also found that
a Fermi surface with nesting between states of different orbital character (as realized in FeSe) favors nematic order, while magnetism is usually favored by nesting between states of the same orbital character,  suggested as a factor in the absence of magnetic order in FeSe.
It is unclear at present whether the proposal of changes in the orbital content  \cite{Kang2018} of the Fermi surface in the nematic state, together with strong $d_{xy}$ decoherence, can account for
the extreme spin fluctuation anisotropy unless tuned very close to the magnetic instability  \cite{Kreisel15}. Models of localized spins are capable of describing the magnetic phase diagram of FeSe and predicting $C_4$ symmetric spin fluctuations  \cite{Wang_Z_16} (Figure~\ref{fig:localized_theory}b) or a strongly anisotropic fluctuation spectrum  \cite{She2018} (see Figure~\ref{fig:localized_theory}c--e), resembling the INS data on twinned crystals presented in Figure~\ref{fig:neutron_FeSe}b.

\subsection{Superconducting Gap}
\label{subsec:gap_expt}
The superconducting transition temperature of bulk FeSe is about \mbox{8--9 K}.
The  symmetry and structure of the superconducting order parameter, in particular the existence of minima or nodes in the gap, determine the density of low energy quasiparticle excitations,  and thereby the form of low-temperature power laws in thermodynamic and transport properties. The gap structure also indirectly reflects the form of the pairing interaction.  Here we review various measurements that provide information on gap structure, theories of $T_c$ and pairing, and what conclusions may be drawn.

\subsubsection{Thermodynamic Probe of Quasiparticle Excitations}
\label{subsubsec:thermo}

\begin{figure*}[tb]

\centering
{\includegraphics[width=\textwidth]{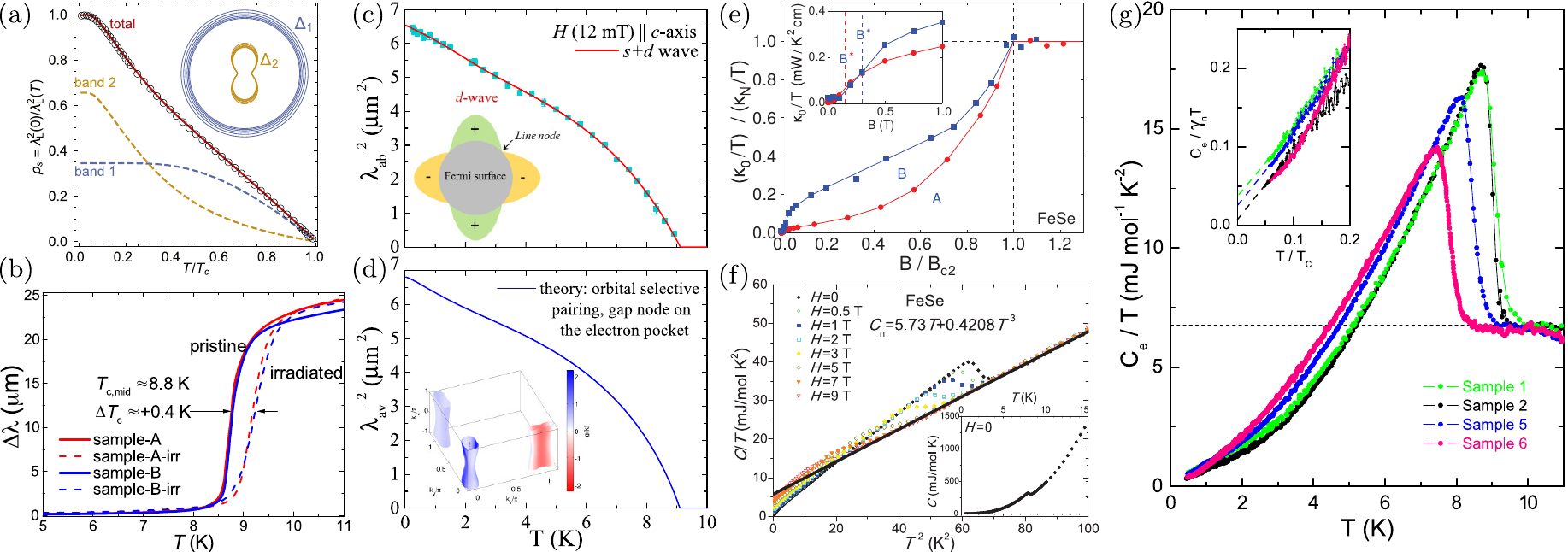}}
\caption{Thermodynamic probes of the superconducting state in FeSe.
(\textbf{a}) Superfluid density
$1/\lambda_L^2$ from microwave conductivity measurements (dots), together with a calculation using a two-band model.
Reproduced from   \cite{Li2016}. \ccthree.
(\textbf{b}) London penetration depth  $\Delta\lambda(T)$ before and after electron irradiation showing an increase of $T_c$ upon introduction of disorder.
Reprinted figure with permission from   \cite{Teknowijoyo2016}, copyright by the American Physical Society (2016).
(\textbf{c}) Temperature dependence of the inverse penetration depth $\lambda^{-2}$ FeSe in the ab plane (symbols) together with a fit using a two gap $s+d$ wave model.
Reprinted figure with permission from   \cite{Biswas2018}, copyright by the American Physical Society (2018).
(\textbf{d}) Calculation of the averaged penetration depth $\lambda_{av}^{-2}$ from
a microscopic model for the order parameter and the electronic
structure. (inset: the gap structure for the microscopic model)~\cite{Biswas2018}.
(\textbf{e}) Field dependence of the residual linear term of the thermal conductivity in FeSe
comparing two samples (A/B) with the conclusion of absence of nodes of the superconducting order parameter. The~main panel shows the normalized conductivity with the normal state value $\kappa_N$ and the inset the raw data.
Reprinted figure with permission from   \cite{Bourgeois-Hope2016}, copyright by the American Physical Society (2016).
(\textbf{f}) Specific heat $C/T$ measured for different fields $H$ from 0 to 9 T revealing the nonlinear function of the field
as evidence for the presence of multiple order parameters. (Solid line: normal state contribution with $C_n(T)=\gamma T+\beta T^3$, inset: raw $C$ vs T at zero field.)
Reprinted figure with permission from   \cite{Lin2011}, copyright by the American Physical Society (2011).
(\textbf{g}) Electronic specific heat $C_e(T)/T$ of four FeSe samples
with different amounts
of disorder consistently exhibiting the same low temperature dependence resembling the behavior of a nodal superconductor (inset: low-temperature part in normalized units). Reprinted figure with permission from   \cite{Hardy2019}, copyright by the American Physical Society (2019).}
\label{fig:thermodynamic_FeSe}
\end{figure*}

Thermodynamic measurements such as specific heat and magnetic penetration depth  probe low energy excitations, and became
more reliable once high quality crystals of FeSe with large RRR (residual resistance ratio) were available  \cite{Chareev2013,Boehmer2013,Boehmer2016II}. However, the existence of
quasiparticles at arbitrarily small energies, i.e., whether or not bulk FeSe has true gap nodes, as opposed to deep gap minima,
has been answered differently by a number of studies.
Initial measurements of the London penetration depth $\lambda_L$ by Kasahara et al.  \cite{Kasahara2014} reported a quasilinear temperature dependence for $T\ll T_c$, consistent with gap nodes, and  $\mu$SR results were also claimed to be consistent with a
nodal superconductor  \cite{Biswas2018} (Figure~\ref{fig:thermodynamic_FeSe}c). However other  measurements observed a small spectral  gap~\cite{Li2016,Teknowijoyo2016} (Figure~\ref{fig:thermodynamic_FeSe}a,b).

The jump of the specific heat at the transition temperature $T_\mathrm{c}$ of $\Delta C/\gamma_nT_\mathrm{c} = 1.65$ (Figure~\ref{fig:thermodynamic_FeSe}f) seems to indicate a moderate to strong coupling superconductor  \cite{Lin2011} because of the deviation from the expected magnitude
of a BCS superconductor,  particularly since the large gap anisotropy tends to reduce rather than enhance this ratio. Some  more recent investigations on the specific heat tried to  extract the order parameter on the different bands  \cite{Abdel-Hafiez2013,Bourgeois-Hope2016,Naidyuk2017,Yang2017,Hardy2019} by fitting procedures. Measurements of field-angle dependent specific heat  \cite{Sun_2017arXiv} found evidence for three distinct
superconducting gaps, where the two smallest appear to be anisotropic and the smallest possibly nodal.
The specific heat studies of references \cite{Wang2017II,Chen2017,Roessler2018,Hardy2019} tend to assign nodal superconductivity to FeSe (Figure~\ref{fig:thermodynamic_FeSe}g), while reference \cite{Jiao_2017} comes to the conclusion that the system is fully gapped.  Reports of thermal conductivity are similarly split on the issue of a true gap: some observe  fully gapped~\cite{Bourgeois-Hope2016,Watashige_2017} and other claim nodal   \cite{Kasahara2014} behavior.

Several authors have attempted to grapple with these apparent conflicts  \cite{Boehmer2016,Roessler2018}, by pointing out differences in low-temperature behavior according to small variations in growth techniques.  In~reference~\cite{Watashige2015}, the authors performed an STM study close to and far away from twin boundaries, pointing out that a full gap existed over rather large distance scales near the boundary, while the bulk was nodal. The authors attributed the full gap to the onset of a time-reversal symmetry breaking mixture of two irreducible representations in the pairing near twin boundaries.   While this behavior is not reproduced in all STM studies  \cite{Roessler2018}, it suggests that the density of twins, which in turn depends on sample preparation, could control thermodynamic properties at very low temperatures.

In any case, the theoretical implication is fairly clear: the observed strong sensitivity of the low-energy gap  to disorder and twin structure almost certainly reflects an order parameter with accidental nodes or near nodes, i.e., not enforced by symmetry.
The nematic phase of FeSe exhibits an orthorhombic crystal symmetry; thus the superconducting order parameter
is  necessarily a mixture of the corresponding tetragonal Brillouin zone harmonics, e.g., $s$ and $d$. Thus, no symmetry protected nodal positions  are expected, and small variations of the gap structure are possible due to differences in the sample preparation such as the local Fe:Se ratio, twin density, or internal stress.

Shallow nodes or near-nodes are then consistent, crudely speaking, with a near-degeneracy of $s$- and $d$-wave pairing in the reference tetragonal system.    It is important to note that the existence or nonexistence of a true spectral gap in the system is perhaps not the most important issue.   On the other hand, if the gap is indeed formed due to the growth of a second irreducible representation near defects, this could be an important hint to the structure of the intrinsic pairing interaction.

One way to decide this issue is to probe the superconducting state with controlled disorder.
A~rapid suppression of $T_c$ upon introduction of (nonmagnetic) impurities or detecting a bound state in STS close to such
an impurity is usually taken as evidence for a sign change of the order parameter.  At present it is not clear if  nonmagnetic impurities in FeSe are pairbreaking or not.  In Figure~\ref{fig:thermodynamic_FeSe}g, specific heat data on four samples from the Karlsruhe group are shown.  If one interprets the lower $T_c$ sample as the most disordered, as would be usual in an unconventional superconductor, the anticorrelation of $T_c$ with the residual Sommerfeld coefficient at $T\rightarrow 0$ could be interpreted as a node-lifting phenomenon, where the spectral gap opens as disorder averages the order parameter; this effect has been established in Fe-pnictides as characteristic of accidental nodes  \cite{Mishra2009}. A proton irradiation study also claimed to observe node-lifting induced by disorder  \cite{Sun2017II}.
Investigations on the field-dependence of the thermal conductivity  \cite{Bourgeois-Hope2016, Kasahara2014} and specific heat  \cite{Chen2017,Hardy2019} came to similar conclusions; for details see reference~\cite{Shibauchi2020}.  On the other hand, penetration depth measurements by Teknowijoyo et al.  \cite{Teknowijoyo2016} with controlled low-energy electron irradiation that creates Frenkel pairs of defects provided evidence that $T_c$  {\it increases }
with increasing disorder (See Figure~\ref{fig:thermodynamic_FeSe}b).  These authors considered various explanations for this remarkable result, including local enhancement of spin fluctuation pairing by impurities  \cite{Roemer2018} and competition of superconductivity and nematic order  \cite{Mishira2016}, but this question remains open.

In summary, the numerous studies agree on the point that FeSe exhibits a strongly anisotropic superconducting order parameter. Nodes, if these are detected, might
be lifted easily by external manipulations or due to disorder  \cite{Sun2018} although the critical temperature does not seem to be very sensitive to such effects  \cite{Roessler2018}.
In addition, the appearance of a feature in the specific heat at very low temperatures  \cite{Roessler2018} seems to be present in some samples only.

\subsubsection{STM/ARPES Measurements of Gap Structure}
\label{subsubsec:STM_ARPES}

Measurement of the superconducting gap in momentum space is possible using ARPES, where~the pullback of the spectral function in the superconducting state is used to obtain maps of the  gap function $\Delta_\k$ on the Fermi surface. STM measurements and the subsequent analysis of the quasiparticle interference makes use of the large partial density of states at saddle points of the Bogoliubov dispersion, and allows one to trace back the Fermi surface and measure the spectroscopic gap  in the superconducting state. The latter experimental technique relies on the interference of  quasiparticles scattered by disorder, and is additionally capable of detecting the  phase of the superconducting order parameter   \cite{Sprau2017}.

From a theoretical perspective, spectroscopy  near impurities can reveal the properties of the superconducting order parameter. A non-magnetic scatterer in a superconductor with no sign change of the order parameter cannot create impurity resonances within the superconducting
gap  \cite{BalatskyRMP,Korshunov2016}. In~a system with a sign-changing gap, the spectral position of the bound state depends on the specific value of the impurity potential of the scatterer and might not be easy to  detect.
More recently a method that does not rely on bound states has been proposed  \cite{HAEM2015,Martiny2017}. It relies on the analysis of the antisymmetrized and (partially) integrated QPI 
signal, i.e., the Fourier transformed conductance maps
as measured in STM. Other approaches to detect the sign change of the order parameter are based on similar mathematical properties of the tunneling conductance and interference effects  \cite{Chi_2017arXiv171009089C,Chi_2017arXiv171009088C,DallaTorre2016, Chen_sign_2019}.

\begin{figure*}[tb]
\centering
{\includegraphics[width=\textwidth]{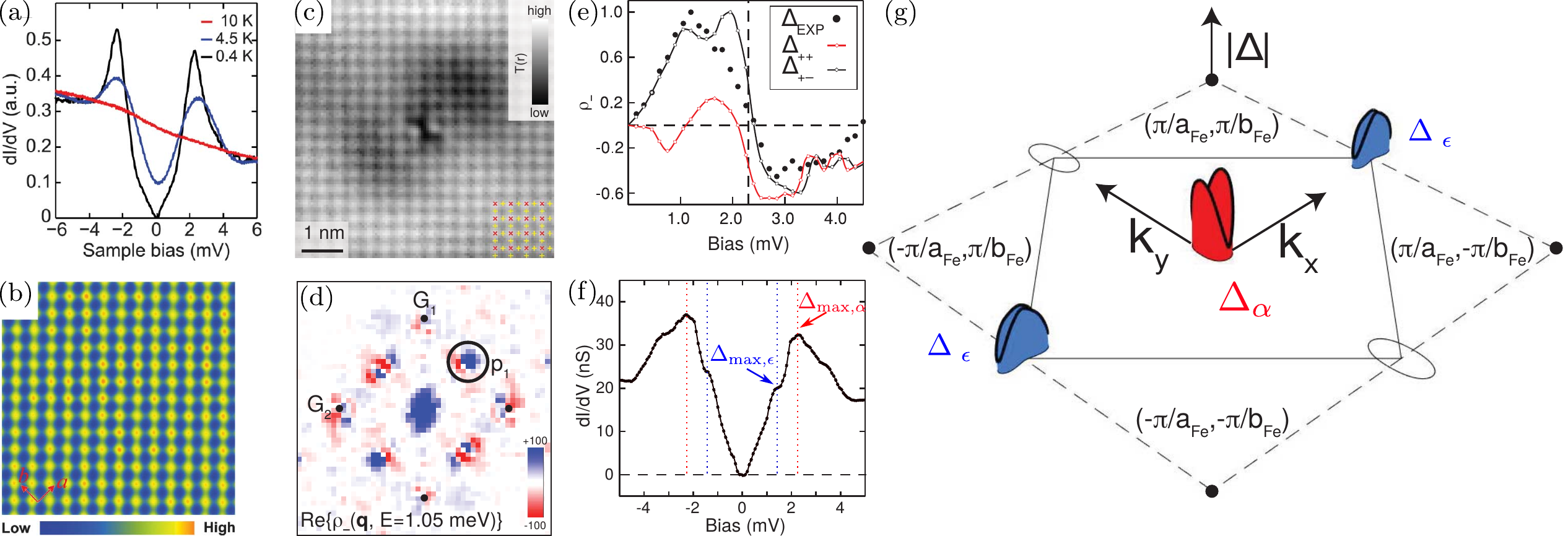}}
\caption{Measurements on FeSe in the superconducting state using STM.
(\textbf{a}) Differential conductance on FeSe (film) at different temperatures  \cite{Song2011}.
(\textbf{b}) Atomic-resolution STM topography of FeSe (film) with bright spots as Se atoms on the top layer; the $a$ and $b$ axes correspond to Fe--Fe bond directions.
From   \cite{Song2011}. Reprinted with permission from AAAS.
(\textbf{c}) Topograph centered on a typical impurity with overlay of local structure (red x: Se atoms, yellow + Fe sites)~\cite{Sprau2017}.
(\textbf{d}) Typical $\rho_-$ QPI map with integration area (black circle) corresponding to interpocket scattering that
has been used to obtain the momentum integrated $\rho_-$ in (\textbf{e}) showing clean signature of a sign changing
order parameter and no agreement to simulation data for non sign-changing order parameter (red curve)  \cite{Sprau2017}.
(\textbf{f}) High resolution differential conductance spectrum $dI/dV(E)$ exhibiting two energy scales of the maximum
energy gap two bands $\Delta_{\mathrm{max},\alpha}$ and $\Delta_{\mathrm{max},\epsilon}$  \cite{Sprau2017}.
(\textbf{g}) Summary of the measured $k$-space structure of the energy gaps of FeSe for the two pockets $\Delta_\alpha$ and $\Delta_\epsilon$ which exhibit a strong anisotropy and a sign change (red/blue color). From   \cite{Sprau2017}. Reprinted with permission from AAAS.
}
\label{fig:gap_STM_FeSe}
\end{figure*}

\begin{figure*}[tb]
\centering
{\includegraphics[width=\textwidth]{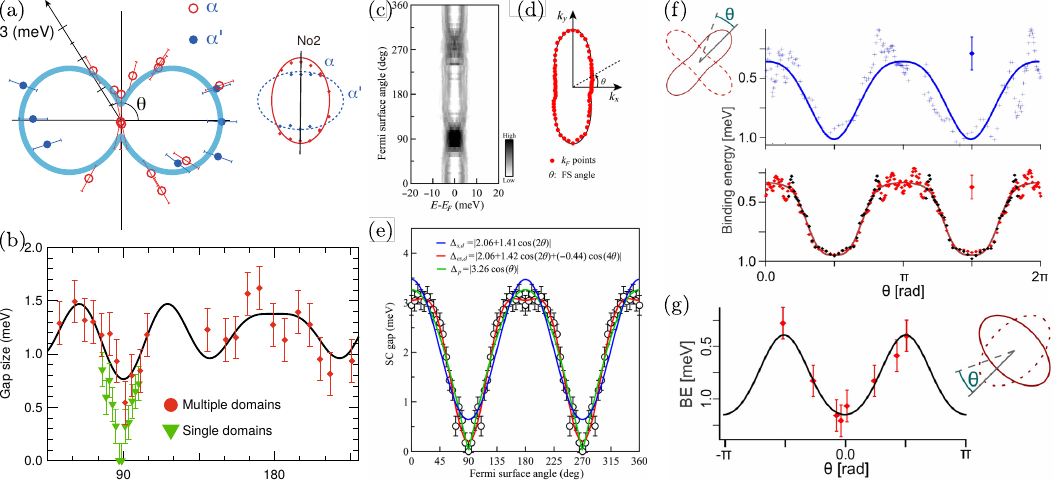}}
\caption{Measurements of the superconducting gap using ARPES: holelike pocket---(\textbf{a}--\textbf{e},\textbf{g}); electronlike pocket---(\textbf{f}).
(\textbf{a}) Superconducting gap as a function of angle around the hole-pocket $\alpha$ and $\alpha^\prime$ in FeSe$_{0.93}$S$_{0.03}$ (twinned crystals).
Reprinted figure with permission from   \cite{Xu2016}, copyright by the American Physical Society (2016).
(\textbf{b}) Superconducting gap anisotropy on the hole pocket of multi- and single-domain FeSe. Reproduced from   \cite{Hashimoto2018}. \ccfour.
(\textbf{c}) Energy distribution curves (EDCs) on the holelike pocket plotted as a  grayscale image such that the variation of the superconducting peak position can be directly visualized~\cite{Liu2018a}.
(\textbf{d}) Location of the Fermi momentum on the holelike pocket~\cite{Liu2018a}.
(\textbf{e}) Momentum dependence of the superconducting gap derived from fitting the symmetrized EDCs.
Reproduced from~\cite{Liu2018a}. \ccfour.
(\textbf{f}) (top) Binding energy of the leading edge of $k_F$ EDCs on the electronlike pocket and (bottom) symmetrized data
by also taking into account the second elliptical pocket (twinned crystals)~\cite{Kushnirenko2018}.
(\textbf{g}) Binding energy of the leading edge of the $k_F$ EDCs from the holelike pocket.
Reprinted figure with permission from   \cite{Kushnirenko2018}, copyright by the American Physical Society (2018).}
\label{fig:gap_ARPES_FeSe}
\end{figure*}

The tunneling spectra, as measured on pristine surfaces of thin films  \cite{Song2012,Song2012II} and crystals~\cite{Kasahara2014}, exhibit a V shaped structure revealing
a strongly anisotropic superconducting order parameter; see~Figure~\ref{fig:gap_STM_FeSe}a,f. A full, but small gap has recently been detected with high resolution measurements~\mbox{\cite{Sprau2017,Jiao_2017}}.
Bulk FeSe is orthorhombic,
and the concomitant anisotropy of  electronic structure and superconducting order parameter has been revealed by the observation
of elongated vortices in thin films   \cite{Song2011} and bulk single crystals   \cite{Watashige2015} of FeSe. A detailed mapping of the order parameter on the $\Gamma$-centered
and X-centered Fermi surface  was performed with high-resolution Bogoliubov quasiparticle interference   \cite{Sprau2017}; these authors find a highly anisotropic gap which has
deep gap minima along the $k_x$ axis for the $X$ pocket and the $k_y$ axis for the $\Gamma$ pocket. This result  is consistent with an ARPES measurement reporting a significant 2-fold anisotropy on the $\Gamma$ pocket
in lightly sulfur doped FeSe  \cite{Xu2016},  expected to have very similar properties as  pristine FeSe since it is still deep in the nematic phase. The same findings for the $\Gamma$ pocket  \cite{Rhodes2018,Hashimoto2018,Kushnirenko2018} and the X-pocket  \cite{Kushnirenko2018} were also reported subsequently by
ARPES measurements on bulk FeSe itself.  The data are summarized in Figure~\ref{fig:gap_ARPES_FeSe}, which shows that the ARPES measurements consistently find strongly anisotropic gaps with maxima on the flat sides of the elliptic $\Gamma$ pocket.
The order parameters between the holelike and electronlike Fermi surface sheets  are also of opposite sign, as evidenced by the antisymmetrized tunneling conductance  \cite{Sprau2017}; see Figure~\ref{fig:gap_STM_FeSe}e.
This is consistent with the observation of nonmagnetic impurity bound states in this system  \cite{Jiao2017}.

\subsubsection{Orbital Selective Pairing}
\label{subsubsec:orb_selective_pair}

Theoretical investigations into the superconducting paring interaction  and  ground state order parameter for bulk FeSe suffer from several problems:
As outlined in the previous section,
FeSe~seems to be more correlated than other
FeSC (such as the Fe pnictides, Figure~\ref{fig:DMFT}); thus reliable methods or phenomenological models describing the
strongly correlated electronic structure are needed. Moreover, FeSe is highly nematic which also strongly modifies the superconducting state, i.e., a theoretical calculation needs to
also include effects of the nematic order parameter. However, models for the electronic structure
based on ab-initio methods are either derived from the tetragonal state, or~need to  begin with a stripelike magnetic
ground state, which differs from the true low energy state of FeSe at ambient pressure.
A way out it to use a model-based approach which starts from the electronic structure that
agrees with the experimentally observed one, i.e.,  a tight binding parametrization with as few
hopping elements  \cite{Eschrig09} as possible, and fit these to agree with the positions and orbital
content of the electronic structure as observed by ARPES and STM. Such an approach has been used
already in the context of the cuprates  \cite{Norman95},  for the Fe-pnictides  \cite{Wang13,Ahn14, Saito14}
and was also adopted for the case of FeSe  \cite{Sprau2017,Kreisel2017}. Starting from an electronic
structure including a nematic distortion in the form of an orbital ordering term  (see Equation~(\ref{eq_oo})) and examining superconducting
instabilities within the spin-fluctuation pairing approach (see Section~\ref{subsec:theory_FeSC}) yields a strong mixture of harmonics
of $s$-wave and $d$-wave character  \cite{Mukherjee2015}.

Examining the effects of electronic correlation in more detail, it has been established that the FeSC (and therefore also FeSe) can be understood
as  ``Hund's metals'',  as first proposed in reference \cite{Yin2011}. The presence of the Hund's coupling in the interaction Hamiltonian, Equation~(\ref{H_int}), leads to enhanced correlations and effective masses, tendencies for
electronic configurations with high local spin and (most importantly for what follows here) for a differentiation or ``selectivity'' of electronic correlation strength depending on the orbital character  \cite{deMedici_review,Lanata2013,deMedici2014}.
This renormalization is gradually enhanced as the electronic filling (in the corresponding orbital channel) approaches
half filling  (5~electrons/iron) where in a one band system the Mott transition would occur, but strongly modifies properties of the metallic state even for most of the Fe-based systems discussed here, which are quite far from this doping. This physics can be understood theoretically
with the slave-spin mean field approach~\mbox{\cite{deMedici_review}} or DMFT~\mbox{\cite{Yin2011,Aichhorn2010,Haule16,Evtushinsky2016}}.  Orbital selectivity is clearly manifest in the FeSC  \cite{deMedici2014,Yi2015}, including the Fe-chalcogenides,  and has a clear connection to the nematicity in FeSe  \cite{Fanfarillo2017,Kostin2018}.
Most of the theoretical approaches to pairing in FeSe which will be reviewed in the following incorporate
the basic fingerprints of the Hund's metal state and are therefore connected to the normal state electronic
properties in this system as well.

In the QPI and subsequent ARPES analysis of the gap structure of FeSe,  a strongly anisotropic
order parameter was observed; it was pointed out by Sprau et al.  \cite{Sprau2017} that its magnitude as a function of Fermi surface angle follows the orbital content of the $d_{yz}$ orbital,
suggesting the conclusion that the superconducting pairing is dominated by electrons in this orbital. Given the
small nematic splitting of the electronic structure, this effect cannot a priori be  explained by a pure spin-fluctuation
scenario, i.e., such a calculation would give small anisotropy and/or small magnitude of the gap  \cite{Kreisel2017,Benfatto2018}, unless~one tunes extremely close to the magnetic instability in an RPA approach where the spin fluctuations
acquire a very nonlinear dependence on the interaction parameters  \cite{Kreisel15}.
Electronic correlations  parameterized by a self-energy in Equation~(\ref{eq_int_G}) lead to band renormalizations and broadening of the spectral function, but also to reduced coherence of the electronic states.
Usually, the self-energy is expanded in powers of frequency near the Fermi level by introducing  the quasiparticle weight as given in Equation~(\ref{eq_def_Z})
such that the
interacting Green's function on the real axis can be parameterized at small frequencies as
\begin{equation}
G^R(\k,\omega)=\frac{Z_\k}{\omega-Z_\k E_{\k}-i\Gamma_{\k}}\label{eq_param_G}\,,
\end{equation}
with the quasiparticle weight $Z_\k$ (Equation~(\ref{eq_def_Z})) and a broadening $\Gamma_{\k}$ which is given by the imaginary part of the
self-energy. In a multiorbital, multiband system, there are effects which cannot be described with the parameterization in Equation~(\ref{eq_param_G}). First, the intrinsic momentum dependence of the self-energy can  induce non-local effects such as relative band shifts which turn out to be important for the electronic structure of FeSC  \cite{Ortenzi2009,Zantout2019,Bhattacharyya2020}, and second, in general $\Sigma(\k,\omega)$ is a matrix in orbital space which can induce different  quasiparticle weights for different states at the Fermi level. For such an  orbitally selective electron gas  \cite{deMedici2014,Medici_review, Yin2011},  where the quasiparticle
weights of the different orbitals are not identical, one expects that the quasiparticle weight at the Fermi surface of band $\mu$
acquires a ``trivial'' momentum dependence due to the matrix elements $a_\mu^\ell(\k_F)$ connecting orbital and band space, as well as one arising through correlations reflected in the orbital quasiparticle weight $Z_\ell$, such that on the Fermi surface $Z_{\k_F,\mu}=\sum_l Z_l |a_\mu^\ell(\k_F)|^2$.

The shifts of the eigenenergies can be captured in a phenomenological model that matches the band energies of the real material
(as found experimentally), but the
orbitally selective reduction of quasiparticle coherence  \cite{deMedici2014,Medici_review, Yin2011} also needs to be incorporated.   In the nematic state, this
can also lead to a distinction of the $d_{yz}$ orbital and the $d_{xz}$ orbital correlations.
To achieve the strongly anisotropic order parameter in FeSe from a theoretical calculation, one needs
(1) strongly reduced coherence of the $d_{xy}$ orbital, as expected from many theoretical
investigations within dynamical mean field theory (DMFT) or slave spin calculations for FeSC in general and FeSe in particular~\cite{Yu2018}.  In~addition,
one needs in order to be consistent with the results of Ref. \cite{Sprau2017} (2) some way to suppress the $xz$ component of the pairing interaction.  However, a simple suppression of the $xz$ quasiparticle weight is apparently contradicted by the fact that both orbital components $xz$ and $yz$ are simultaneously detected at the Fermi level on the hole-like pocket \cite{Rhodes2018, Liu2018a}.
The second effect can
be achieved by
making the assumption that the $d_{xz}$ states are much less coherent than the $d_{yz}$ states~\mbox{\cite{Sprau2017, Kreisel2017}}
and doing a modified spin-fluctuation pairing calculation. One should in principle calculate
quasiparticle weights in each orbital channel,  requiring an approximation to the full self-energy,  or self-consistency within renormalized mean field theories  such as DMFT or slave-spin theory \cite{Hu2018,Yu2018}, but in practice the $Z_\ell$ have been mostly free (fit) parameters so far (see reference \cite{Bhattacharyya2020_II} for an exception). However, in the spirit of Fermi liquid theory, one might employ an approach where one considers one
experiment to fix the phenomenological parameters and then uses the same model to predict other observables such as
penetration depth \cite{Biswas2018,Cercellier2019} and specific heat \cite{Cercellier2019}.
It~turns out that the low-$T$, low-energy susceptibility as calculated from such a correlated model, where the $Z_\ell$ are chosen to fit the QPI-derived gap structure in reference \cite{Sprau2017} yields
a strong $(\pi,0)$ contribution, but~essentially no $(0,\pi)$ contribution at low energies \cite{Kreisel2018} (Figure~\ref{fig:itinerant_theory}c). The low energy, low-$T$  magnetic excitations have recently been measured in
detwinned FeSe via neutron scattering, finding no $(0,\pi)$ over a low energy range \cite{Chen2019} (see Figure~\ref{fig:neutron_FeSe}f--k), in agreement with the prediction.
As the temperature is raised, nematic order vanishes at $T_s$, such that $Z_{xz}=Z_{yz}$.  Such a theory then
naturally explains the observed transfer of spectral weight \cite{Kreisel2018, Wang2015GS} from $(\pi,0)$ to both $(0,\pi)$ and $(\pi,\pi)$; see Figure~\ref{fig:neutron_FeSe}c.

At the same time
the decoherence, as parametrized by orbitally distinct quasiparticle weights  $Z_{xy} < Z_{xz} < Z_{yz}$ can account for (a) the strongly anisotropic scattering properties on impurities in the nematic state \cite{Kostin2018} (Figure~\ref{fig:QPI}) and
(b) the difficulties to detect the $Y$ pocket in spectroscopic probes~\mbox{\cite{Kostin2018,Watson2017II}} because the corresponding $d_{xy},d_{xz}$ states would be incoherent
and thus the spectral  weight, as calculated from Equation~(\ref{eq_param_G}),
would be small as well. Similar arguments should hold for the  QPI data, since the scattering amplitude from an impurity may be expressed via  the Fourier transform of the modulations of the density of states,
calculated using the T-matrix formalism by~\mbox{\cite{BalatskyRMP, Kostin2018}}
\begin{align}
\delta N({\q},\omega)=-\frac 1 \pi \mathop{\text{Im}}\mathop{\text{Tr}}\sum_{\k} \hat G^R(\k,\omega)\hat T(\omega)\hat G^R(\k+{\q},\omega)\,.\label{eq_Nq}
\end{align}

Again, the quasiparticle weight enters quadratically  through the Green's functions $\hat G^R(\k,\omega)$ characterizing the homogeneous system, which in the multiband case are  matrices, as is the impurity ${\hat T}$-matrix.  The original choice of $Z_\ell$ used to fit the gap structure worked well to explain the evolution of the normal state QPI at temperatures just above $T_c$ \cite{Kostin2018}; see Figure~\ref{fig:QPI}a.

\begin{figure*}[tb]

\centering
{\includegraphics[width=\textwidth]{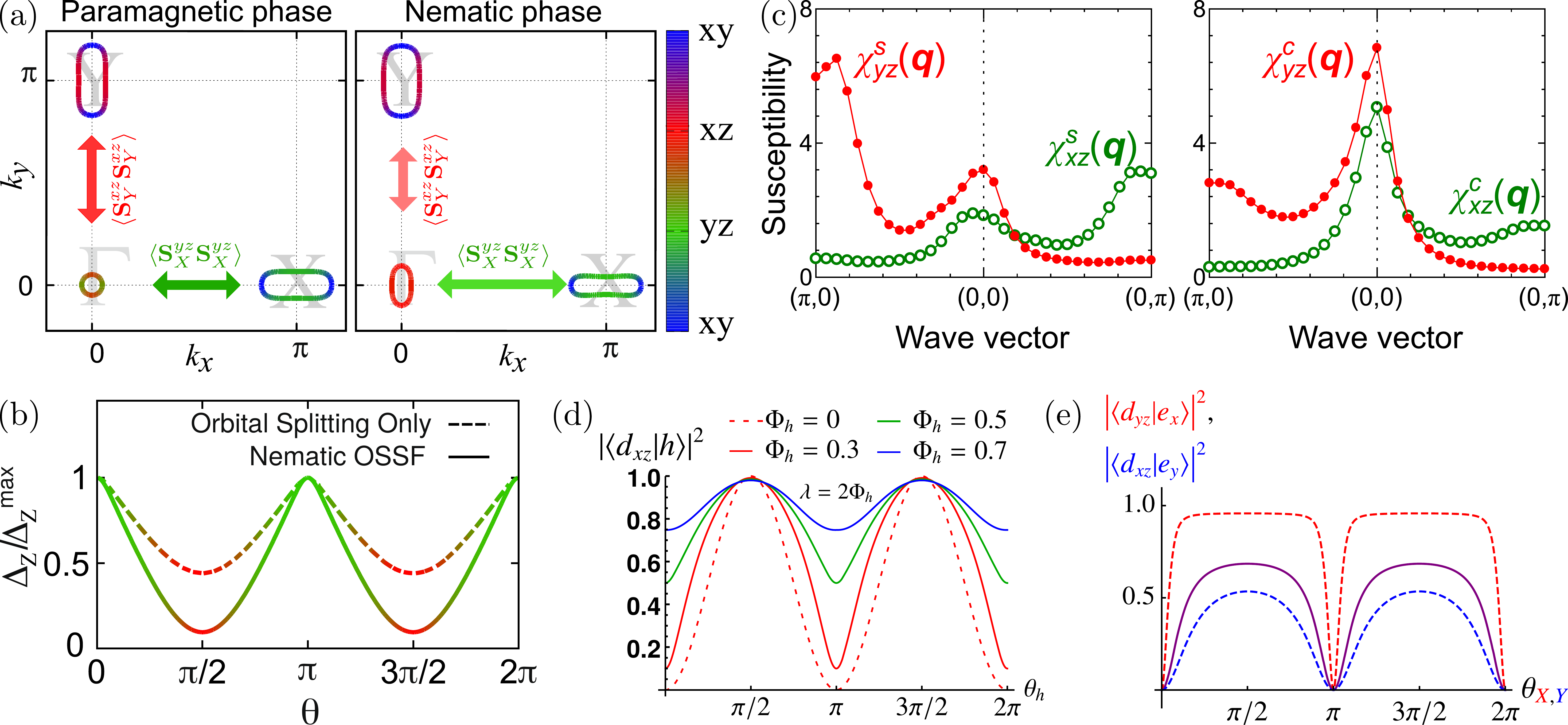}}
\caption{Theoretical proposals for superconducting pairing in FeSe.
(\textbf{a}) Fermi surface in the paramagnetic and nematic phase. Driven by orbital selective spin fluctuations (red and green arrows), the electronic structure exhibits shifts; pairing mainly mediated by $d_{yz}$ spin fluctuations yields a gap structure comparable to experiment, as shown in (\textbf{b}), while orbital splitting in the electronic structure only will lead to much smaller anisotropy.   Note that the theory of reference \cite{Kang2018} is based on  a  Fermi surface at $k_z=\pi$ only, allowing for a much larger variation of the orbital content, such that a strongly anisotropic gap is easier to achieve. Reproduced from   \cite{Benfatto2018}. \ccfour.
(\textbf{c}) Calculating the spin susceptibility and the charge susceptibility in a framework for higher order many-body effects yields a strong orbital dependence and favors an anisotropic $s_{\pm}$ state for FeSe at ambient pressure.
Reprinted figure with permission from   \cite{Yamakawa2017}, copyright by the American Physical Society (2017).
(\textbf{d}) Modifications of the orbital content on the holelike pocket due to nematic order $\Phi_h$, (\textbf{e}) same on the electronlike pocket at the $X$ point, leading to a strongly anisotropic gap structure when pairing in the $d_{xy}$ channel is suppressed.
Reprinted figure with permission from   \cite{Kang2018}, copyright by the American Physical Society~(2018).}
\label{fig:theory_sc}
\end{figure*}

Recently, alternative explanations of the latter two experimental results were brought
forward in terms of effects of the three dimensional electronic structure giving rise to some averaging in the sum over $\k$ on the r.h.s of Equation~(\ref{eq_Nq}) \cite{Rhodes2019arXiv} or possible lifting
of the $Y$ pocket due to band hybridizations~\mbox{\cite{Yi2019arXiv190304557Y, Huh2019arXiv190308360H}}.
However, the latter explanation has been questioned
by a theoretical model of the expected spectral functions accounting for an electronic structure that
has a nematic order parameter in the $d_{xz}/d_{yz}$ channel and the $d_{xy}$ channel, and additionally hybridizations
due to spin-orbit coupling are taken into account \cite{Christensen2019arXiv190804889C}.
In the quest to find the microscopic origin of the strongly anisotropic order parameter, there have also been other theoretical
proposals: Kang et al. examined the effects of a modified orbital content on the Fermi surface which, together with
strongly correlated states in the $d_{xy}$ orbital channel and an induced anisotropy in the pairing interaction from small nematic splitting
of the electronic structure, can also account qualitatively for  the observed  structure of the order parameter \cite{Kang2018}; see Figure~\ref{fig:theory_sc}d--e. As mentioned in the previous
section, a phenomenological model of strongly anisotropic spin fluctuations, i.e., orbitally selective spin fluctuations can explain the modifications
of the electronic structure in FeSe at low temperatures, thereby giving rise to the Fermi surface shrinkage and nematic distortion \cite{Fanfarillo2016, Fanfarillo_mismatch2018}; see Figure~\ref{fig:theory_sc}a.  Subsequently, it was  shown that the same spin fluctuations can also lead to a strongly anisotropic superconducting
order parameter since these provide the strongly anisotropic pairing interaction \cite{Benfatto2018}; see Figure~\ref{fig:theory_sc}b. It is our belief that ultimately these two alternative approaches \cite{Benfatto2018,Kreisel2017} are quite similar in spirit to the orbitally selective $Z$-factor approach described above; the equations solved are ultimately the same, but the required  anisotropic pairing interaction incorporated in somewhat different ways. These~approaches differ from that of reference \cite{Kang2018}, where anisotropy in the spin fluctuation spectrum must arise entirely from $d_{xz}/d_{yz}$ orbital content anisotropy in the nematic state.  This effect seems unlikely to explain the dramatic measured anisotropy in the spin fluctuation spectrum reported in reference~\mbox{\cite{Chen2019}}.

Including vertex corrections in a calculation of the superconducting instabilities, a strongly anisotropic
order parameter on the electron and hole pockets was obtained \cite{Yamakawa2017}, while the pairing interactions
become anisotropic due to the creation of the orbital order from the intraorbital vertex corrections; see Figure~\ref{fig:theory_sc}c.
Another proposal assumes the existence of a nematic quantum spin liquid in FeSe that exhibits strongly anisotropic spin fluctuations.
Taking additionally into account the spectral imbalance between $d_{xy}$ orbitals and $d_{xz}/d_{yz}$   via orbital dependent Kondo-like couplings, one indeed finds a strongly anisotropic superconducting order parameter comparable to experiment \cite{She2018}.

\subsubsection{BCS-BEC Crossover Scenario}
\label{sec_crossover}
Finally, let us remind the reader that the Fermi surface pockets in FeSe are particularly small, such~that the maxima or minima of the band dispersions, i.e., the Fermi energies, are nearly comparable to the energy scale of the order parameter, as directly measured by ARPES and QPI~\mbox{\cite{Hashimoto2018,Kasahara2014}}.
Other~experiments arrive at similar conclusions about these energy scales: for example, the Fermi temperature as obtained from the value of the penetration depth is for FeSe closer to the critical temperature of the BEC than for many other high temperature superconductors \cite{Uemura1989,Shibauchi2020}.
The~observation of oscillations in the local density of states of vortices is another piece of evidence that FeSe exhibits a small $E_F/\Delta$, such that individual Caroli--de Gennes--Matricon states can be detected~\cite{Hanaguri2019}.
Therefore, the experimental results have been interpreted in terms of the  system's proximity  to the BCS-BEC crossover regime where the Cooper pairs form already at $T_{\text{pair}}>T_c$ and then eventually condense into a superfluid state at $T_c$. In this limit, the size of the Cooper pairs is much smaller than the average distance between electrons, i.e.,   the product of the Fermi wavevector and the coherence length is small $k_F\xi\ll 1$.  Consequences of proximity to the crossover  regime in a one-band model are well known: that the chemical potential $\mu$ becomes negative and larger in magnitude than the energy gap, such that the band shows a back-bending instead of an opening of a gap when entering the superconducting state. As mentioned in the Introduction, however, the generalization to electron-hole multiband systems  is not obvious because the chemical potential may be pinned with temperature, or nearly so, due to the presence of a band with opposite curvature.   Thus, the pseudogap due to  preformed pairs  above $T_c$ that is characteristic of    the one-band crossover regime, as well as the characteristic broadening of thermodynamic transitions, may not occur \cite{Chubukov2016}.  This indeed seems to be the case for FeSe and Fe(Se,S) \cite{Hanaguri2019}. However, the occurrence of a large diamagnetic response in weak magnetic fields above $T_c$ beyond the expected signature of Gaussian superconducting fluctuations yielding the Aslamasov-Larkin susceptibility might be a signature of preformed Cooper pairs \cite{Kasahara2016}. For~a more detailed discussion of these intriguing observations and the behavior of superconducting FeSe in high magnetic field, the reader is referred to a recent review \cite{Shibauchi2020}.

In fact, the prospect of observing phenomena characteristic of the BCS/BEC crossover was first raised in the
Fe$_{1+y}$Te$_{1-x}$Se$_x$ system in measurements of the electronic structure using ARPES that found a very shallow holelike band that just crosses the Fermi level \cite{Rinotte1602372,Lubashevsky2012} and evidenced a gap opening at higher temperatures \cite{Lubashevsky2012}.
Note that the superconducting gap in this system is larger than in FeSe and additionally, the chemical potential can be tuned by chemical doping using excess Fe \cite{Rinotte1602372}. Interestingly, also an electronlike band above the Fermi level at the $\Gamma$ point has been detected which participates in pairing and shows effects of a pseudogap  at temperatures of few Kelvin above $T_c$ \cite{Okazaki2014}.  The consequences of small Fermi energies in this system have been less comprehensively explored than in FeSe and FeSe,S due to the materials difficulties, and also because of the overwhelming interest in the topological properties of Fe$_{1+y}$Te$_{1-x}$Se$_x$; see Section~\ref{sec:topo}.

\subsubsection{Spin Fluctuations in Superconducting State}
\label{subsubsec:sf_SC}

In the superconducting state, a clear spin-resonance is observed around the stripe-type wave vectors \cite{Wang2015} (see Figure~\ref{fig:neutron_FeSe}a,d,  later confirmed in an experiment on detwinned FeSe; Figure~\ref{fig:neutron_FeSe}g), showing a consistent behavior in energy and momentum with a spin-fluctuation-mediated superconducting pairing mechanism \cite{Korshunov2018, Kreisel2018,Maier09}.
Polarized neutron measurements found that low-energy magnetic fluctuations, including the superconducting resonance, are mainly along the $c$-axis \cite{Ma_2017}.
The~spin resonance in the superconducting state just at the $(\pi,0)$ momentum transfer vector was detected~\mbox{\cite{Kreisel2018,Chen2019}} (Figure~\ref{fig:neutron_FeSe}g) and its dependence on field analysed \cite{Chen2020}, pointing towards a sign-changing order parameter.
From the perspective of the theoretical conclusions of this experimental finding, there are a number of proposals that actually
predict or assume strongly anisotropic spin fluctuations to mediate superconducting pairing: the proposal of an orbitally selective spin-fluctuation mechanism \cite{Benfatto2018}; the nematic quantum spin liquid \cite{She2018}; more exotic proposals on quadrupolar magnetic order which exhibits magnetic fluctuations in the dipole channel \cite{Yu2015,Wang_Z_16} (Figure~\ref{fig:localized_theory}b);  and~the picture of itinerant electrons exhibiting
orbitally selective decoherence \cite{Kreisel2017}; see Figure~\ref{fig:itinerant_theory}.

Many of these theoretical proposals are effective low-energy theories; it is important to have a phenomenology that can demonstrate a good agreement with inelastic neutron data, including the temperature---and energy---dependendent transfers of spectral weights and evolution of the strong spin fluctuation anisotropy up to the nematic temperature $T_s$ and at least over energies scales of $\sim$50~meV or so in order to have some predictive power for superconductivity. To our knowledge only the last of the examples given \cite{Kreisel2017,Kreisel2018} has been compared sufficiently closely with experiment to make this claim.

\section{Effects of Physical and Chemical Pressure\label{sec:pressure}}

\subsection{FeSe under Pressure}
\label{subsec:pressure}

\begin{figure*}[tb]

\centering
\includegraphics[width=\textwidth]{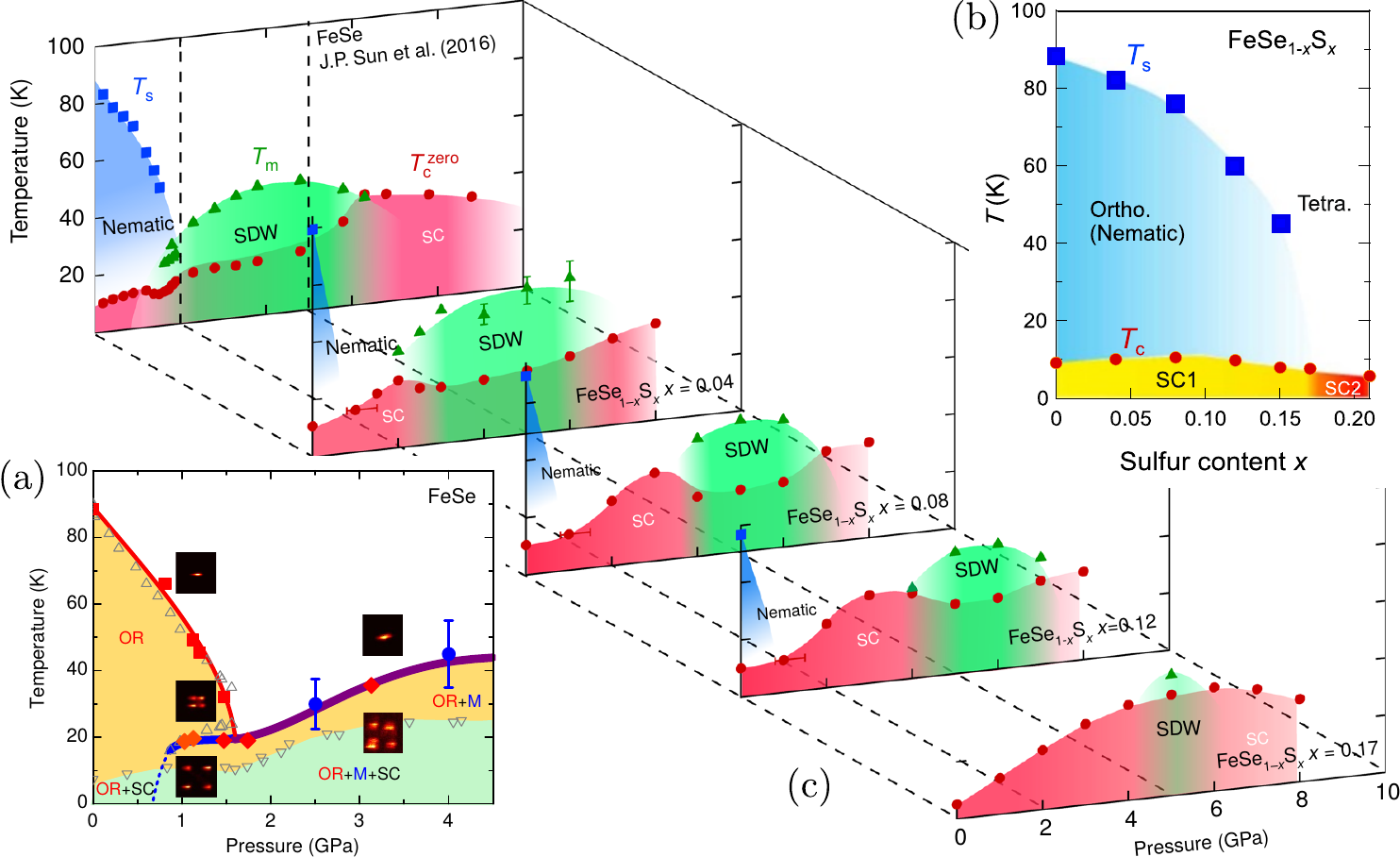}
\caption{Comparison of phase diagrams of FeSe under pressure. (\textbf{a}) X-ray diffraction and M\"ossbauer study showing orthorhombic (OR), superconducting (SC), and magnetic (M) boundaries vs. pressure. Reproduced from  \cite{Kothapalli2016}. \ccfour.
Inserts indicate typical X-ray patterns showing splitting in OR phase. (\textbf{b}) Specific heat and thermal conductivity study of S doping (``chemical pressure'') \cite{Sato2018}.  SC1 and SC2 delineate two distinct superconducting phases identified in this study and confirmed by reference~\cite{Hanaguri474}. (\textbf{c}) Variation of both S doping and pressure, indicating magnetic (SDW), superconducting (SC), and~nematic phases (orthorhombic) phases.
Reproduced from  \cite{Matsuura2017}. \ccfour.}
\label{fig:phasediags}
\end{figure*}

As discussed above, interest in FeSe among all Fe-based superconductors was initially muted because of the difficulty of preparing stoichiometric crystals, and the relatively low 8 K $T_c$.
However it was soon realized that large changes in $T_c$ can be obtained by applying pressure~\mbox{\cite{Margadonna2009,Medvedev2009,Kumar2010}}, up to a maximum of nearly
$40\,\mathrm{K}$  at $p$ of order $10\,\mathrm{GPa}$.  This remarkable enhancement was not initially associated with any change in magnetic order, but there were early hints from NMR that pressure strongly enhanced spin fluctuations \cite{Imai2009}, consistent with the enhanced $T_c$.  Subsequently, \mbox{Bendele et al. \cite{Bendele2010}} used AC magnetization and muon spin rotation measurements to argue that the kink in $T_c$ vs. $p$ that had been observed earlier at $\sim$1 GPa was in fact due to the onset of some kind of magnetic order.  Conceptually the discovery of magnetism ``nearby'' the ambient pressure point was important, because it rendered less likely claims that FeSe was dramatically different from other Fe-based superconductors due both to the absence of   magnetic order  and to the lack of significant rise of spin fluctuation intensity in $(T_1T)^{-1}$ at the nematic transition (as, e.g., in BaFe$_2$As$_2$).  With this and subsequent measurements of magnetism in the pressure phase diagram \cite{Terashima2015,Kothapalli2016,Sun_2016,Matsuura2017,Gati2019}, it became clear that special aspects of the FeSe electronic structure were frustrating or suppressing long-range magnetic order of the usual type \cite{WangLee2015NatPhys_FeSe-Para-Nematicity,Glasbrenner2015} in the parent compound.   With a small amount of pressure, this special frustrating condition is relieved, allowing spin fluctuations and $T_c$ to grow, suppressing nematic order and leading eventually to a magnetically ordered state.  The situation is summarized in the phase diagram of Figure~\ref{fig:phasediags}a.

\subsection{FeSe under Chemical Pressure: S Substituion}
\label{subsec:chem_pressure}
The similarity of phase evolution under chemical pressure to physical hydrostatic or near-hydrostatic pressure was noted already in the BaFe$_2$As$_2$ system using P substitution for As.  Similarly, in FeSe, we may expect that substituting the isovalent but smaller S atom on the Se site will act as chemical pressure.  Indeed, as seen in Figure~\ref{fig:phasediags}b, the nematic order is supressed with S doping more or less as with pressure, with an apparent vanishing of nematic order at around S concentration of $x=0.17$  accompanied with unusual properties in the resistivity \cite{Licciardello2019,Hosoi2016} close to the quantum critical point (QCP).  On the other hand, substitution of S does not appear to stabilize any long-range magnetic order.  Perhaps equally interestingly, the superconducting critical temperature is not enhanced at the critical point, suggesting that nematic fluctuations themselves are irrelevant for superconductivity in this particular system,  a conclusion that has also been drawn recently from an ab-initio study of FeSe~\cite{acharya2020role}.  Paul and Garst have pointed out that lattice effects should cut off the divergence of the nematic susceptibility, so this absence of a peak in $T_c$ is, in that light, not surprising \cite{Paul2017}.

\begin{figure*}[tb]

\centering
{\includegraphics[width=\linewidth]{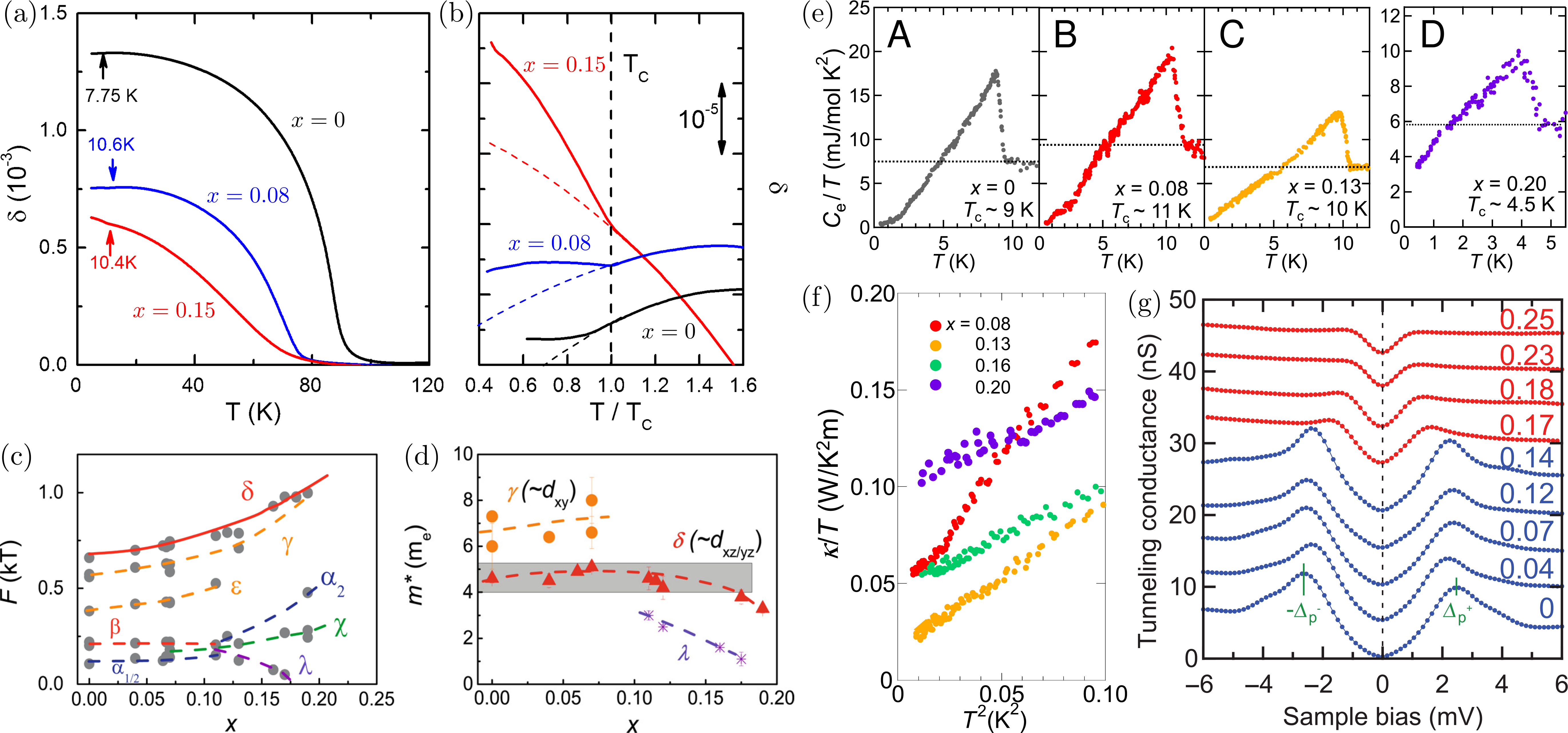}}
\caption{(\textbf{a}) Temperature dependence
of the orthorhombic distortion $\delta=|a-b|/(a+b)$ of the orthorhombic lattice parameters $a$ and $b$ in FeSe$_{1-x}$S$_x$ for $x=0, 0.08, 0.15$; and (\textbf{b})  enlarged view of the distortion as a function of $T/T_c$ in revealing a cooperative effect upon entering the superconducting state. Data are shifted relatively; dashed lines are extrapolations
of the normal-state data.
From~\cite{Wang2017II}, copyright 2016 WILEY‐VCH.
(\textbf{c}) Evolution of the quantum oscillations frequencies as a function of sulfur doping $x$. Solid lines are the calculated frequencies for the large outer hole band based on ARPES data.
Reprinted figure with permission from  \cite{Watson2015III}, copyright by the American Physical Society (2015).
(\textbf{d}) Quasiparticle effective masses of the frequencies as labeled in panel (\textbf{c}).
Reproduced from~\cite{Coldea_Watson2018II}. \mbox{\ccfour.}
Evidence for abrupt change in superconducting order parameter at or near nematic transition in FeSe,S:  (\textbf{e}) specific heat (A--D are different S concentrations $x$ as indicated) 
and (\textbf{f})~thermal conductivity at four different S concentrations crossing the nematic transition at $x=0.17$~\cite{Sato2018}; (\textbf{g}) confirmation of finite DOS onset near $x=0.17$.
Reproduced from  \cite{Hanaguri2018}. \ccfour.}
\label{fig_FeSeS_delta_FS_size_m}
\end{figure*}

In general, the interplay of nematic and superconducting effects are subtle in Fe-based systems and may be dependent on details.  In FeSe,S,  the Karlsruhe group reported a surprising lack of coupling between the orthorhombic $a,b$ axis lattice constant splitting ($\propto$ nematic order) and superconductivity in FeSe, in stark contrast to Ba-122, where the splitting was suppressed below $T_c$, indicating competition of the two orders. In FeSe there was no effect at all on $a-b$ \cite{Boehmer2013} at $T_c$. When  FeSe was doped with S by the same group in reference \cite{Wang2017II}, $a-b$ was found to increase as $T$ was lowered below $T_c$, indicating a cooperative effect of superconductivity and nematicity in these samples; see Figure~\ref{fig_FeSeS_delta_FS_size_m}a,b. The reason for this difference between the two canonical Fe-based families is not clear at this writing \cite{labat2020};  from a theoretical point of view, details of the relative orientation of Fermi surface distortion and gap function, along with orbital degrees of freedom, may govern this behavior \cite{Chen_Maiti2020}.

To compare how chemical pressure and external pressure affect the FeSe system, several groups have subjected FeSe,S samples at different dopings to external pressure \cite{Matsuura2017,Reiss2020}.
For example, the~Tokyo--Kyoto group has measured the comprehensive phase diagram shown in Figure~\ref{fig:phasediags}c.  The~apparently inimical effect of the S-substitution on magnetism already mentioned above was shown explicitly, with the shrinking of the pressure-induced SDW phase.  The similar effects of pressure and S-substitution on nematic order are also confirmed: the weaker the nematic transition temperature $T_s$ in $x$, the smaller its extent in pressure.

\subsection{Diminishing Correlations}
\label{subsec:correlations}
Given the large discrepancies discussed above between DFT and DMFT electronic structure calculations in the parent compound FeSe, it is interesting to investigate the evolution of the correlations assumed responsible for these dramatic renormalizations with physical and chemical pressure.  In~both cases, one naively expects that the compression of the lattice should result in a decrease  of the effective degree of correlation, simply because the kinetic energy (hoppings) in such a situation should be enhanced due to the decreased distance between the ions, whereas the effects on local interaction parameters like $U$ and $J $ should be smaller.

The first systematic study of the electronic structure changes with S doping was performed with ARPES by Watson et al. \cite{Watson2015III}, who reported results roughly in line with expectations.  Increasing S substitution was found to enlarge the Fermi surface pockets, and increase the Fermi velocities, albeit by a relatively small amount; see Figure~\ref{fig_FeSeS_delta_FS_size_m}.  While the pocket size thus changes to become closer to (but still smaller than) DFT predictions, the $d_{xy}$ band  predicted in DFT studies at the Fermi level remained 50 meV below in the ARPES study, and thus never plays a role in the chemical pressure phase diagram.  At the same time, these authors pointed out a systematic decrease in the $d_{xz}/d_{yz}$ orbital ordering as the structural transition temperature fell (Figure~\ref{fig:phasediags}).  Finally, the apparent weakening of correlations also led to a Lifshitz transition around $x=0.12$ as the inner $d_{xz}/d_{yz}$ hole pocket, pushed below the Fermi level in the parent compound, reappeared.  While this evolution was largely confirmed in a subsequent Shubnikov-de Haas study by Coldea et al. \cite{Coldea_Watson2018II}, these authors reported the disappearance at $x=0.19$ of the small oscillation frequency associated with the outer $d_{xz}/d_{yz}$ hole pocket (see Figure~\ref{fig_FeSeS_delta_FS_size_m}c), and~interpreted it as a second Lifshitz transition where the outer cylinder pinched off at $k_z=0$ to form $Z$-centered 3D pockets.

\subsection{Abrupt Change in Gap Symmetry in Tetragonal Phase}
The evolution of the superconductivity with S substitution is clearly of great importance, since it provides a clue to how the changing electronic structure, including the disappearance of the nematic order, affects the pairing.  While $T_c$ itself is relatively insensitive to these changes, Sato et al. \cite{Sato2018} reported evidence that the superconducting gap undergoes a dramatic change at a concentration around $x=0.17$, very close to the disappearance of nematic order.

We begin by summarizing what is known about the gap structure in the nematic phase of FeS$_{1-x}$S$_x$. At $x=0$, as discussed in Section~\ref{sec:properties}, the gap structure measured on the  hole and electron pocket detectable by spectroscopy at the Fermi surface is highly anisotropic and nematic.  As shown in Figure~\ref{fig:gap_ARPES_FeSe}a, 3\% sulfur did not change the gap structure significantly, at least on the $\alpha$ pocket where it was measured,  so it seems likely that the gap structure for small nematicity is quite similar to FeSe.  There are no other direct measurements of the gap in FeSe,S of which we are aware at higher doping, so information about gap structure has been deduced mostly from thermodynamics.

As shown in Figure~\ref{fig_FeSeS_delta_FS_size_m}e, the superconducting state specific heat is consistent with a zero residual value in the nematic phase, while jumping to a rather large residual value in the tetragonal phase.  Since the gap in FeSe is known to be highly anisotropic, with nodes or near-nodes, it is tempting to attribute any such residual term to disorder.  There are several reasons to reject this explanation, however.
First,  STM topographs  on these samples suggest that they are very clean, inconsistent with residual Sommerfeld coefficients $\gamma_s/\gamma_n$ of ${\cal O}$(1) \cite{Hanaguri2018}.  Second, there is no reason to expect a discontinuous change in the disorder itself at the nematic transition, so one would have to postulate that the superconducting state undergoes a transition making it fundamentally more sensitive to disorder.  Since the system evolves already out of a state with
nodes or near-nodes, and an established gap sign change \cite{Sprau2017}, it is far from  compelling that a transition, e.g., from $s_\pm$ to $d$-wave would cause such an abrupt enhancement of the residual density of states, which is clearly present, as shown also directly by STM \cite{Hanaguri2018} (Figure~\ref{fig_FeSeS_delta_FS_size_m}g). Finally, it is expected that the S dopants, away from the Fe plane, act as relatively weak scatterers, and could in any case not give rise to a  $\gamma_s/\gamma_n$ of ${\cal O}(1)$.

Traditionally, the existence of a finite residual $\kappa/T$ as $T\rightarrow 0$ is taken as an indication that the unconventional superconductor in question has line nodes.  However, this is normally a signature of quasi-universal transport \cite{DurstLee2000}, as in the canonical $d$-wave case, where $\kappa/T$ in the limit of weak disorder is a constant independent of the quasiparticle relaxation time (the limiting low-$T$ value is not quite universal in an $s_\pm$ state, but the nonzero $\kappa/T$ does remain even in the limit of vanishing disorder~\cite{Mishra2009II}).     This helps explain why for low $x$, the FeSe,S material displays a small thermal conductivity (Figure~\ref{fig_FeSeS_delta_FS_size_m}f) that stays roughly constant over several low dopings.  On the other hand, it~is also seen that the $x$ = 0.20 doping residual $\kappa/T$ jumps significantly to a large fraction of the normal state value, inconsistent with the usual ``universality'' arising from superconducting gaps with line nodes.

In the STM data reproduced in Figure~\ref{fig_FeSeS_delta_FS_size_m}g, it is clear that changes in the superconducting state are not confined to the finite value of the residual DOS that occurs near $x=0.17$.  In addition, it is seen from the drop in coherence peak energies that the gap is becoming abruptly smaller in magnitude. This,~together with the fully developed Volovik effect observed in reference \cite{Sato2018} in the tetragonal phase, led the authors of references \cite{Sato2018,Hanaguri2018} to conclude that as the system becomes tetragonal, the~gap structure beyond the nematic critical point was becoming even more anisotropic than it was known to be for FeSe.

\subsection{Bogoliubov Fermi Surface Scenario}
One intriguing solution to this puzzle was put forward in reference \cite{Setty2019}, where it was suggested that the system might naturally make a transition into a topological state that manifested a so-called Bogoliubov Fermi surface, a locus of points in $k$-space that supported zero energy excitations at low temperature in the superconducting state. Note this manifold has the same dimension as that of the underlying normal state Fermi surface, i.e., a 2D patch in a system of three spatial dimensions, etc., as distinct from an unconventional superconductor with line or point nodes.   This state was a generalization to spin-1/2 multiband systems of an idea of Agterberg and collaborators \cite{Timm2017,Brydon2018} for a system of paired $j=3/2$ fermions.  The conditions for the existence of this topological transition are that the pairing be in the even parity channel, with dominant intraband spin singlet gaps (e.g., $\Delta_1$ and $\Delta_2$ for a 2-band systems) together with SOC-induced triplet interband component $\delta$.  The latter amplitude is assumed to spontaneously break time reversal symmetry in spin space, analogous to the A1 phase of the $^3$He superfluid \cite{VW_book}.    The authors show that   the Pfaffian of the system is proportional
to $|\Delta_1(${\bf k}$)||\Delta_2(${\bf k}$)|-\delta^2$, such that the change from trivial to topologically nontrivial, accompanied by the creation of the Bogoliubov Fermi surface, occurs when the Pfaffian changes sign from positive to negative.  Although $\delta$ is expected to be small, the topological transition can be achieved due to the nodal (or near-nodal) structure of the intraband gaps.

The existence of a Bogoliubov Fermi surface in this system would naturally explain why a relatively clean superconductor can support a finite density of quasiparticles, as reflected in the residual Sommerfeld coefficient and the differential conductance seen in STM \cite{Setty2019,Setty2020}.  Clearly, independent verification of the assumed time reversal symmetry breaking is needed before such an explanation can be accepted above more conventional ones, but the idea is intriguing.  Very recently, a~signature of TRSB in tetragonal FeSe,S from $\mu$SR was reported \cite{Shibauchi2020}, but details are not yet available.

\section{FeSe/STO Monolayer + Dosing of FeSe Surfaces\label{sec:monolayer}}

\subsection{Single Layer Films of FeSe on \texorpdfstring{\STO}{STO}}
\label{subsec:monolayer_intro}
\begin{figure*}[tb]

\centering
\includegraphics[angle=0,width=\textwidth]{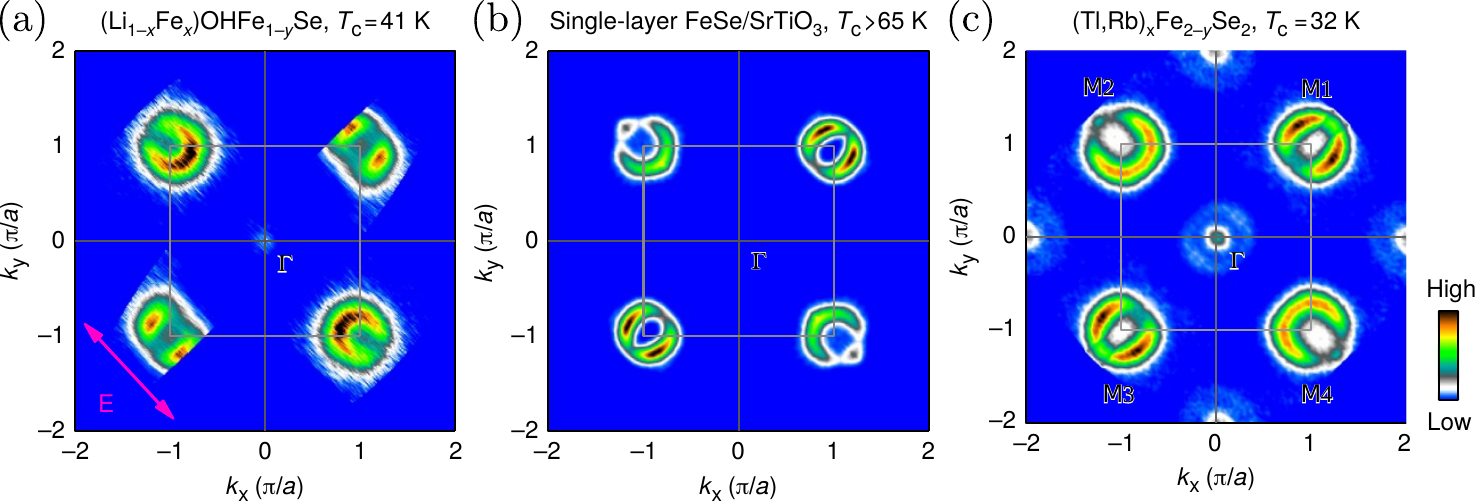}
\caption{Fermi surfaces of (\textbf{a}) (Li$_{0.8}$Fe$_{0.2}$)OH FeSe; (\textbf{b}) FeSe monolayer on STO \cite{Liu2012}; (\textbf{c})~(Tl,Rb)Fe$_2$Se$_2$~\cite{Mou2011} measured by ARPES.  Note the apparent weight of the $\Gamma$-centered hole band in (\textbf{a}) is due to a broad band whose centroid disperses below the Fermi surface, whereas in (\textbf{c}) it was attributed to
electron-like bands crossing the Fermi level.
Reproduced from \cite{Zhao2016}. 
}
\label{fig:3Fermisurfaces}
\end{figure*}

Despite the unusual features of the  FeSe alkali-intercalates (see Section \ref{sec:intercalates}), interest in the FeSe system itself was relatively muted until the discovery
in 2012 of high temperature superconductivity in monolayer FeSe films epitaxially grown on \STO ~by the group of Qi-Kun Xue \cite{Wang12}.  Within a relatively
short period, it~was established (a) that the superconducting gap magnitude was much larger than bulk FeSe or FeSe films grown on other substrates \cite{Song2011II}, such as graphite; (b) that two monolayers were either not or rather weakly superconducting \cite{Wang12,Liu_bilayer};
(c) that the gap closing temperature of the best films (according to ARPES) was in the neighborhood of 65 K \cite{Zhou_monolayer}, the highest $T_c$ measured in the Fe-based systems to that date; and~(d)~that the electronic structure was  similar to the alkali-doped intercalates (and LiOH intercalates, discovered later; see Figure~\ref{fig:3Fermisurfaces}), in the sense that the $\Gamma$-centered hole band had a maximum $\sim$80 meV below the Fermi level, such that the Fermi surface consisted of electron pockets only.
Figure~\ref{fig:FeSe_mono}b shows the epitaxial
structure of the film used originally to obtain a transport $T_c$ four times higher than bulk FeSe (8 K); and (c)  the subsequent
ARPES gap closing temperature of 65 K measured on similar samples \cite{Zhou_monolayer}.  The $T_c$ enhancement in monolayer FeS films on STO is not observed~\cite{Shigekawa2019} and the same was found when FeSe is deposited on graphene \cite{Song2011II} or Bi$_2$Se$_3$ \cite{Eich2016}.
Taken together, these observations point to a unique high-$T_c$ superconducting system based on FeSe where the substrate \STO\,plays an essential role.  Exactly what that role is, is still being debated.  Here~we sketch and extend the discussion of this question in the excellent review of Huang and Hoffman \cite{Huang2017}.

\subsubsection{Electronic Structure and Electron Doping}

\label{subsubsec:band_structure_e-doping}

ARPES measurements have elucidated in great detail how the
high-$T_c$ superconducting state evolves out of the as-grown sample, and how the requisite electronic structure evolves with it \cite{Zhou_monolayer}.
When one starts the annealing process, the $\Gamma$-centered  hole band is at the Fermi level, as in a typical Fe-pnictide,  but by the final stage
it has been pushed 80 meV below, and the electron pockets have correspondingly enlarged.   The Fermi surface therefore consists only of  electron pockets at the M points (Figure~\ref{fig:3Fermisurfaces}).     Note that the Fermi surface obtained by standard DFT calculations is significantly different from ARPES, even if the system is electron doped ``by hand'' using rigid band shift or virtual cluster methods, and even accounting for the strained lattice constant imposed by the STO  \cite{Wang12,Huang2017,Tikhonova2020}.
However, strain-modified hopping parameters were proposed to qualitatively account for the band structure in the monolayer FeSe \cite{Hao_PRX2014}.

Initially it was believed that Se vacancies created in the film itself during  the annealing process might electron-dope the film, but these appear to induce a hole doping effect instead  \cite{Berlijn_Sevacancy}.  More~recently, attention has focused on the doping of the STO layer by O vacancies in various configurations.  In DFT studies of the FeSe-STO interface, it has been speculated that the O vacancies give rise to a $T_c$ enhancement due to a surface reconstruction \cite{Bang2013}  or suppression of
an incipient monolayer spin density wave \cite{Gong2014}.  A problem with such calculations is that unphysically large  O vacancy concentrations appear to be required to suppress the position of the hole band sufficiently, suggesting that electron correlation effects play an important role, consistent with conclusions for the bulk FeSe material.  Indirect evidence against O vacancy doping scenarios comes also from measurements of FeSe on anatase TiO$_2$ (001) surfaces, where high-$T_c$ superconductivity was deduced by large STM gaps similar to FeSe/STO (001).
In this system, direct imaging of O vacancies was shown to give a concentration much too small to account for the doping level observed \cite{Ding2016}, and variation of O vacancy content did not affect the gap.

Other theoretical approaches to the charge transfer problem focus on the novel properties of the STO itself, in particular due to large work-function mismatch. In this picture, the strong coupling to long-wavelength polar phonons generated in the depletion region by the nearly ferroelectric character of STO can enhance superconductivity \cite{Millis2016}.   Charge transfers of the required magnitude can be obtained by this mechanism, but the details of the renormalized band structure of the interface was not addressed in this work.

\subsubsection{Structure of the Interface}
\label{subsubsec:interface}
The calculations and analyses above assumed a single layer of FeSe deposited on the TiO$_2$ terminated layer of \STO.  However, as pointed out by Huang and Hoffman \cite{Huang2017}, the fabrication process does not necessarily result in such a simple structure.  Several groups provided evidence for a reconstructed interface that creates a TiO$_x$ double layer at the interface \cite{Li2016II,Zou2016}.  This may certainly aid in the charge transfer process to the FeSe, but suffers from the same requirement as discussed above that the absolute number of O vacancies required, of ${\cal O}(50\%)$, appears to be too large. Zhao~et~al.~\cite{Chan2018} performed scanning electron transmission (STEM) imaging together with electron energy loss spectroscopy (EELS) and concluded that,  in addition to the double TiO$_x$ layer, a Se layer in proximity to the FeSe was necessary to explain observations.  This is difficult to understand because the annealing process is intended specifically to remove excess Se.  A further proposal came from Sims~et~al.~\cite{Sims2019}, who suggested on the basis of STM, STEM and DFT calculations that an interlayer close to Ti$_{1.5}$O$_2$ (excess Ti) ``floats'' between the FeSe and the STO, weakly van der Waals coupled to both, and provides part of the requisite doping to shift down the central hole band.

\subsubsection{Transition Temperature}
\label{subsubsec:monolayer_Tc}
\begin{figure*}[tb]

\centering
{\includegraphics[width=\linewidth]{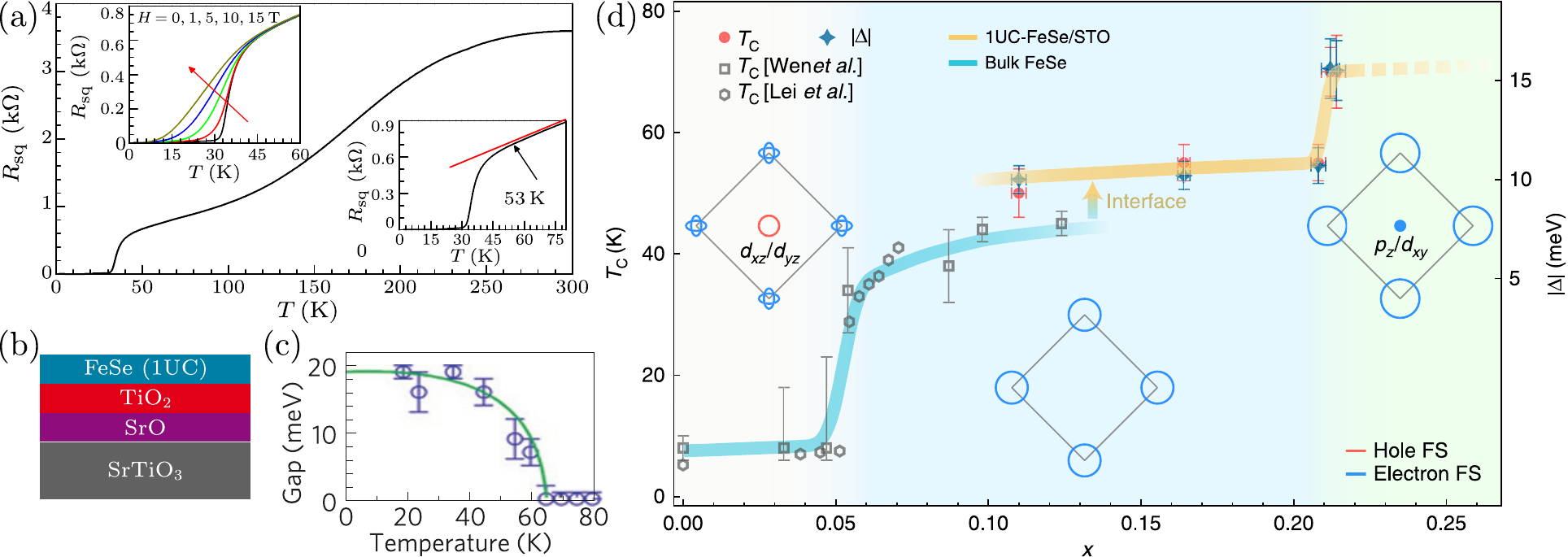}}
\caption{(\textbf{a}) Resistivity of original monolayer FeSe films on STO \cite{Wang12} with layered structure, as~shown in (\textbf{b}). (\textbf{c}) Spectral gap measured by
ARPES on such films.
Reprinted with permission from Springer~\cite{Zhou_monolayer}, copyright (2013).
(\textbf{d})  Effects of surface dosing on $T_c$ of both FeSe bulk crystals and monolayer films on \STO, as~determined by gap closing.
Reproduced from  \cite{Shi_Ding2017}. \ccfour.}
\label{fig:FeSe_mono}
\end{figure*}

While monolayer FeSe on STO has frequently been cited as exhibiting the highest critical temperature in the FeSCs, it is important to examine this claim critically.  Normal published accounts of superconductivity require proof of (a) zero resistance and  (b) the Meissner effect, and ideally, (c) other measures of an energy gap closing.  Most of the measurements of $T_c$ in these systems have been of type (c), rather than (a,b), due to the difficulties inherent in the low dimensionality and air sensitivity of the samples.  ARPES, which measures the onset of an energy gap in the one-particle spectrum rather than a phase coherent state, has  generally  reported the highest $T_c$s \cite{Wang12,Zhou_monolayer,Tan13,Shen_FeSe_STO,Peng2014}, between 60 and 70 K; see Figure~\ref{fig:FeSe_mono}c.  Most attempts to cap the samples  to avoid the air sensitivity problem have apparently led to sample degradation, such that ex situ transport measurements have generally yielded considerably lower zero resistance $T_c$s in the 20--30 K range, with 40--50 K onset values. Very~recently, a~zero resistance $T_c$ of 46 K was reported for monolayer films on LaAlO$_3$ substrates \cite{Shikama2020}. The~single in situ transport result, reporting a $T_c$ of 109 K with a 4-probe ``fork'' measurement \cite{Ge2015}, has not been~reproduced.

Magnetization measurements \cite{Zhang_Meissner2014,Sun_Meissner2014,Zhang_Meissner2015} generally report high onset temperatures consistent with ARPES, but have very broad transitions, and significant suppression of the magnetization does not occur until lower temperatures near the zero resistance $T_c$s of the ex situ transport zero resistance measurements.  Still, it is possible that these lower $T_c$s are due to extrinsic experimental difficulties. An~outstanding problem is therefore to prove that long-range or quasi-long range superconducting phase coherence really does set in at the higher (60--70 K) temperatures with probes of true superconducting order rather than gap closing.

\subsection{Dosing of FeSe Surface}
\label{subsec:dosing}

The question of why $T_c$, interpreted optimistically, is so much higher for the FeSe monolayers on STO than either FeSe itself, or, for that matter, all the other Fe-pnictides and chalcogenides, led~to speculation that the high levels of electron-doping, perhaps related to the special Fermi surface structures shown in Figure~\ref{fig:3Fermisurfaces} might be responsible.  This led to new attempts to  enhance $T_c$ via doping by novel means.  Utilizing a technique pioneered by Damascelli for the cuprates \cite{Damascelli2008}, potassium atoms were deposited on the surface of films or crystals of FeSe \cite{Miyata2015,Wen_DL_Feng_2016,Song_16}, sufficient to increase the electron doping of the surface layers.  Critical temperatures of the surface layer as high as 40 K were reported, roughly the same as bulk FeSe maximum $T_c$ under pressure, and similar to FeSe intercalates (see below). Similar results were reported utlilizing ionic liquid gating \cite{XHChen2016}.  Alkali dosing experiments were performed also on FeSe/STO monolayers, with about 10\% additional electron dosing leading to a jump of about 10 K to about 70 K \cite{Shi2017}.  The authors of reference \cite{Shi2017} associated the initial rise of $T_c$ in crystals upon electron doping to a Lifshitz transition when the central hole pocket disappeared, and the rise in the more highly doped FeSe/STO monolayers as due to a second Lifshitz transition when a central electron pocket appeared, as shown in Figure~\ref{fig:FeSe_mono}d.

\subsection{Replica Bands and Phonons}
\label{subsec:replica}
A potentially  important clue to the physics of these systems, and the influence of the substrate, was found in ARPES measurements \cite{Shen_FeSe_STO}, which identified in second derivative spectra clear ``replica bands'',  ``shadow'' copies
of bands both at $\Gamma$ and M shifted rigidly downward in energy by $\sim$100~meV.  These experiments, along with others reviewed recently in reference \cite{Xu_2020}, were interpreted by the authors of reference \cite{Shen_FeSe_STO} as implying the presence of a strong  interaction of FeSe electrons with phonons, probably originating from the substrate.  It was furthermore argued  that the electron--phonon interaction must be rather strongly peaked in the forward direction $ q=0$ to explain this observation.  Theories of the influence on electron pairing of forward electron--phonon scattering in cuprates \cite{Kulic1994} and Fe-pnictides~\cite{Aperis2011} had been elaborated earlier.  The connection with the unusually high $T_c$ in the monolayer system made by the authors of reference \cite{Shen_FeSe_STO} was more or less circumstantial,  but~shortly thereafter it was proposed theoretically that coupling primarily to such a forward scattering phonon could provide a natural explanation for the high $T_c$, since $T_c$ was found to vary linearly rather than exponentially with the electron--phonon coupling in this
extreme case \cite{Johnston_FeSe-STO,Kulic2016}.  A multi-band version of this proposal~\cite{Aperis2018} reached similar conclusions but pointed out the important contribution of incipient bands to the interaction.

It is important to note that such forward scattering phonons are potentially interesting for two reasons.  First, electron--phonon superconductivity and unconventional superconductivity based, e.g., on spin fluctuations, generally  compete, because the latter usually relies on strong repulsive interband interactions.  Conventional attractive electron--phonon couplings vary slowly with momentum transfer, and therefore suppress the interband interaction leading to pairing. Forward scattering phonons, on~the other hand, provide an attractive intraband pairing and therefore should add to the total attraction in the appropriate channel.
Secondly, the forward scattering electron--phonon interaction  by itself may lead to an unusually large
$T_c$ due to the linear dependence on the coupling \cite{Johnston_FeSe-STO}; however~this requires that the repulsion seemingly present in all other FeSC would be negligible in this system, which seems unlikely.   In other words, forward scattering allows phonons to assist spin fluctuations, but may not necessarily amplify the usual electron--phonon mechanism in any qualitative way.

\begin{figure*}[tb]

{ \includegraphics[width=\linewidth]{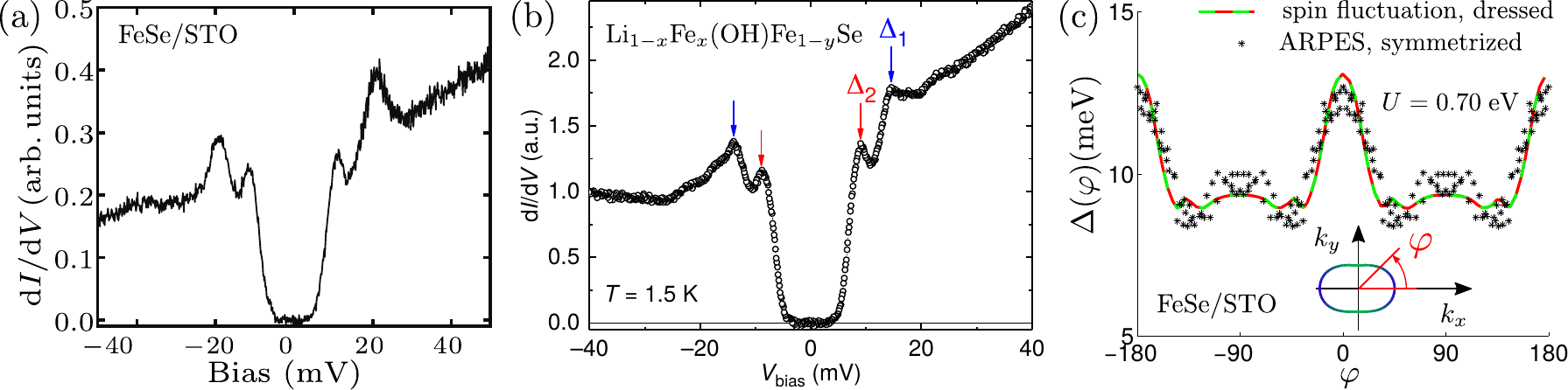}}
\caption{STM spectra of (\textbf{a}) the FeSe monolayer of reference \cite{Wang12}; and (\textbf{b}) of Li$_{1-x}$Fe$_x$(OH)Fe$_{1-y}$Se, both exhibiting two-peak structures at both positive and negative bias.~Reproduced from  \cite{Du2016}.~\mbox{\ccfour}.
(\textbf{c}) Fit to ARPES-determined gap on electron pocket from reference \cite{Zhang16}  within orbitally-selective spin-fluctuation theory;
the two maxima in $\Delta$ give rise to two saddle points in the quasiparticle dispersion, consistent with two coherence peaks in the spectra.
Reprinted figure with permission from~\cite{Kreisel2017}, copyright by the American Physical Society (2017). }
\label{fig:LiOH_STM}
\end{figure*}

STM and ARPES measurements have both reported   a full superconducting gap in the monolayer system, of order 20 meV, drastically different both in magnitude and in isotropy   relative to bulk FeSe.
There are two coherence-like peaks in the STM spectrum \cite{Wang12}  (see Figure~\ref{fig:LiOH_STM}a), not unlike other Fe-based systems; however, in this case, the hole band is presumed not to participate in
superconductivity.   Furthermore, the two electron pockets at $M$ in ARPES do not appear to  hybridize~\cite{Shen_FeSe_STO}, so that the double peak is unlikely to be explained by two isotropic gaps on these bands. The~most likely scenario is that the two energy scales  indicate two independent maxima on the electron pockets (minima do not lead to peaks in the STM spectrum, at least within BCS theory), as~indeed measured by ARPES \cite{Zhang16}.

Despite the apparent effect of phonons on the ARPES measurements, electron--phonon interactions in the FeSe are likely too weak to by themselves explain a $T_c$ of 70 K or above \cite{Li_etal_2014,Linscheid2016A,YWang2016} (a~calculation that  finds a much higher $T_c^{ph}$ than others under some
rather generous assumptions is given in  reference~\cite{Cohen15}).
Thus a ``plain'' $s$-wave from attractive interactions alone seems improbable, even if  soft STO phonons play a role \cite{DHLee2012,Shen_FeSe_STO}.   The forward scattering scenario for electron--phonon processes~\cite{Shen_FeSe_STO,Johnston_FeSe-STO} then implies that phonons cannot  contribute significantly to the interband~interaction.

\subsection{Pairing State in Monolayers\label{subsec:monolayer_gap}}

\subsubsection{e-Pocket only Pairing: \texorpdfstring{$d$}{d}- and Bonding--Antibonding \texorpdfstring{$s$}{s}-Wave}
\label{subsubsec:e-pocket_gaps}
By itself, the interband  spin fluctuation interaction due to pair scattering between electron pockets, considered in the 1-Fe zone,  should lead  to nodeless $d$-wave (since $\chi({\bf q},\omega)$ will be roughly  peaked at the momentum connecting the electron pockets) \cite{Fangdwave,Maierdwave}.
The double maximum in the gap function found by ARPES \cite{Zhang16} does not arise from conventional spin fluctuation pairing theory for this system~\cite{Kreisel2017}, however, so forward scattering phonons could potentially not only boost this mechanism, but~also contribute to the observed anisotropy of the $d$-wave gap on the electron pockets.

An alternative explanation entirely within the spin-fluctuation approach, but incorporating the orbital-selective renormalizations of the dynamical susceptibility as described in Section~\ref{subsubsec:orb_selective_pair} was given in reference \cite{Kreisel2017}.  Simply suppressing the $d_{xy}$ orbital weight $Z_{xy}$, with nearly negligible renormalizations of the $d_{xz/yz}$ weights, consistent with the nearly absent evidence for nematicity in this system, was sufficient to produce the double maximum of the $d$-wave gap on the electron pockets at the correct energies (Figure~\ref{fig:LiOH_STM}c).
Recently, spin-fluctuation mediated pairing was examined using a full-bandwidth Eliashberg approach, finding a $d$-wave instability as well, but claiming that the high critical temperature cannot be obtained \cite{Aperis2020}.
The existence of such a $d$-wave gap was also deduced phenomenologically from STM measurements on thin films of FeSe on \STO ~with a step edge \cite{Ge2019}.

In the 2-Fe zone, the two electron ellipses overlap and may hybridize due to orbital mixing or SOC. The former is forbidden by symmetry in the monolayer, but SOC may play a role.  In the case of large hybridization from either source, the bonding--antibonding $s$-wave state between two electron pockets \cite{HKM_ROPP} is expected to be stabilized by repulsive interactions \cite{Khodas2012}, but ARPES has not observed any hybridization of the two electron pockets, suggesting that these effects are small \cite{Shen_FeSe_STO}.  This point is also relevant for the discussion of nodes on a $d$-wave gap.  A $d_{x^2-y^2}$ state defined in the 1-Fe zone has nodes along the (0,0)-$(\pi,\pi)$ direction, which does not intersect the electron pockets.  However,  to~the extent the two elliptic pockets hybridize and split, nodes will be forced on the hybridized inner and outer electron sheets.   Note that these nodes on the Fermi surface are not required by symmetry as in the 122 crystals \cite{Mazin2011}, and have a narrow angular range proportional to the magnitude of the hybridization, and hence are sometimes referred to as ``quasinodes'' \cite{Kreisel13}. In the monolayer system, ARPES and STM report a full gap, as discussed above; so if such quasinodes exist, they must contribute a negligible amount of phase space for low-energy excitations.

\subsubsection{ SOC Driven Pair States}
\label{sec:SOC_driven}

The spin-orbit coupling is not particularly strong in the 3d Fe-based superconductors, but~it is sufficient to create band splittings of order tens of meV in the low-energy band structure, which~can lead to some important effects.  Furthermore, it has been found to be particularly strong in the \mbox{11 materials \cite{BorisenkoSO,Day2018}}.
Direct evidence for the significance of SOC for superconductivity comes from
the  discovery of magnetic susceptibility anisotropy ($\chi_{xx}\ne \chi_{yy}\ne \chi_{zz}$) in the neutron spin-resonance~\cite{Lipscombe2010,Ma_2017,Scherer_NS2018}, which is generally understood to exist due to a coherence factor depending on the sign change of the superconducting gap below $T_c$.    This type of spin response anisotropy in the superconducting state can in principle be captured qualitatively merely by incorporating SOC in the electronic structure but ignoring it in the superconducting pairing itself~\cite{Korshunov_SOC_2013}. Similar approaches to incorporating SOC in  superconductivity  were adopted in treatments of spin fluctuation pairing in 122 and 111 systems, to the extent that the SOC-induced hybridization of bands at high-symmetry points on the electron pockets  was included \cite{Kreisel13,Wang13}.  This effect was in fact found to be rather small.  However, it has been argued that in the strongly electron-doped systems,  the hybridization of the electron pockets is more significant due to stronger SOC \cite{Kreisel13,Cvetkovic2013,Agterberg2017,Eugenio2018}.

It  is clear in general that the pairing problem itself is also influenced by SOC.  Since the ${\bf L}\cdot {\bf S}$ interaction preserves time reversal symmetry,  one should pair states in the pseudospin basis \`a la Ng and Sigrist \cite{NgSigrist00}.  Including these effects in realistic multiband models with repulsive interactions can be quite cumbersome, but was implemented by Scherer and Andersen, who also found that  within the traditional RPA approach, the effects of SOC on the gap structure were actually  negligible \cite{Scherer2019}.  Recall, however, that this method considers pairing only between time-reversed states at the Fermi level belonging to the same band.

In an alternate approach, SOC is included in the one-body Hamiltonian treated in the $\bf k\cdot p$ approximation, which preserves crystal symmetry near the $M$ points, and the Hubbard--Kanamori interaction is projected onto these low-energy states and decomposed in mean field theory.  In one-band interacting Hamiltonians with repulsive interactions, such a procedure cannot lead to a stable pair state, and one is forced to compute the effective interaction that leads to unconventional pairing in the usual way \cite{MaitiChubukovreview}.
In multiband models, although all bare interactions
$U,J,U',J'$ are repulsive,  attractive interactions are found in certain channels for some choices of the interaction parameters.  For example, in the Vafek--Chubukov model \cite{VafekChubukov2017} for two orbitals and two hole pockets, the ``$A_{2g}$''  spin-triplet state
\begin{equation}
\Delta \sim \langle \psi_\alpha^T(\rr{}) \tau_2 (i s_zs_y)_{\alpha\beta}\psi_\beta(\rr{})\rangle,
\end{equation}
corresponds to interaction constant $(U'-J)/2$, which may under physically reasonable circumstances become negative.
Here $\psi_{\rr,\sigma}=(d^\dagger_{yz,\sigma}(\rr),-d^\dagger_{xz,\sigma}(\rr))$, $\tau_i,s_i$ are Pauli matrices in orbital and spin space, respectively.    It is easy to check that there is no Cooper logarithm driving a superconducting instability in such cases, so a finite threshold value is required for the coupling.   However, the effect of even infinitesimal SOC is found to induce a Cooper log \cite{VafekChubukov2017}, leading to the suggestion that such exotic pair states occur ``naturally'' in Fe-based superconductors despite
interorbital pairing, which would normally be suppressed because it generically requires significant pairing of electrons of states away from the Fermi energy.
In the case of electron pocket only systems, the even parity states stabilized in mean field \cite{Eugenio2018,Boeker2019} are essentially those proposed by Khodas and Chubukov \cite{Khodas2012}, but of course contain admixtures of spin-triplet components due to SOC.

Agterberg and co-workers have followed a similar approach, considering a phenomenological pairing  interaction purporting to describe spin-fluctuations of N\'eel type, which in the 2-Fe zone correspond to small-$\bf q$ scattering processes \cite{Agterberg2017,Nakayama2018}.  These then scatter pairs between folded electron ellipses.  Spin-orbit coupling hybridizes these ellipses as expected, but for values consistent with upper bounds set by ARPES \cite{ZhangY15} may produce ``naturally'' a true nodeless $d$-wave state, consistent with experiment. A disadvantage is that the theory, which studies only two pockets, does not apparently distinguish the  $d$-wave from the bonding--antibonding $s$-state that should occur at sufficiently large SOC.
In contrast, a theory of pairing by nematic fluctuations finds a gap with the same sign on both electron pockets, stabilized by SOC over the competing $d$-wave state \cite{Kang2016}.
A final proposal for e-pocket only systems studies the exchange of spin and orbital fluctuations, going beyond the usual RPA to include Aslamosov--Larkin type vertex corrections, claiming a conventional $s$-wave ground state \cite{Yamakawa2017a}. As discussed in Section~\ref{subsec:impurity}, there is significant evidence against $s_{++}$ scenarios from the observation of nonmagnetic impurity bound states in this system.

\subsubsection{\texorpdfstring{Incipient Band $s_\pm$ Pairing}{incipient band s+- pairing}}
\label{subsubsec:incipient}
Although the $\Gamma$-centered hole band observed by ARPES lies $\sim$80 meV below the Fermi level and is usually neglected in pairing studies, a few authors have discussed the possibility of pairing in the  ``incipient $s_\pm$'' state, driven by the conventional spin fluctuation interaction between the hole band and the electron band centered at $M$. Naively, such a state is  disfavored by energetic arguments \cite{HKM_ROPP}.  On the other hand,  Bang \cite{Bang2014} and Chen et al. \cite{Xiao_incipient2015} revisited these arguments and found that in the presence of robust Fermi surface-based superconductivity (e.g., a phonon attraction in the electron pockets of the monolayer) this state could be strongly favored, consistent with the findings of Lee and co-workers~\cite{DHLee2012,Shen_FeSe_STO,DHLee2015}.

This scenario does not address the central question of why superconductivity appears to be stronger in situations with electron-like Fermi surface pockets  only. However, Linscheid et al. \cite{Linscheid2016} pointed out that if one includes the dynamics of the spin fluctuation interaction, high-$T_c$ pairing in a traditional $s_\pm$ state with incipient hole band could be understood.  Consider a situation with a constant interband pairing interaction between an electron band crossing the Fermi surface and an incipient hole band; without an intraband attraction, no robust superconductivity can be produced \cite{Xiao_incipient2015,Chubukov2016A}.  On~the other hand, if the interaction is calculated self-consistently, moving the hole band below the Fermi level can enhance the interband pairing because the paramagnon spectrum is peaked at a finite energy $\omega_{sf}\sim$ 50--100 meV.  Thus within the incipient band $s_\pm$ picture, a ``sweet spot'' in the pairing interaction can be obtained where the hole band extremum is a comparable distance below the Fermi level \cite{Linscheid2016}. This scenario has not yet been confirmed within a realistic multiorbital framework;
Eliashberg spin fluctuation calculations using multiple bands find only a weak enhancement of $T_c$ due to incipient pairing \cite{Aperis2020}.

\subsection{Impurity Experiments}
\label{subsec:impurity}
The previous section mostly reviews proposals for pairing states in the FeSe monolayer that involve sign changes of the order parameter over the Fermi surface, ultimately due to the repulsive electronic interactions.
On the other hand, there is some evidence that the system does not have a sign-changing gap. In  STM measurements by Fan et al.  \cite{Fanetal_impurities_Fe-STO15} $T_c$ and the gap were reported to be   suppressed only by magnetic impurities, as one might indeed expect from a ``plain'' $s$-wave SC.  These arguments, if correct, would also rule out states of the ``bonding--antibonding $s$-wave'' type~\cite{HKM_ROPP}.  However, the impurities in these studies were adatoms on the surface of the monolayer rather than atoms substituting in the layer itself, and it is possible that the potentials produced in the Fe plane were simply too weak to produce bound states.   In a subsequent study where various impurities were incorporated
into the monolayer, bound states were observed for certain nominally nonmagnetic atoms \cite{Liu2018}, suggesting a sign change in the superconducting order parameter.

\section{FeSe Intercalates\label{sec:intercalates}}

\subsection{Alkali-Intercalated FeSe}
\label{subsec:alkali_interc}

The apparent paradox of high-temperature spin fluctuation driven  pairing in electron-pocket only systems was raised first in the context of the alkali-intercalated FeSe materials, discovered in 2010 \cite{Guo2010},  which nominally correspond to the chemical formula AFe$_2$Se$_2$, with A = K, Rb, Cs.  To~this date,  the superconducting samples of these  materials are available only in mixed-phase form and have relatively low superconducting volume fractions.  There is considerable evidence from STM and X-rays that the superconductivity exists only in 3D filamentary form \cite{XDing2013}.    These systems nevertheless excited considerable interest both because of their proximity to an unusual high-moment block  antiferromagnetic phases, and because    ARPES \cite{KFe2Se2ARPES} measurements on KFe$_2$Se$_2$ reported that there were no $\Gamma$-centered hole pockets at the  Fermi level (although a small electron pocket pocket is found near the $Z$ point near the top of the Brillouin zone). At  $\Gamma$, the hole band maximum is $\sim$50 meV below the Fermi level.  An~example of one of the ARPES-determined Fermi surfaces of these materials is shown in Figure~\ref{fig:3Fermisurfaces}c.

Several workers recognized early on that despite the missing hole pockets, repulsive interactions at the Fermi level existed between  electron Fermi surface pockets, which could lead to $d$-wave pairing with
critical temperatures \cite{Fangdwave,Maierdwave} of roughly the same order as in the usual $s_\pm$ hole-electron pocket scattering scenario.  As discussed above,  the expected hybridization of the two electron bands in the proper 122 body-centered tetragonal crystal symmetry leads to two roughly concentric electron Fermi surface  sheets at the $M$ point in the 2-Fe Brillouin zone, leading also to the possibility of the bonding--antibonding $s$ wave state, as in the monolayer.  In the 122 structure, however,
such~a state was  found to be subdominant to $d$-wave pairing due to the relatively weak hybridization found in first principles calculations for KFe$_2$Se$_2$ \cite{Kreisel13}. However,  the bonding--antibonding $s$-wave remains an interesting candidate in part because these systems are  apparently intrinsically multiphase; therefore~one may question  the conventional  electronic structure derived from  ARPES results on a metallic, filamentary phase embedded in an insulating background, which apparently depend on 
the assumption that averaging over a micron size domain provides a reliable description of the intrinsic properties of this phase.

Inelastic neutron scattering measurements \cite{Inosov2012} agree rather well with the wave vector $\sim (\pi,\pi/2)$ for
the neutron resonance found in calculations of reference \cite{Maierdwave},  corresponding  to scattering between the
sides of the electron pockets centered at $(\pi,0)$ and $(0,\pi)$ in the 1-Fe zone.  Such an interorbital scattering process was found to lead to a $d$-wave ground state, which appears to disagree with the absence of nodes on the small $Z$-centered pocket observed by ARPES \cite{Xu2012}.  It may be difficult given current momentum and energy resolution to reliably measure the spectral function pullback below $T_c$ on an extremely small Fermi pocket, however. In addition,
Pandey et al. \cite{Pandey2013} argued that the bonding--antibonding $s$ wave state would also support a resonance at roughly the same wave vector observed in experiment.  This explanation is  natural in 2D, since the bonding--antibonding state represents a folded $d$-wave state in the 2-Fe zone stabilized by hybridization, it is less obvious in 3D for the bct crystal structure of these systems.  Furthermore, such a scenario requires significant hybridization \cite{Khodas2012} that is not present, at least in DFT \cite{Kreisel13}.

The uncertainties associated with this system, particularly the materials issues, have left the question of pairing unresolved. Taking the nodeless Z-pocket as a given,  another solution was proposed by Nica et al. \cite{Nica2017} as a way of understanding both neutron and ARPES experiments, by~constructing a pair function that builds both $s$ and $d$-wave symmetry into different orbital channels, such that different symmetry channels  effectively dominate different Fermi surface sheets. The resulting orbitally mixed state was shown to be the leading candidate within a $t-J_1-J_2$ mean field calculation over some range of parameters.

\subsection{Organic Intercalates}
\label{subsec:organic_interc}
The origin of the higher $T_c$ in the alkali-intercalated FeSe  remains unclear, but after their discovery one obvious explanation was simply the  enhanced  FeSe layer spacing, possibly by  enhancing two-dimensionality and therefore Fermi surface nesting.
Intercalation of larger spacer molecules between the layers, initially organic molecular complexes,   was achieved shortly thereafter~\cite{Burrard-Lucas2013,Scheidt2012,Sedlmaier2014,
Noji2014,
Krzton_Maziopa2012}.   These materials indeed had higher $T_c$s, up to 46 K, but were  extremely air-sensitive and only powders were available, so there is no ARPES data on either
Li$_{0.56}$(NH$_2$)$_{0.53}$(NH$_3$)$_{1.19}$Fe$_2$Se$_2$ with $T_c =
39~\mathrm{K}$~\cite{Sedlmaier2014} or
Li$_{0.6}$(NH$_2$)$_{0.2}$(NH$_3$)$_{0.8}$Fe$_2$Se$_2$ with $T_c =
44~\mathrm{K}$~\cite{Burrard-Lucas2013}.
Noji et al.~\cite{Noji2014} reported a wide variety of FeSe intercalates, along with a strong
correlation of $T_c$  with inter- FeSe layer spacing,  with a quasilinear increase between 5 and $9\,$\AA,   after which $T_c$ saturated   between 9 and $12\,$\AA.
This tendency was plausibly explained by Guterding et al. \cite{Guterding2015} within spin-fluctuation pairing theory as due to a combination of doping and changes in nesting with increasing two-dimensionality.  Recently, Shimizu et al. \cite{Shimizu2020}   extended this work with a detailed discussion of the doping dependence of $T_c$ in the organic intercalate Li$_x$(C$_3$N$_2$H$_{10}$)$_{0.37}$FeSe. While the ammoniated FeSe intercalates are fascinating, their air sensitivity prevented many important experimental
probes and limited their utility.

\subsection{LiOH Intercalates}
\label{subsec:lioh_interc}
As mentioned above, there is some similarity between the low-energy band structures of several electron-doped FeSe materials a Fermi surface without $\Gamma$-centered hole pockets, including the monolayers on STO and the alkali-doped intercalates, already discussed above.  There is a third class of air-stable FeSe intercalates that fits into this category \cite{Zhao2016}, the lithium iron selenide hydroxides,
reported in references \cite{Sun_LiOH2014,Lu2014} (Figure~\ref{fig:3Fermisurfaces}a).

While the surface of the alkali-intercalated FeSe does not cleave easily, and aside from a full gap it is difficult to discern distinct features \cite{Li2012}, the STM spectra of  FeSe monolayers and LiOH-intercalated FeSe show a striking similarity, with both exhibiting  double coherence peaks with roughly the same large gap/small gap ratio \cite{Du2016}, both with extremely large inferred gap-$T_c$ ratios of order 8 (see Figure~\ref{fig:LiOH_STM}).  Du et al. \cite{Du2016} attributed the two peaks to gaps on two hybridized electron pockets, in~contrast to the interpretation of reference \cite{Zhang16}, which proposed two separate maxima on each unhybridized electron pocket  in the case of FeSe monolayer (see Figure~\ref{fig:LiOH_STM}c).  Note that within BCS theory, only gap maxima, not minima, produce peak structures in the density of states.

The further observation of an in-gap impurity resonance at a native (Fe-centered) defect site~\cite{Du2016} suggests that the gap is sign changing.  This conclusion was bolstered by a subsequent study of phase-sensitive quasiparticle interference (QPI) by the same group \cite{Du2018}, who found a strong single-sign antisymmetrized conductance between the two gap energies, a signature of sign-changing gap structure~\cite{HAEM2015}.
They concluded  that the LiOH intercalate system has a sign-changing gap, but could not distinguish reliably between a bonding-antibonding $s_\pm$ state and a nodeless $d$-wave state.

As discussed in Section~\ref{sec:SOC_driven}, treatments of pairing in the orbital basis including SOC have come to the conclusion that additional superconducting states beyond the $d$-wave and bonding--antibonding s-wave may play a role in the electron-pocket only systems if SOC is important \cite{Eugenio2018,Boeker2019}.  These exotic pair states are either not present in the conventional Fermi surface based approaches, or correspond to strongly subdominant pairing channels.  Specifically, in the context of LiOH-intercalated FeSe, Eugenio and Vafek \cite{Eugenio2018} proposed that the ground state of this system could be an interorbital spin triplet state, citing as evidence the two-peak structure seen in STM at positive bias (see Figure~\ref{fig:LiOH_STM}; this~structure has alternative explanations, as discussed in Section~\ref{sec:monolayer}).  Gaps away from the Fermi surface are characteristic of interband pairing amplitudes induced by the interorbital interaction \cite{Eugenio2018}.

\begin{figure*}[tb]

\centering
{\includegraphics[width=\linewidth]{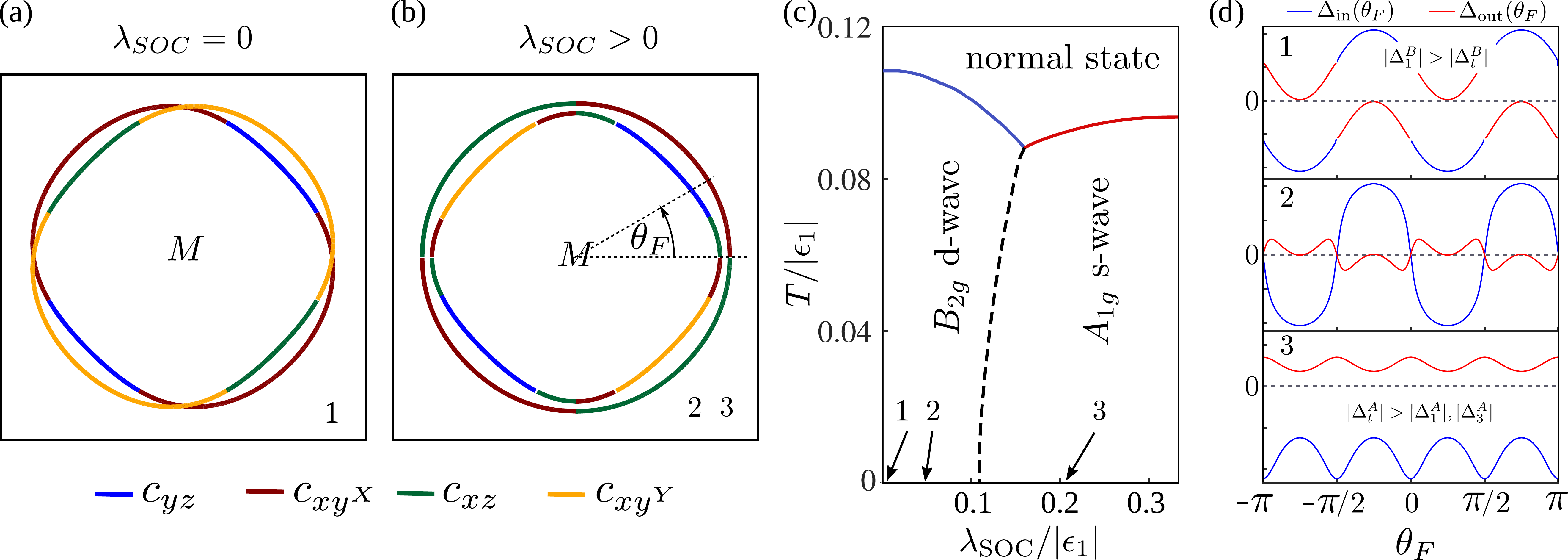}\vspace{-.15cm}}
\caption{(\textbf{a}) Electron pockets at the M point without SOC corresponding to is X-pocket and Y-pocket in the 1 Fe BZ; compare Figure~\ref{fig:FS_orbitals}. (\textbf{b}) Electron pockets with finite SOC leading to lifted degeneracy at the Fermi angles $\theta_F=0,\pi/2,\pi,3\pi/2$.
(\textbf{c}) Phase diagram as a function of temperature and interband SOC in units of $|\epsilon_1|=45$ meV, the energy  at the M-point of the two electron bands.
(\textbf{d}) Corresponding superconducting gap projected on the inner (blue) and outer (red) Fermi-surface as function of the Fermi angle $\theta_F$ at three different values of $\lambda_{\text{SOC}}$.
Reproduced from  \cite{Boeker2019}. \ccthree.\vspace{-.2cm}
}
\label{fig:MF_SOC}
\end{figure*}

The full mean-field phase diagram in the presence of SOC was worked out by B\"oker et al. \cite{Boeker2019}, and is partially displayed in Figure~\ref{fig:MF_SOC}c together with the order parameters at selected values of the SOC (d).
For nonzero SOC, the
superconducting order parameter is a definite parity combination of spin singlet and spin triplet states. In the weak SOC limit,  a dominant spin singlet and small spin triplet gap combination is stable, yielding a state essentially equivalent either to the bonding--antibonding $s$ or quasinodeless $d$
identified in the usual spin fluctuation approach. For stronger SOC, however, the superconducting order
parameter evolves into a combination of spin singlet and dominant spin triplet gaps in each state. In the
(Li$_{1-x}$Fe$_x$)OHFeSe system,  the even parity $A_{1g}$ and $B_{2g}$ pairing states with dominant spin triplet
component appear to be consistent with available experiments indicating a full gap, including current quasiparticle interference data \cite{Boeker2019}.
The~$A_{1g}$ state shown in Figure~\ref{fig:gaps}c
is slightly favored.
The~spin-singlet dominated $A_{1g}$ and
$B_{2g}$-states in this scenario without strong spin fluctuations (mean field approximation) are not consistent with at least one of the existing experiments.
The states with dominant spin triplet pairing may be, and B\"oker et al. proposed ways to detect them with spin polarized quantum interference measurements \cite{Boeker2019}.

The end result of this analysis is not clear: on the one hand, it is striking that these exotic states requiring interorbital pairing are obtained in a ``natural'' way from mean field theory.  On the other hand, on general grounds the conventional effective spin fluctuation interaction should be significantly stronger and drive pairing in the
nodeless $B_{2g}$, incipient $A_{1g}$, or (singlet) bonding--antibonding $A_{1g}$ channels (Figure~\ref{fig:gaps}b--d).
Experiments probing the presence of a spin triplet component would therefore be of the greatest interest.  At the same time, it should be possible to make progress theoretically to treat the ``exotic'' states on the same footing with the ``conventional'' ones within a generalized spin fluctuation pairing theory that incorporates the SOC into the pairing interaction, and allows for pairing away from the Fermi surface.  Only such an approach will be able to ultimately decide whether the exotic pair states are energetically favorable.

\section{Topological Phases of Matter in Iron-Based Superconductors\label{sec:topo}}

\subsection{Basic Properties of \texorpdfstring{Fe$_{1+y}$Te$_{1-x}$Se$_x$}{Fe1+yTe1-xSex}}

Studies of possible topological states of matter in iron-based superconductors are mainly concentrated on the material Fe$_{1+y}$Te$_{1-x}$Se$_x$. Before embarking on a discussion of potential topologically nontrivial effects, it is worth briefly reviewing the basic properties of this material, since~they are not particularly ``basic'', and hence important to keep in mind. Fe$_{1+y}$Te$_{1-x}$Se$_x$  has played a prominent role throughout the ``iron age'', despite the need to invest considerable effort to control and optimize sample quality, and to decipher the  roles of  the structural and magnetic order, possible  phase separation, and excess Fe ions. These issues, and many others, have been reviewed in several review papers dedicated specifically to this material \cite{Wen_2011,Sun_2019,Tranquada2019}. The excess Fe ions, indicated by $y$ in Fe$_{1+y}$Te$_{1-x}$Se$_x$, lead to a partial occupation of the second Fe site in the crystal, and complicate the characterization of the material since these sites disorder, dope, and locally magnetize the system. For~a recent review of the role of Fe non-stoichiometries and annealing effects in Fe$_{1+y}$Te$_{1-x}$Se$_x$, we~refer to reference \cite{Sun_2019}.

\begin{figure*}[tb]

\centering
\includegraphics[width=\linewidth]{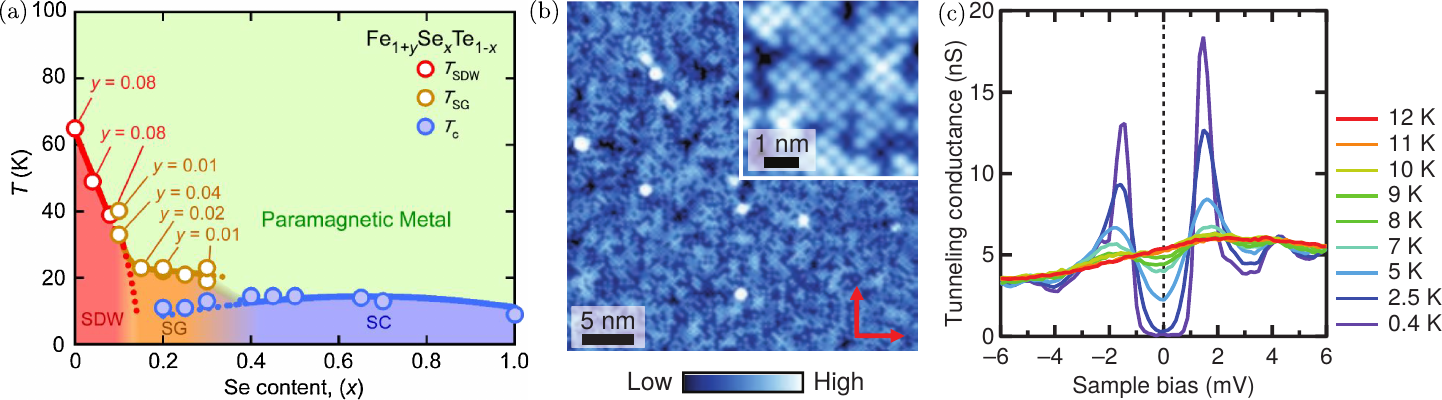}
\caption{(\textbf{a}) Phase diagram of Fe$_{1+y}$Te$_{1-x}$Se$_x$ versus temperature $T$ and Se content $x$ \cite{Katayama2010}. The~excess Fe content $y$ is zero unless explicitly specified. The  blue (red) circles indicate long-range ordered superconducting critical temperature $T_c$ (magnetic SDW critical temperature $T_{\rm SDW}$), whereas SG refers to spin-glass ordering. (\textbf{b},\textbf{c}) STM topographic image and temperature-dependent tunneling conductance for Fe$_{1+y}$Te$_{1-x}$Se$_x$ samples with $T_c=13.0$ K and $T_c=14.5$ K, respectively~\cite{Hanaguri474}. \mbox{In (\textbf{b}) the} bright white spots reveal the excess Fe ions on the surface. From \cite{Hanaguri474}. Reprinted with permission from AAAS.
}
\label{FeTeSe_phasediagram}
\end{figure*}

In Figure~\ref{FeTeSe_phasediagram}a, we display the phase diagram of Fe$_{1+y}$Te$_{1-x}$Se$_x$
as mapped out by bulk magnetization measurements in conjunction with (elastic/inelastic) neutron scattering techniques~\cite{Katayama2010}. The compound stays metallic for all $x$. At $x=0$, (non-superconducting) Fe$_{1+y}$Te supports a bicollinear antiferromagnetic phase for $y\lesssim 0.12$, where magnetic helical order exists for $y$ beyond $12\%$. For a detailed recent discussion of the magnetic properties and the associated spin excitations of FeTe, we refer to reference \cite{Tranquada2019}. Here we focus on the superconducting properties, where as seen from Figure~\ref{FeTeSe_phasediagram}a, with~enough Se substitution for Te, superconductivity emerges with maximum $T_c\sim 14.5$ K at optimal doping near the $50/50$ composition, FeTe$_{0.5}$Se$_{0.5}$. We stress that even at this composition level, excess Fe ions can play an important role in suppressing $T_c$ and inducing local magnetism~\mbox{\cite{Liu2009,Roessler2010,Sun_2019,Tranquada2019,Thampy2012}}. Unless explicitly addressed, the discussion below relates to nominally excess-Fe-free (annealed) samples. Even for such samples, however, it is well-known, e.g., from STM studies, that significant electronic and superconducting spatial inhomogeneity remains since Se and Te sit at random lattice positions, as seen from the STM topograph in Figure~\ref{FeTeSe_phasediagram}b \cite{He2011,SinghWahl2013,Masseee1500033,Wahl2020,Cho2019}.

From a theoretical perspective, FeTe$_{1-x}$Se$_x$ is a challenging material to address. Overall there is substantial evidence that FeTe is among the strongest correlated materials of the FeSCs. DFT + DMFT calculations comparing local moments and mass renormalization across the ``iron family'' locates FeTe as the most strongly correlated compound; see Figure~\ref{fig:DMFT}d \cite{Yin2011}. The mass renormalization is orbital selective due to the Hund's coupling, featuring the strongest correlations in the $d_{xy}$-dominated bands. Such results seem consistent with a non-nesting-driven magnetic ordering, bad metal behavior, local moment behavior, and the detection of orbital selectivity \cite{Yi2017,Si2016}; more details are discussed in Section~\ref{subsec:different}. As discussed above, remnants of this orbital selectivity seem to survive all the way to FeSe, and it appears therefore likely that similar correlated physics is present throughout the phase diagram, Figure~\ref{FeTeSe_phasediagram}a, of this fascinating material.

The electronic structure of Fe$_{1+y}$Te$_{1-x}$Se$_x$ has been thoroughly investigated, for example, by~various spectroscopic probes \cite{Xia2009,Tamai2010,Nakayama2010,Chen_2010_ARPES,Rinotte1602372,Lubashevsky2012,Liu2015_arpes}. Focusing on the composition close to $x\sim 0.5$, and~minimal amount of excess Fe ions, the Fermi surface consists of two small hole pockets around $\Gamma$ and two (also small) electron pockets around the $M$-point of the BZ. ARPES studies indicate that whether the inner (smallest) hole pocket crosses the Fermi level or not, depends sensitively on the exact values of $x$ and $y$ \cite{Rinotte1602372,Okazaki2014,Liu2015_arpes}.

At low temperatures, the superconducting density of states spectrum in optimal $T_c$-samples features a fully gapped state with prominent coherence peaks located close to $\pm2$ meV, as seen from Figure~\ref{FeTeSe_phasediagram}c \cite{Hanaguri474,Yin2015}. From STM measurements, additional shoulders can be identified in the tunneling conductance at lower energies between 1 and 2 meV, presumably related to the detailed gap structure around the largest Fermi sheets \cite{Sarker2017}. However, a consensus regarding the detailed momentum structure of the superconducting gaps around both hole and electron pockets has not yet been achieved. An early ARPES report claimed an isotropic 4 meV gap on the electron pockets \cite{Miao2012}, whereas most other spectroscopic probes point to a  maximum gap of around 2 meV. The hole pocket at $\Gamma$ is known to host a $2$ meV gap \cite{Lubashevsky2012,Rinotte1602372,Miao2012,Okazaki2014}. Early STM studies using magnetic field dependence of the QPI concluded that FeTe$_{0.55}$Se$_{0.45}$ displays sign changes in the superconducting gap between electron and hole pockets \cite{Hanaguri474}, which was recently confirmed by a phase sensitive measurement \cite{Chen_sign_2019}.
Finally, we note that the small Fermi energy $E_F$ of order a few meV, and the correspondingly large ratio for $\Delta/E_F\sim$ 0.1--0.5, has given rise to several studies of potential BCS-BEC crossover physics in Fe$_{1+y}$Te$_{1-x}$Se$_x$ \cite{Rinotte1602372,Lubashevsky2012,Okazaki2014}; see Section~\ref{sec_crossover}.

\begin{figure*}[tb]
\centering
\includegraphics[width=0.7\textwidth]{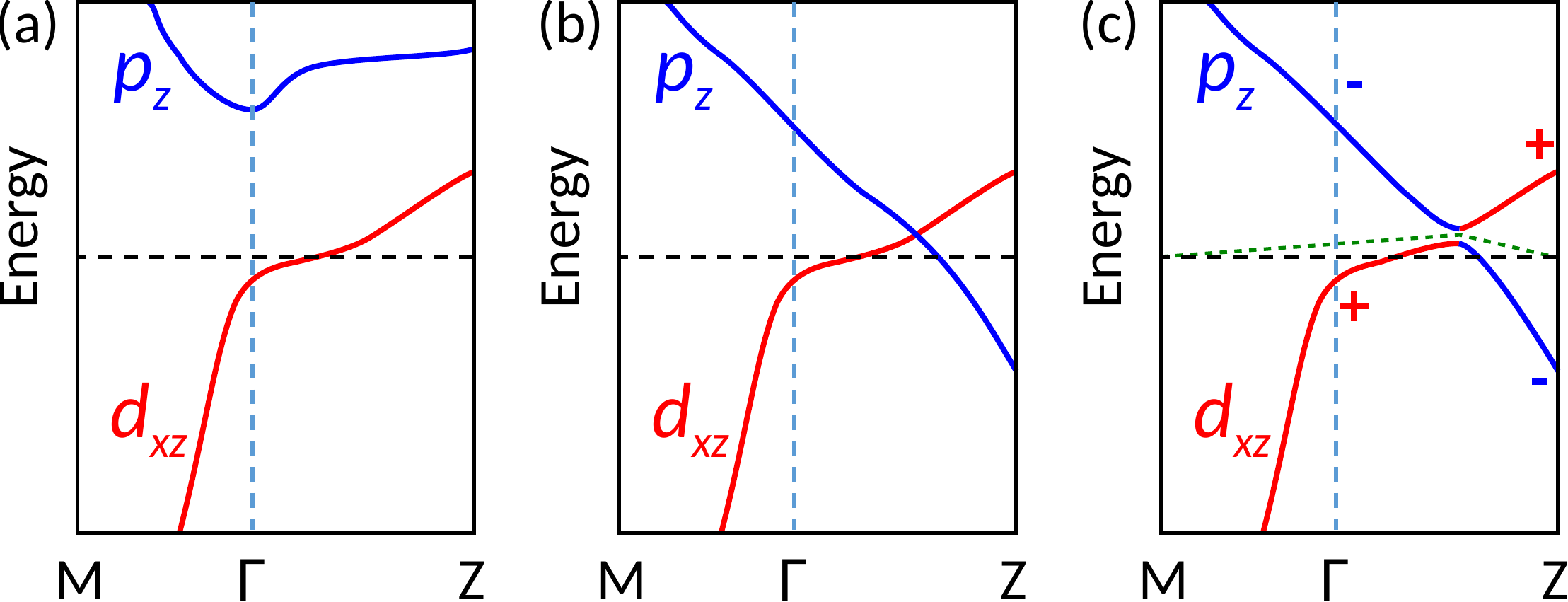}
\caption{Schematic of the band crossing leading to a topological state: (\textbf{a}) Band structure with the $p_z$ band exhibiting  limited dispersion such that only the $d_{xz}$ band crosses the Fermi level. (\textbf{b})~Situation where the $p_z$ band is pushed below the Fermi level at the $Z$ point and crosses the $d_{xz}$ band. (\textbf{c})~A~hybridization gap opens between the two bands such that a $\mathbb{Z}_2$ invariant can be constructed by following the ``curved chemical potential'' (green-dashed line) which passes by the hybridization gap. Classification of the filled states at time-reversal invariant momenta (TRIM) according to the eigenvalue of the parity operator $+/-$ and multiplication of the eigenvalues for filled states yields the invariant.}
\label{Topobands_schematic}
\end{figure*}

Below, we focus on recent nontrivial topological aspects relevant for FeTe$_{0.55}$Se$_{0.45}$ and related compounds. Growing evidence points to these materials belonging to a rare class of intrinsic topological superconductors, and in fact may constitute the first known high-temperature topological superconductors. The fact that electron correlations are also substantial makes the matter even more intriguing. We stress that in the current context, topological superconductivity refers mainly to superconducting surface states able to host Majorana zero modes (MZMs). The~surface states consist of a single Dirac cone generated by the nontrivial bulk band structure, and superconductivity is generated on the surface by proximity to the bulk superconducting electrons. The resulting surface state is topologically nontrivial in the sense, that, it should host a single zero energy Majorana fermion in the core of every vortex present. These MZMs are characterized by being their own antiparticles, and in addition  obey non-Abelian quantum statistics \cite{Majorana1937,Wilczek2009,Elliott2015}; they~could in principle be useful for fault-tolerant topological quantum computing.~MZMs have been previously realized in, for example, spin-orbit coupled semiconductor nanowires \cite{Mourik1003,Deng1557}, topological insulators~\cite{Sun_PRL_TI}, and~ferromagnetic adatom chains \cite{Nadj-Perge2014}, all, however, proximitized to conventional $s$-wave superconductors. These~platforms are required to operate at very low temperatures because the superconducting gap is small, and the relevant topological gap which protects the MZMs from random external perturbations is tiny. Therefore FeSC, with their relatively large $T_c$ and superconducting gap, could provide a superior platform. On the other hand, for quantum computation applications, one~needs to manipulate the MZMs, especially exchange the position
of two of them (braiding) to make use of the non-Abelian statistics. For the low dimensional systems with MZMs at the endpoints of chains, this might be done, for example, through a so-called $T$-junction \cite{Halperin2012} if the non-topological and topological phase can be tuned, while for MZMs in vortex cores of unconventional superconductors one is faced with the considerable challenge of moving sizable vortex objects to achieve braiding \cite{november2019scheme}.

\subsection{Theoretical Proposals for Topological Bands}
\label{subsec:band_inversion}

\begin{figure*}[tb]

\centering
\includegraphics[width=\linewidth]{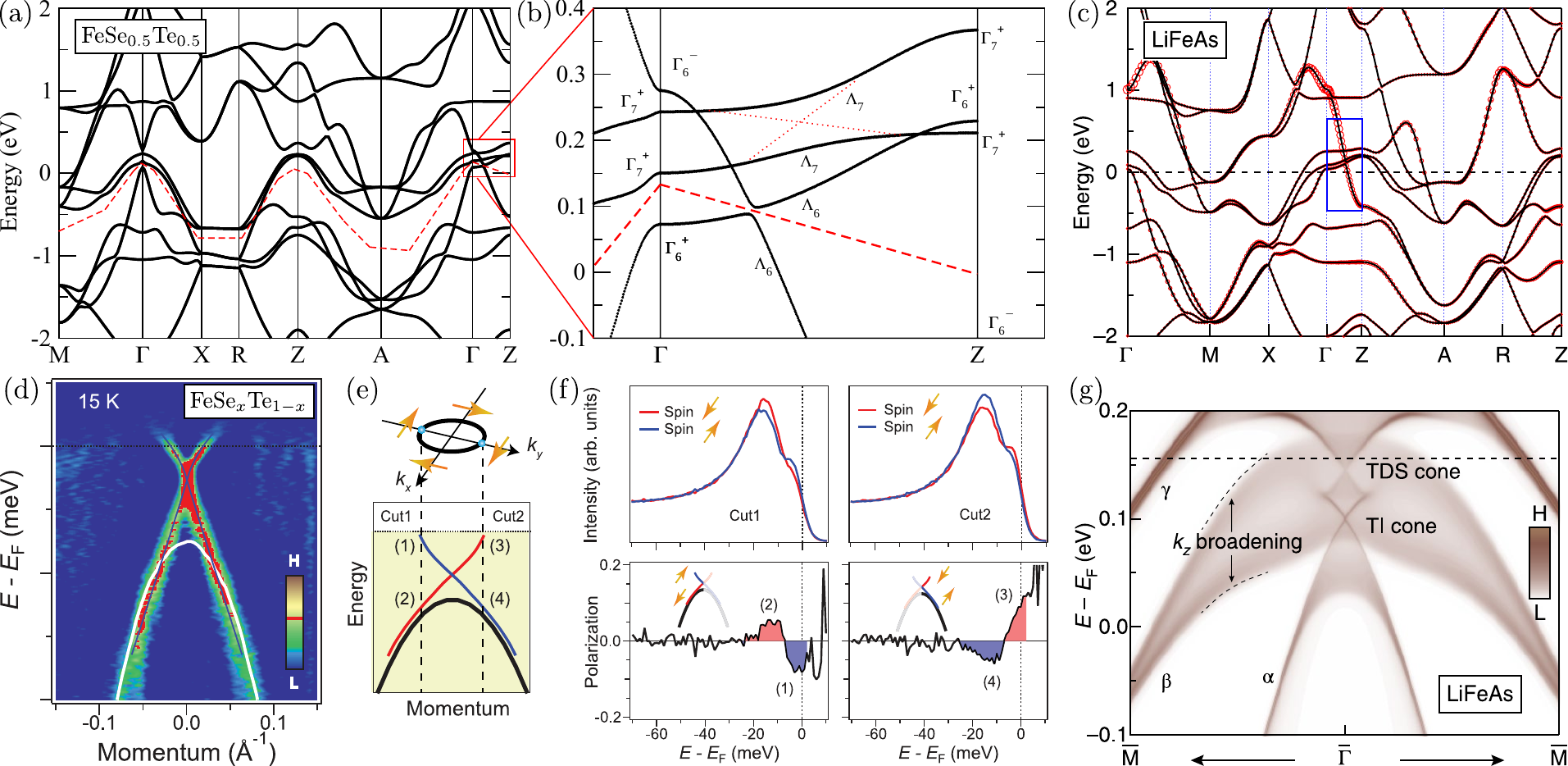}
\caption{Electronic band structure from DFT calculations relevant for FeTe$_{0.5}$Se$_{0.5}$ (\textbf{a},\textbf{b}) and LiFeAs~(\textbf{c}). Panel (\textbf{b}), displaying a zoom-in to the $\Gamma-Z$ direction of the BZ, reveals several band crossings leading to nontrivial topological band structures. The red dashed line outlines the ``curved chemical potential'' used to define the $\mathbb{Z}_2$ invariant.
Reprinted figure with permission from \cite{Wang_PRB_2015}, copyright by the American Physical Society (2015).
(\textbf{d}) Momentum distribution curvature plot displaying the dispersions of a quadratic holelike bulk band (white line) and a linear Dirac-like band close to the Fermi level. (\textbf{e})~Sketch of helical Fermi surface and the Dirac-like bands with spin polarization (red blue) together with the cuts at which the spin-resolved ARPES (\textbf{f}) was performed to reveal the helical spin structure.
From~\cite{Zhang_Science_2018}. Reprinted with permission from AAAS.
(\textbf{g}) Calculated (001) surface spectrum relevant for LiFeAs. Both~the topological insulator-like (TI) surface states, and the topological Dirac semimetal-like states (TDS) are indicated.
Reprinted with permission from Springer \cite{Zhang2019}, copyright (2019).}
\label{Topobands}
\end{figure*}

From DFT calculations it was discovered that several bulk FeSC materials could exhibit topologically nontrivial band structures due to band inversion along the $\Gamma-Z$ direction of the Brillouin zone \cite{Wang_PRB_2015,Zhang2019}. Wang et al. \cite{Wang_PRB_2015} first compared the band structure of FeTe$_{0.5}$Se$_{0.5}$ to FeSe, and found that the more extended $5p$ orbitals of the Te atoms lead to a stronger $pp$ bonding which enhances the interlayer hybridization and thereby increases the dispersion of the anion Te/Se $p_z$-dominated band along $\Gamma-Z$, pushing it close to the Fermi level. Upon hybridization via atomic SOC (boosted by the heavier Te ions) of the $p_z$ band with the opposite parity $d_{xz}/d_{yz}$-dominated band, a single band inversion takes place near the $Z$ point, characterized by a nontrivial $\mathbb{Z}_2$ topological invariant. The band-inversion process is shown in the schematic illustration of Figure~\ref{Topobands_schematic}. The result of the DFT calculations are shown in Figure~\ref{Topobands} \cite{Wang_PRB_2015}. Here panels (a) and (b) reveal the strongly dispersing $p_z$-band along the $\Gamma-Z$ direction for FeTe$_{0.5}$Se$_{0.5}$, and the resulting band-inversion near 0.1 eV in as seen from Figure~\ref{Topobands}b. The generated bulk band-inversion supports  topological spin-helical Dirac surface states protected by time-reversal symmetry, positioned inside the spin-orbit-induced gap and centered at the $\Gamma$ point of the BZ for (001) surfaces, as seen from Figure~\ref{Topobands}g. This process of generating nontrivial topological surface states is similar to the generation of topological surface states in 3D strong topological insulators, but for metallic FeTe$_{0.5}$Se$_{0.5}$ the surface states necessarily overlap with bulk states. DFT calculations point to similar nontrivial topological electron states being present also in LiFeAs, as seen from Figure~\ref{Topobands}c \cite{Wang_PRB_2015,Zhang2019}.

Nontrivial topological band structures have also been proposed for monolayers of FeSe on STO~\cite{Hao_PRX2014} and thin films of FeTe$_{1-x}$Se$_{x}$ \cite{Wu_PRB_2016}.
Hao and Hu performed a theoretical study of the band structure of single-layer FeSe, including the effect of lattice distortion from substrate strain \cite{Hao_PRX2014}. It was found that in principle a parity-breaking substrate can both suppress the holelike band at $\Gamma$ and induces a gap at the $M$ point. Provided that the SOC is large enough compared to the tensile strain-induced gap at $M$, a topological nontrivial $\mathbb{Z}_2$ phase can be stabilized from a band-inversion at $M$, with associated helical edge states \cite{Hao_PRX2014}. The band structure of FeSe/STO was also theoretically~studied under the additional assumption of checkerboard antiferromagnetic Fe moment ordering \cite{ZWang2016,Chen_SR_2019}. In this case, SOC induces a topological gap centered at $M$ slightly below the Fermi level, which supports quantum spin-Hall edge states protected by the combined symmetry of time-reversal and a discrete (primitive) lattice translation \cite{ZWang2016}.

In addition to the above topological ``$M$-point''-scenarios, Wu et al. \cite{Wu_PRB_2016} proposed a ``$\Gamma$-point''-scenario for the generation of nontrivial topological bands in monolayer FeTe$_{1-x}$Se$_x$. In~this case, by adjusting lattice constants, particularly the anion height (with respect to the Fe plane), it~was shown theoretically how a nontrivial $\mathbb{Z}_2$ topological phase can arise by band inversion at the $\Gamma$ point for FeTe$_{1-x}$Se$_x$ monolayer films, $x<0.7$ \cite{Wu_PRB_2016}. In this mechanism, it is a smaller hybridization between Fe $d_{xy}$ orbitals and Se/Te $p_z$ orbitals caused by an enhanced anion height, that leads to a band-inversion at $\Gamma$ by pushing the $p_z/d_{xy}$-dominated electron band far enough down in energy to mix with the hole bands. In the resulting band structure, the inverted parity-exchanged hole (electron) bands acquire $p_z/d_{xy}$ ($d_{xz}/d_{yz}$) orbital weight. For further details about the generations of topological bands in FeSCs monolayers, we refer to references \cite{Hao_PRX2014,Wu_PRB_2016} and recent reviews in  references \cite{Hao_review,Liu_2020}.

While the scenarios for band-inversion discussed above are intriguing and important,
the~predictions from DFT studies deserve further scrutiny when applied to iron-chalcogenides. This is due to the significant electron interactions and their associated band renormalizations. Standard methods assuming momentum independent self-energies describe the orbital-dependent band squeezing, but in reality nonlocal self-energy effects will further distort the DFT band structure and the final dispersion \cite{Ortenzi2009,Zantout2019,Bhattacharyya2020}.
Certainly for FeSe, as discussed at length in Section~\ref{subsec:bandstructure}, the link between the experimentally extracted low-energy band structure and the DFT-derived bands, remains~unclear at present. In this light, for the proposal of nontrivial band topology, e.g., in bulk systems, it~seems particularly crucial to determine whether the relevant $p_z$-dominated band indeed disperses enough between $\Gamma-Z$ to instigate a band-inversion, and whether the induced surface states can be relevant near the Fermi level. Thus, it is of crucial importance at this point to turn to experiments, and~check for experimental evidence for band inversion and topologically nontrivial surface states.

\subsection{Experimental Evidence for Topological Bands}
\vspace{-0.3cm}
\label{subsec:topo_band_expt}
In this section we turn to the experimental evidence for nontrivial topological bands in FeSCs. As discussed above, several ARPES studies \cite{Okazaki2014,Wang_PRB_2015,Shi2017} have addressed the band structure and the superconducting gaps in FeTe$_{0.55}$Se$_{0.45}$, but the acceleration of research in topological aspects of this material was kick-started by the work of Zhang et al. \cite{Zhang_Science_2018} published in 2018. By use of high-resolution ARPES and spin-resolved ARPES, Zhang et al. succeeded in detecting Dirac cone dispersive surface states near the Fermi level, with associated momentum-dependent spin polarization. This result is shown in Figure~\ref{Topobands}d--f. This discovery is consistent with the theoretical prediction of spin-helical (001) surface states near the $\Gamma$ point of the surface BZ, as shown in the surface projected band structure of Figure~\ref{Topobands}g~\mbox{\cite{Wu_PRB_2016,Wang_PRB_2015,Xu2016}}.
The Fermi level is not guaranteed to fall inside the band gap near $\Gamma$, and additional bulk states reside at the Fermi level in other regions of the BZ, complicating the experimental and theoretical analysis. Zhang et al. additionally focused on the EDC spectra near $\Gamma$ at $T<T_c$, and found an isotropic 1.8 meV superconducting gap on this band. This was interpreted as superconductivity induced on the
helical surface states by the bulk trivial bands, providing an example of self-proximitized topological superconductivity \cite{Zhang_Science_2018}.

The discovery of spin-polarized topological surface states in FeTe$_{0.55}$Se$_{0.45}$, has been followed up by proposals of additional nontrivial topological states in related systems. For example, LiFeAs was argued theoretically to possess similar topological insulator-like surfaces states as discussed above, and ARPES studies of Co-doped LiFeAs have found evidence thereof \cite{Zhang2019}. Reference \cite{Zhang2019} pointed out the additional existence of topological Dirac semimetal states from DFT studies. The bulk 3D Dirac cone associated with the semimetal states remains ungapped due to crystal symmetry, and produces surface states detectable, e.g., on (001) surfaces. Laser ARPES experiments on LiFeAs with varying degrees of Co doping to tune the Fermi level, found evidence for both topological insulator-like surface states and Dirac semimetal-like surface states, both with some degree of spin-polarization as determined from spin-resolved ARPES, and in agreement with their proposed nature of topological surface states~\cite{Zhang2019}. Topological Dirac semimetal states are also proposed to exist in FeTe$_{0.55}$Se$_{0.45}$, and could lead to novel types of bulk topological superconductivity \cite{Zhang2019,sato1,Hashimito2016}.

As stated above, the identification of low energy states observed in ARPES
as protected topological surface states is complicated by the fact that they invariably overlap with bulk states, and often exist in bands with tiny Fermi surfaces. Additionally, one has to rely on comparison to DFT studies, which are challenged in obtaining quantitatively correct band structures, particularly for the iron-chalcoginides.
In this respect, a recent ARPES study on FeTe$_{0.55}$Se$_{0.45}$ investigated the dependence of the photocurrent on incident photon energy \cite{Kanigel2019}. In contrast to DFT results, this photoemission study did not detect any highly dispersive band along the $\Gamma-Z$ direction, but nevertheless did find evidence for band-inversion along this direction. Specifically, the normalized ARPES intensity as a function of $k_z$ revealed pronounced oscillations with maxima (minima) at $\Gamma$ ($Z$). From an analysis of the relevant dipole selection rules, this behavior is consistent with a change of the parity eigenvalue (i.e., a change from $d$-like to $p$-like orbital character) along the $\Gamma-Z$ path \cite{Kanigel2019}. Thus, the $k_z$-dependence of the ARPES intensity points to bulk band inversion, and an overall picture consistent with a topological nature of the low-energy surface states in FeTe$_{0.55}$Se$_{0.45}$ \cite{Kanigel2019}.

Regarding the proposal of nontrivial band topology in FeSe/STO monolayers only a few experimental studies have hinted at the possible existence of the required edge states. For example, Wang et al. \cite{ZWang2016} using both STS and ARPES measurements studied the spectral weight inside the SOC gap for both [100] and [110] edges in the FeSe/STO monolayer, and reported evidence for dispersing modes near both edges, consistent with the existence of topological edge states. These,~however,~were~relatively far below the Fermi level. A more recent study focused on domain walls between different nematic domains in multi-layer FeSe/STO \cite{Yuan2018}.

Two experimental studies have addressed the electronic spectroscopic properties of FeTe$_{1-x}$Se$_x$ monolayers grown on STO, and interpreted their measurements in terms of possible topological bands~\mbox{\cite{Shi2017,Peng_PRB_2019}}. Shi et al. \cite{Shi2017} performed an ARPES study of monolayers of FeTe$_{1-x}$Se$_x$/STO for varying $x$, and found a systematic decrease of the band gap at $\Gamma$, i.e., holelike (electronlike) pockets that move up (down) with increasing Te content $x$. As usual, ARPES can only detect the unoccupied bands through possible thermally excited electrons, visualized via division of the measured data by the Fermi-Dirac distribution function. Through this procedure, Shi et al. \cite{Shi2017} found evidence for gap closing around $x\sim 33\%$, a prerequisite for possible nontrivial band inversion, but no direct experimental evidence for topological electronic states. More recently, Peng et al. \cite{Peng_PRB_2019} continued this line of research by a comprehensive energy- and polarization dependent ARPES study of FeTe$_{1-x}$Se$_x$/STO monolayers for samples in a wide range of $x$. Overall, for samples with $x\lesssim 0.21$ the location and polarization dependence of the bands are consistent with the ``$\Gamma$-point'' topological band-inversion scenario discussed above, with a band gap approximately $20$ meV below the Fermi level \cite{Wu_PRB_2016,Peng_PRB_2019}. Convincing evidence for nontrivial topological states, however, requires also, e.g., detection of symmetry-protected edge states. Using STS measurements near both [100] and [110] edges, some LDOS enhancement could be identified around $-$50 meV near the edges \cite{Peng_PRB_2019}. This result is consistent with associated DFT calculations including topological edge states, but still inconclusive in terms of the topological nature of the edge states.

\subsection{Topological Superconductivity}
\label{subsec:topo_sc}
The possibility of topological nontrivial band inversion in bulk FeSCs along the $\Gamma - Z$ direction close to the Fermi level discussed above, raises the possibility of surface-induced topological superconductivity arising from the proximity effect between bulk superconductivity and spin-helical topological surface states. This idea originates from Fu and Kane, who proposed that topological superconductivity can be realized on the surface of a topological insulator in proximity to a conventional $s$-wave superconductor \cite{Fu2008}. In this setup, superconductivity induced in the Dirac cone spin-helical surface states may resemble a two-dimensional $p_x+ip_y$-like pairing state, which~preserves time-reversal symmetry and exhibits topological characteristics, including MZM bound states in the center of vortex cores \cite{Read2000}. A bulk 2D $p_x+ip_y$ superconductor is well known to support a Chern number which, when nonzero, ensures the presence of chiral edge modes and the possibility of trapping a single MZM per superconducting vortex. Notably, the MZMs in vortices follow a $1+1=0$ rule, since they are protected by a $\mathbb{Z}_2$ invariant given by the product of the Chern number and the value of the vorticity modulo 2 \cite{teokane2010}.

\begin{figure*}[tb]

\centering
\includegraphics[width=\textwidth]{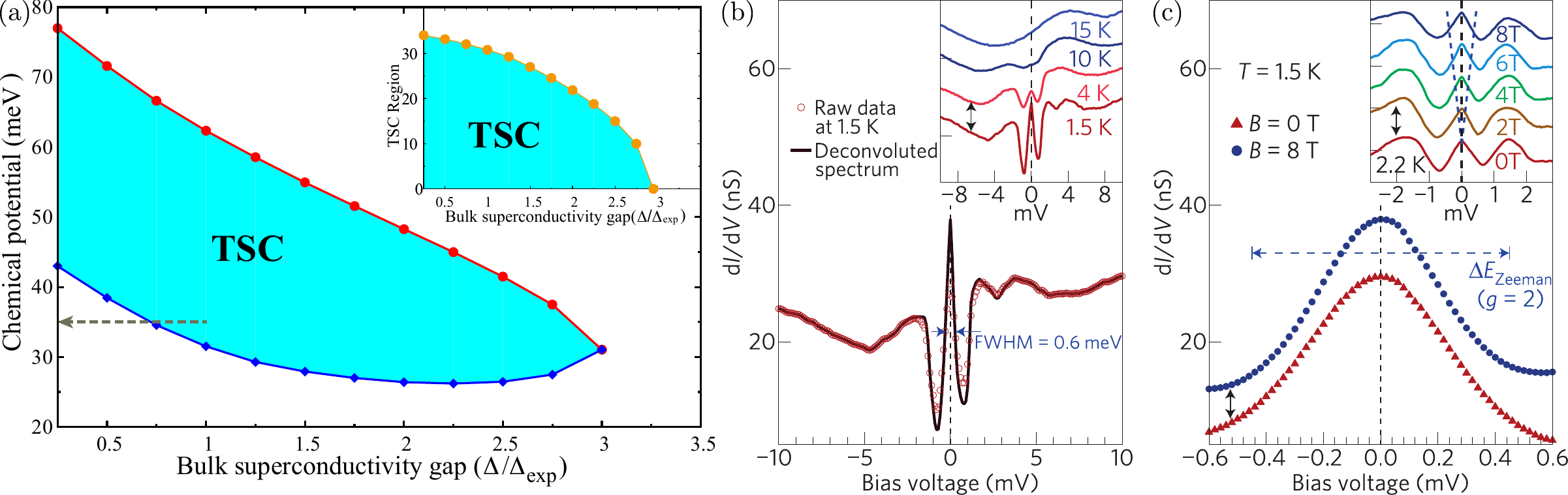}\vspace{-0.15cm}
\caption{(\textbf{a}) Region of stability for topological superconductivity (TSC) as a function of bulk superconducting gap amplitude and chemical potential for an effective 8-band model including spin orbit coupling,
containing both topological surface states and bulk trivial bands. The inset shows the width of the TSC region, i.e., red line minus blue line, as a function of the bulk gap. Reprinted figure with permission from \cite{Xu2016_topo}, copyright by the American Physical Society (2016).
(\textbf{b}) Zero-energy bound state as seen from the STM conductance near interstitial iron ions in near-optimally doped Fe$_{1+y}$Te$_{1-x}$Se$_x$. The inset displays the $T$-dependence of the spectrum. (\textbf{c}) High-resolution data taken in zero external magnetic field (red symbols) and $B=8$ T (blue symbols). As seen also from the inset, there is no evidence for Zeeman splitting of the zero-energy peak. Reprinted with permission from Springer  \cite{Yin2015}, copyright (2015).\vspace{-0.15cm}}
\label{TSC_theory_Yin_MZM}
\end{figure*}

For the current discussion, mainly focused on FeTe$_{0.55}$Se$_{0.45}$, and, e.g., Co-doped LiFeAs, where~any topological surface states necessarily overlap with bulk metallic bands, important questions arise concerning the detailed self-proximity mechanism and stability of the possible resulting topological surface superconductivity. Experimentally, superconductivity in the surface states seems confirmed in the sense that a full momentum-independent gap was detected on the surface band of FeTe$_{0.55}$Se$_{0.45}$~\cite{Zhang_Science_2018}.  Theoretically, the stability of topological surface superconductivity was investigated by Xu et al.~\cite{Xu2016_topo} by studying the nature of superconductivity on (001) surfaces within an effective low-energy eight-band model relevant to the $\Gamma$ and $Z$ points, with input parameters based on a fit to DFT calculations.  From this model, the phase diagram shown in Figure~\ref{TSC_theory_Yin_MZM}a of the stability of topological surface superconductivity could be established as a function of chemical potential and the amplitude of the bulk superconducting gap. As expected, the topological superconductivity is stable in a finite range of chemical potential, but reference~\cite{Xu2016_topo} also pointed out two additional properties both evident from Figure~\ref{TSC_theory_Yin_MZM}a: (1) surface topological superconductivity is suppressed if the bulk superconductivity becomes too strong due to pairing of surface states with the bulk states, and (2) for a regime of parameters, a topological surface phase transition can take place as a function of temperature with topological superconductivity only at the lowest $T$ and normal (nontrivial) superconductivity at higher $T$.

The bulk Dirac point pointed out by Zhang et al. \cite{Zhang2019} could also have interesting consequences for superconductivity, if properly tuned near the Fermi level, for example, doping of LiFeAs with cobalt.
By analogy to theoretical studies of possible superconducting pairing states in other (doped) Dirac semimetals \cite{Hashimito2016} it was suggested that semimetal Dirac cones could give rise to topological superconductivity. Both bulk and surface topological superconductivity has been considered, depending on the nature of the pairing on the bulk Dirac semimetal Fermi surfaces \cite{Zhang2019}. Furthermore, the presence of the bulk 3D Dirac cones led to a recent theoretical proposal of dispersive 1D helical MZMs inside the vortex cores of (singlet, $s$-wave) superconducting systems doped close to such Dirac semimetal points, protected by $C_4$ crystalline symmetry \cite{KonigPRL2019}. This scenario is qualitatively distinct from the 0D MZMs localized at the ends of $c$-axis aligned flux lines discussed further below. Two~recent theoretical works are also relevant for this discussion, by providing a topological classification of vortex Majorana modes in doped ($s$-wave) superconducting 3D Dirac semimetals \cite{QinPRL2019,QinSciBull2019}. A recent STM experiment on Co-doped LiFeAs did not observe pronounced zero-energy bound state in the vortex cores, which puts constraints on the existence of MZMs in this material, but does not completely eliminate the possibility of the existence of vortex MZMs (perhaps extended along the flux line) as there is still a finite density of states at zero energy \cite{Yin2019arXiv191011396Y}.

Finally, a theoretical study has investigated the possibility of intrinsic topological superconductivity in FeSe/STO monolayers \cite{Hao_PRB_2015}. Specifically, Hao and Shen performed a classification of the allowed pairing symmetries relevant to this system, and computed the superconducting phase diagram based on a phenomenological attractive pairing model \cite{Hao_PRB_2015}. In short, this allowed to theoretically identify a leading odd-parity topological $s$-wave  pairing state with spin-triplet, orbital-singlet structure.

\subsection{Experimental Evidence for Majorana Zero Modes: Defect States}
\label{subsec:expt_majorana}

The initial experimental observation of robust zero-energy states in FeSCs came from STM measurements near interstitial iron impurities on the surface of near-optimally doped Fe$_{1+y}$Te$_{1-x}$Se$_x$. Within samples containing $~0.5\%$ ($T_c=12$ K) and $~0.1\%$ ($T_c=14$ K) excess Fe ions, Yin et al. \cite{Yin2015} located individual interstitial Fe ions on the surface and reported a zero-bias-centered conductance peak at these impurity sites; see Figure~\ref{TSC_theory_Yin_MZM}b,c. Within experimental resolution, the peak was found to remain centered at zero bias as a function of both STM tip position and application of $c$-axis applied magnetic fields up to 8 T. Furthermore, it was measured to extend uniformly in space with a length scale of order 3--4 \AA,  and decrease in amplitude (but not split) when proximate to other impurity bound states \cite{Yin2015}. These peculiar properties are not characteristics of standard in-gap bound states of FeSCs arising from magnetic or nonmagnetic impurities \cite{Gastiasoro2013,Mukherjee_2013_imps,Kreisel2016_imp,Martiny_2019_imp,JungpilSeo2020}. Intriguingly, similar robust zero-energy conductance peaks have been recently detected near Fe adatoms, deposited on top of the stoichiometric materials LiFeAs and PbTaSe$_2$ \cite{Songtian2020}, and on monolayers of FeSe/STO and FeTe$_{0.5}$Se$_{0.5}$/STO \cite{Liu_ScAdv_2020}.

What is the origin of these seemingly robust non-split zero-bias conductance peaks discussed above?
A recent theoretical work suggested that interstitial iron ions may induce so-called quantum anomalous vortices, which in conjunction with effective $p$-wave pairing on the surface, host MZMs in their center \cite{Jiang2019}. By including both impurity-induced SOC, and exchange coupling with the magnetic impurity ion, it was shown theoretically that vortices can be nucleated by the iron moment~\cite{Jiang2019}. Notably, the vortices are stabilized even in the absence of external magnetic fields, hence the name ``quantum anomalous''. In the presence of topological surface superconductivity, the quantum anomalous vortices support MZMs at their center, thereby providing a possible explanation of the STM results reported in references \cite{Yin2015,Songtian2020,Liu_ScAdv_2020}.

Recently, Fan et al. \cite{Fan2020} performed a comprehensive STM study of the variability of the conductance near different Fe adatoms deposited on the surface of FeTe$_{0.55}$Se$_{0.45}$. In agreement with the finding by Yin et al. \cite{Yin2015} robust zero-bias conductance peaks exists near some of the Fe adatoms, a finding interpreted in favor of the quantum anomalous vortex scenario \cite{Jiang2019}. In addition, however, a fraction of the Fe adatoms was shown to feature more standard bound states similar to Yu-Shiba-Rusinov (YSR) impurity states with finite energy bound state energies. Interestingly, some~of these YSR bound states can be reversibly manipulated (and irreversibly manipulated by moving the adatom position) into zero-energy peaks (ZEPs) by changing the tip-sample distance. It is known from other STM studies of YSR states, that the tip can exert a force on adatoms, and~thereby alter the exchange coupling between the impurity moment and the conduction electrons, resulting in a tip-induced shift of the YSR bound state energies~\cite{Farinacci}. The data from Fan et al. \cite{Fan2020} reveals the existence of a critical coupling necessary for generating the ZEPs, a result again discussed in reference~\cite{Fan2020} in terms of impurity-induced vortices and MZMs. Lastly, we mention a recent STM study of sub-surface impurity states in FeTe$_{0.55}$Se$_{0.45}$, reporting on another kind of tip-tunable in-gap state~\cite{Damianos_2020}. As shown in reference \cite{Damianos_2020}, some bound states accidentally appear to be located at zero-energy, but ``disperse'' with the tip position, a property shown to be consistent with a local tip-induced gating of the impurity levels in low-density systems.

Zero-energy localized states have also been recently detected by STM at the ends of 1D atomic line defects in  2D single unit-cell thick FeTe$_{0.5}$Se$_{0.5}$ films grown on STO(001) substrates \cite{Chen2020}. This system exhibits superconductivity below 65 K, and a fully gapped spectrum with two large identifiable gaps of 10.5 meV and 18 meV. The line defects consist of unidirectional lines of missing Te and/or Se atoms at the top layer, as determined by topographic images~\cite{Chen2020}. Chen et al. \cite{Chen2020} studied line defects of 15 and 8 missing Te/Se atoms, and inferred from their spectroscopic characteristics that the most likely explanation for the emergence of zero-energy end states is topological MZMs. However, since unit-cell thick FeTe$_{0.5}$Se$_{0.5}$/STO is likely topologically trivial, it was suggested that the missing atoms themselves induce the necessary ingredients for local nontrivial topological states~\cite{Chen2020}. Microscopically, this could include locally enhanced Rashba SOC, or induced chain magnetism. Two recent theoretical works explore both possibilities, i.e., local topological ``Rashba-chains'' and associated local chain-induced odd-parity spin-triplet pairing~\cite{zhang2020atomic}, and topological antiferromagnetic line defects~\cite{wu2020_linedefects}.

\subsection{Experimental Evidence for Majorana Zero Modes: Vortex States}
\label{subsec:expt_majorana_vortex}
A clear prediction of proximitized topological surface states is the emergence of MZMs inside the cores of field-induced vortex lines \cite{Fu2008}. A number of experimental groups have performed detailed STM experiments on the surfaces of FeTe$_{0.55}$Se$_{0.45}$ and (Li$_{0.84}$Fe$_{0.16}$)OHFeSe and discussed the possibility of MZMs in the data \cite{Wang333,Chen2018,Machida2019,Liu_ScAdv_2020,chen_2019vortex,HHWen_Majorana,Kong2019,Zhu2020}.
The topic is complicated, e.g., due to the sensitivity to instrumental resolution, unknown effects of impurity pinning sites \cite{Masseee1500033}, contributions from bulk states, quasi-particle poisoning, sample inhomogeneity, disordered vortex lattices, and the presence of other topologically trivial but contaminating low-energy vortex cores states. The latter are the well-known Caroli--de Gennes--Matricon (CdGM) \cite{CAROLI1964307} states which
exist quite generally in the cores of superconductors,
and tend to produce a broad peak centered at zero energy simply because the energy separation between CdGM states, $\Delta^2/E_F$, may be significantly smaller than the instrumental resolution \cite{Hess1989}. Only in the so-called quantum limit where $\Delta/E_F$ becomes large enough, can one expect to see discrete well-separated CdGM states. An advantage of searching for Majorana modes in FeSCs is that indeed for these materials $\Delta/E_F$ can be rather large, and therefore finite-energy CdGM states should be distinguishable from a potential zero-energy MZMs \cite{Hanaguri2012_vortex,Uranga2016}.

\begin{figure*}[tb]

\centering
\includegraphics[width=\textwidth]{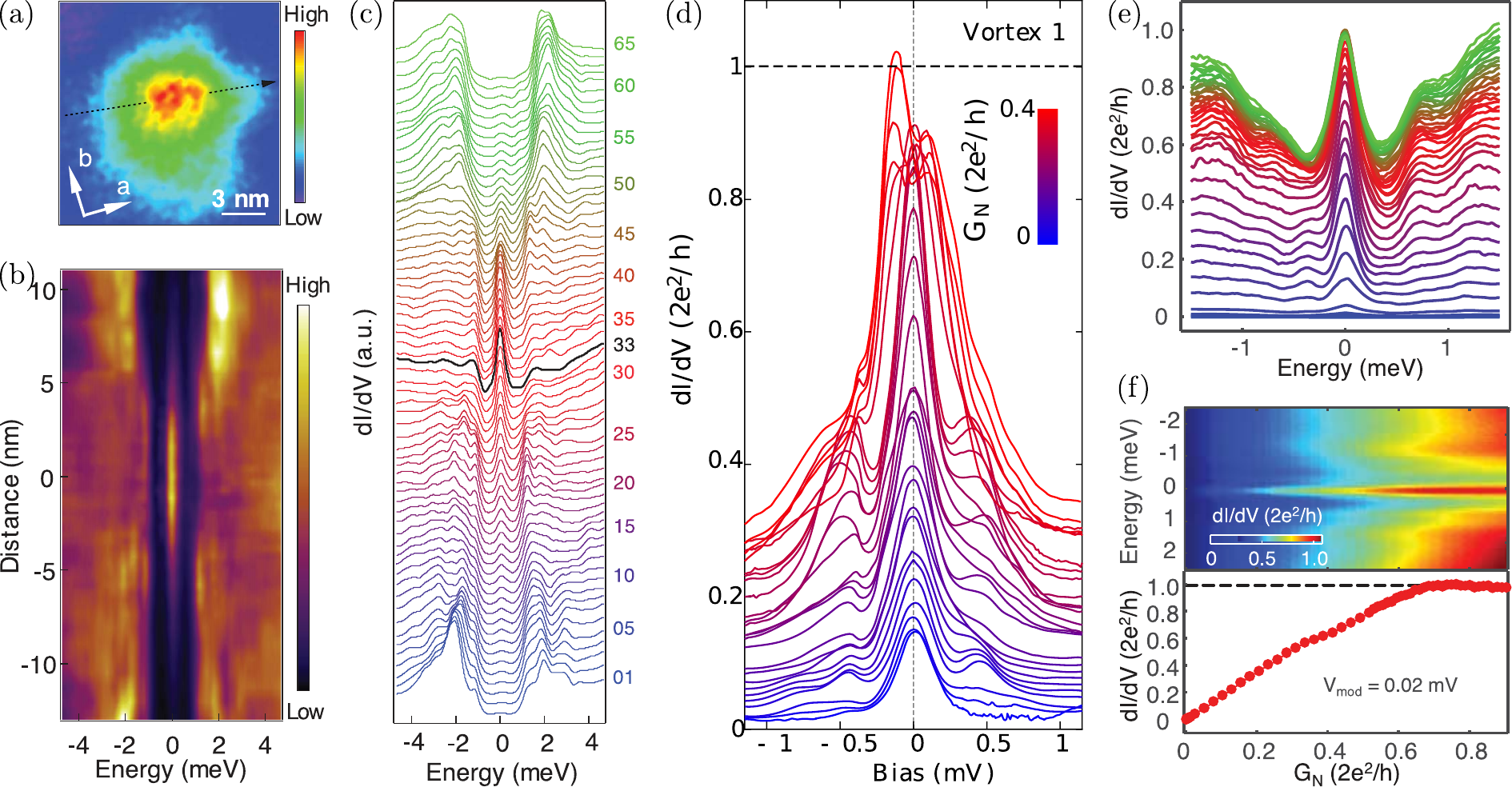}
\caption{(\textbf{a}) Zero-bias conductance image near a vortex core in FeTe$_{0.55}$Se$_{0.45}$.
(\textbf{b}) Conductance versus bias voltage (energy) along the cut indicated by the black dashed line in (\textbf{a}), revealing the spectrally isolated Majorana zero mode (MZM). (\textbf{c}) Waterfall plot of the same data as in (\textbf{b}). The~black curve indicates the vortex core center.
From \cite{Wang333}. Reprinted with permission from AAAS.
(\textbf{d}) Experimental conductance $dI/dV$ as a function of increased tunnel-barrier conductance $G_N$ in (Li$_{0.84}$Fe$_{0.16}$)OHFeSe~\cite{chen_2019vortex}. (\textbf{e}) Same as (\textbf{d}) but for  FeTe$_{0.55}$Se$_{0.45}$ \cite{Zhu2020}. (\textbf{f}) Color-scale plot of the same data as in (\textbf{e}) (upper panel), and a line-cut at zero bias versus $G_N$ (lower panel), reaching a plateau near $2e^2/h$. Most vortices in FeTe$_{0.55}$Se$_{0.45}$ do not reach plateaus at $2e^2/h$, possibly due to thermal smearing or other non-ideal (unresolved) effects. From  \cite{Zhu2020}. Reprinted with permission from AAAS.}
\label{Wang_vortex_topo_conductance}
\end{figure*}

\begin{figure*}[tb]

\centering
{\includegraphics[width=\textwidth]{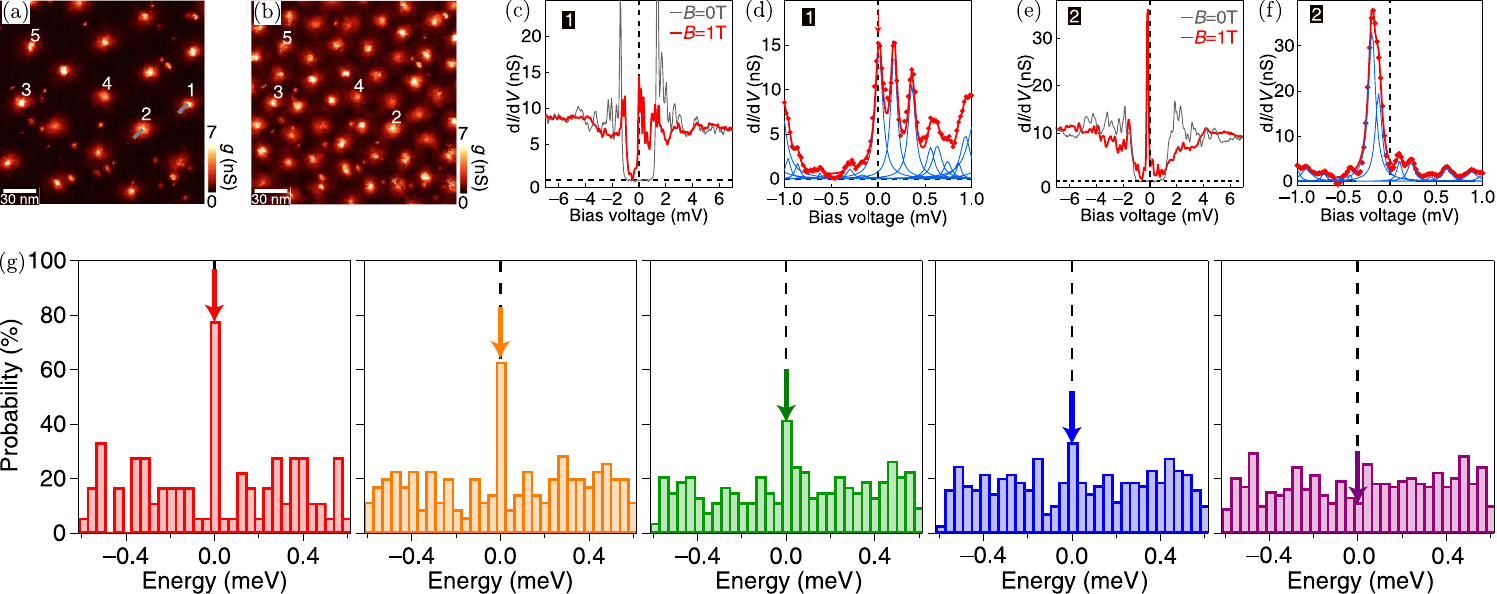}}
\caption{(\textbf{a},\textbf{b}) STM zero-bias conductance maps revealing the vortex lattice at magnetic fields of $B=1$ T (\textbf{a}) and $B=3$ T (\textbf{b}) in FeTe$_{0.55}$Se$_{0.45}$. (\textbf{c}--\textbf{f}) Ultra-high resolution spectroscopy in the~ center of two different vortices; (\textbf{c},\textbf{e}) vortex 1 hosts a MZM, whereas vortex 2 (\textbf{d},\textbf{f}) apparently does not. (\textbf{g})~Histograms of the probability of appearances of conductance peaks at low energies, revealing a clear decrease of MZMs with increasing field of 1 T, 2 T, 3 T, 4 T, and 6 T in the rightmost panel. Reprinted with permission from Springer \cite{Machida2019}, copyright (2019).}
\label{Machida_vortex_topo}
\end{figure*}

In FeTe$_{0.55}$Se$_{0.45}$, STM reports have identified a range of different CdGM states depending on which specific vortex was probed \cite{Chen2018,Chen2020_CdGM,Kong2019}. Interestingly, however, some vortex cores feature a conductance peak centered exactly at zero bias, as shown in Figure~\ref{Wang_vortex_topo_conductance}b,c, and
is pinned to zero energy over a spatial range away from the vortex core center \cite{Wang333,Machida2019,Kong2019,HHWen_Majorana}. This zero-energy peak constitutes an important fingerprint of MZMs associated with topological superconductivity on the surface of  FeTe$_{0.55}$Se$_{0.45}$ \cite{Wang333}. Importantly, it was found from high-resolution STM measurements by Machida et al. \cite{Machida2019} that not all vortex cores host zero modes, and the fraction of those that do
is inversely proportional to the applied magnetic field strength, with approximately 80\% (10\%) probability of detecting zero energy states at $B=1$ T ($B=6$ T), as shown in Figure~\ref{Machida_vortex_topo}. The variability of the low-energy states and the presence of possible MZMs appear to be unrelated to disorder sites or local Te/Se concentration variations  \cite{Machida2019,HHWen_Majorana}. The STM study presented in reference \cite{Kong2019} utilized the spatial variability of the vortex electronic structure to identify two classes of vortices distinguished by a half-integer level shift between the in-gap vortex states. In agreement with model calculations, a~level sequence of $0, 1, 2, 3, \ldots$ versus
$\frac 12, \frac 32, \frac 52, \ldots$ in terms of $\Delta^2/E_F$ is indeed expected for topologically nontrivial and trivial bands, respectively~\cite{Kong2019}.

Recently, based on theoretical simulations of an effective low-energy Hamiltonian it was suggested that Majorana hybridization in conjunction with a realistic distribution of disordered vortex sites, offers an explanation of the decreasing number of ZEPs with increasing magnetic field \cite{Chiu2019,Machida2019}.
Earlier~theoretical studies investigated the role of random Se/Te substitution on the vortex bound states, finding insignificant effects from this kind of isovalent disorder \cite{Berthod2018}. It was shown how disorder in vortex locations is important for smearing out oscillations in the field dependence of the density of zero-energy peaks.  Reference \cite{Chiu2019} also compared the statistics of the lowest energy peaks in vortices without ZEP between experiment and simulations, providing evidence for the scenario of random Majorana hybridizations causing the decrease of ZEP. More recently, an alternate explanation was proposed, namely that the surface hosts two distinct phases competing for Dirac surface states in FeTe$_{0.55}$Se$_{0.45}$ \cite{Wu_2020}.  In this picture, remnant magnetic interstitial moments aligned by an external magnetic field may stabilize regions of half quantum anomalous Hall phases, supporting standard vortices without MZMs, and other regions of topological superconductivity hosting MZMs  in their vortices \cite{Wu_2020}. This scenario offers a prediction of chiral Majorana modes located at the domain wall between these two spatially distinct phases. Another theoretical study suggested that Zeeman coupling from clustering of magnetic impurities may locally induce a superconducting orbital-triplet spin-singlet state near some vortices, thereby rendering those vortices non-topological and hence unable to support zero-energy MZMs \cite{Ghazaryan2020}.


In (Li$_{0.84}$Fe$_{0.16}$)OHFeSe, STM studies have also detected zero-bias-centered conductance peaks inside the vortex cores \cite{Liu_2018vortex,chen_2019vortex}.
In this case, however, zero-energy modes were found only at free vortices, i.e., not pinned by dimer-like impurity sites, on FeSe surfaces. Similar to the discussion above, these zero-energy conductance peaks were interpreted as evidence for MZMs, a conclusion backed up by ARPES measurements and DFT calculations. The photoemission measurements found some spectral weight near the BZ center, which was interpreted as surface Dirac cone states, but no evidence of superconductivity could be detected on these surface bands. However, as further discussed in reference~\cite{Liu_2018vortex}, (Li$_{0.84}$Fe$_{0.16}$)OHFeSe has the favorable property that the FeSe layers are stoichiometric and $T_c$ ($\xi$) is higher (shorter) than those in FeTe$_{0.55}$Se$_{0.45}$ by roughly a factor of four, making the ZEPs (1) correlated to free vortices, and (2) less sensitive to high magnetic fields.

An important characteristic of MZMs is the so-called quantized Majorana conductance of $2e^2/h$~\cite{Law2009,Flensberg2010}, a property recently verified in topological semiconductor nanowires \cite{Nichele2017}. The~quantized property is a direct consequence of the particle-antiparticle equivalence of MZMs. If~Majorana-induced resonant Andreev reflection takes place between the Majorana bound state in the vortex cores and the normal STM tip, the conductance should reach $2e^2/h$ in the ideal $T=0$ case, independent of the tunnel coupling. This conclusion holds for the case of a single conducting contributing channel. Two recent low-temperature experimental STM studies on FeTe$_{0.55}$Se$_{0.45}$ and (Li$_{0.84}$Fe$_{0.16}$)OHFeSe both reported on experimental evidence for such quantized $2e^2/h$ conductance~\cite{chen_2019vortex,Zhu2020}. More specifically, as seen from Figure~\ref{Wang_vortex_topo_conductance}d--f it was observed that upon decreasing the tip-sample distance, and thereby increasing the tunnel conductance, the zero-bias peak appears to reach a saturation plateau close to 60\% or 90\% of $2e^2/h$ for FeTe$_{0.55}$Se$_{0.45}$ and (Li$_{0.84}$Fe$_{0.16}$)OHFeSe, respectively \cite{chen_2019vortex,Zhu2020}. The sub-gap conductance, however, exhibits a significant amount of background weight, and the potential influence of other channels in contributing to the final conductance appears unclear at present.

In addition to the interesting developments summarized above for chalcogenide systems, recent experimental evidence was reported for MZMs in the vortices of iron-pnictide superconductors. Specifically, Liu et al. \cite{liu2019new} used both ARPES and STM to perform a spectroscopic study of CaKFe$_4$As$_4$, and found evidence for superconducting Dirac surface states and vortex core MZMs. In this material, the origin of nontrivial topology is suggested from DFT + DMFT calculations to arise from band inversion along $\Gamma-Z$, catalyzed from an additional folding of the BZ along $z$ due to glide-mirror symmetry breaking along the $c$-axis. \cite{liu2019new} There has also been a recent theoretical proposal that the iron-pnictide 112-material Ca$_{1-x}$La$_x$FeAs$_2$ may be a topological superconductor~\cite{Wu_Thomale_2020}. Thus, while~more experimental studies of topological aspects of CaKFe$_4$As$_4$, Ca$_{1-x}$La$_x$FeAs$_2$, and LiFeAs are clearly desirable, at present it seems possible that the realm of topological high-T$_c$ superconductivity may extend into the iron-pnictides as well.

\subsection{One-Dimensional Dispersive Majorana Modes}
\label{subsec:1d_majorana}

The above discussion of zero-dimensional MZMs hosted by induced topological superconductivity on surfaces can be traced back to the original theoretical proposal by Fu and Kane, also discussed briefly above \cite{Fu2008}. That work, however, suggested another realization of Majorana fermions by use of a $\pi$-junction between two ordinary superconductors deposited on a topological insulator, generating a one-dimensional wire for helical Majorana fermions \cite{Fu2008}. As a consequence, if $\pi$ phase shift domain walls could be generated on the surface of topological FeSCs, 1D Majorana modes could exist along the domain walls. Above, we briefly mentioned an STM study of nematic domain walls in 20 unit-cell thick FeSe on STO, interpreted in the light of 1D dispersing topological edge states \cite{Yuan2018}. Unlike monolayers, the multilayer films are known to feature strong electronic nematicity. Specifically, Yuan et al. \cite{Yuan2018} grew 20 unit cell thick FeSe films on top of STO substrates, and studied the electronic states near domain walls between two distinct nematic regions. The resulting STS data found evidence for edge-induced zero-energy states localized to the domain walls, and interpreted them in terms of topological edge modes \cite{Yuan2018}.

Another recent STM study managed to identify a certain type of crystalline domain walls associated to half-unit cell shifts of the Se atoms on the surface of bulk FeTe$_{0.55}$Se$_{0.45}$, and~measured almost flat $dI/dV$ conductance spectra at low bias (inside the superconducting gap) at the domain wall, as opposed to fully gapped conductance spectra away from the domain wall \cite{Madhavan2019}. This~peculiar conductance behavior was not observed near, e.g., step edges in FeTe$_{0.55}$Se$_{0.45}$, or at twin-domain walls of topologically trivial FeSe. Therefore, it was proposed that the flat $dI/dV$ curves constitute spectroscopic evidence for linearly dispersing helical Majorana modes generated by the half-unit-cell-shifted domain walls \cite{Madhavan2019}.

\subsection{Higher-Order Topological States}
\label{subsec:higher_order_majorana}

Another possibility for generating 1D helical Majorana modes in FeTe$_{0.55}$Se$_{0.45}$ was first discussed theoretically by Zhang et al. \cite{Zhang_hinge}, who predicted the emergence of so-called higher-order superconducting topology with associated 1D localized helical Majorana hinge states between (001) and (100) or (010) surfaces. An $n$'th order topological phase hosted in a $D$-dimensional system features $(D-n)$-dimensional topological edge states \cite{Benalcazar61,Schindlereaat0346,Calugaru_topo_2019,Yan_topo_2019}. For example, a 2nd order topological superconductor hosts Majorana corner and hinge modes as opposed to standard edge or surface modes in 2D and 3D, respectively. Such topological MZMs are only detectable by probes able to selectively pick out sample corners or hinges. The theoretical analysis by Zhang et al. \cite{Zhang_hinge} ``hinges'' on the band inversion along $\Gamma-Z$, and standard $s_{\pm}$ superconductivity in the bulk with opposite sign of the pairing gap between the $\Gamma$ and $M$ points, $\Delta(k)=\Delta_0 + \Delta_1(\cos(k_x)+\cos(k_y))$~\cite{Zhang_hinge}. The~latter property is necessary for generating opposite signs of the order parameters on (001) and (100) surfaces, producing an effective $\pi$-shifted domain wall at the hinge of the two surfaces. Notably, no external magnetic field is required for the generation of higher-order MZMs. A recent experiment probing the edges of exfoliated flakes of FeTe$_{0.55}$Se$_{0.45}$ samples found evidence for such zero energy hinge states \cite{Burch2019}. More~specifically, by draping suitable contacts over the sides of the sample, normal metal/superconductor junctions were created on the hinges between (001) and (100) surfaces, and a pronounced zero-bias conductance peak could be measured only for junctions in direct contact with the putative hinge modes \cite{Burch2019}. Precautions were taken to separate contact effects from intrinsic topological modes existing in FeTe$_{0.55}$Se$_{0.45}$ samples, leading to the conclusion that the zero-bias conductance peak was direct evidence for topological Majorana hinge modes, and thereby higher-order topological superconductivity in FeTe$_{0.55}$Se$_{0.45}$ \cite{Burch2019}.

More recently, higher-order topological phases have also been discussed theoretically in the context of Majorana corner modes in 2D superconductors  coexisting  with suitable magnetic structures~\cite{Zhang_corner}. For example, a proposed theoretical setup consists of a monolayer of FeTe$_{0.55}$Se$_{0.45}$, experiencing the magnetic exchange field in proximity to a layer of FeTe exhibiting bicollinear antiferromagnetism~\cite{Zhang_corner}. This magnetic structure allows for corners of ferromagnetic and antiferromagnetic edges. From this property, in addition to standard $s$-wave superconductivity and within an effective band model that includes only part of the band structure (near $\Gamma$), it was found that indeed the magnetic order instigates topological edges in the form of an effective nanowire Hamiltonian, known to support Majorana end states, resulting in the present case in Majorana corner modes \cite{Zhang_corner}. A related theoretical study also explored conditions for stabilizing Majorana corner modes in FeTe$_{1-x}$Se$_x$ monolayers, but with time-reversal symmetry breaking from an external in-plane magnetic field, creating distinct edges being either parallel or perpendicular to the in-plane field \cite{Wu_corner}. Future experimental studies will hopefully pursue these interesting proposals for higher-order MZMs in FeSC systems.

\section{Conclusions}

We have tried here to review  the background necessary for the reader to understand the debate over superconductivity in the Fe-chalcogenide materials, what is known about the superconducting state, and some theoretical ideas that have been put forward in this context.  It is clear that relatively strong electron correlation  (quite possibly involving nonlocal effects that have not yet been treated systematically) and spin-orbit coupling play important roles and distinguish these materials, at least in degree, from their pnictide counterparts.  We have abandoned the attempt to cover many important and fascinating aspects of the normal states of these materials in favor of doing a reasonably thorough job on the superconducting states.  Even with our more limited goals, we have necessarily been forced to leave out many important contributions, an omission for which we apologize to the authors so~neglected.

Here we attempt to summarize our personal view of the important open questions in this field.  First, let us assume that the standard paradigm is correct; that spin fluctuations due to repulsive Coulomb interactions are primarily responsible for pairing in both Fe-pnictides and Fe-chalcogenides; and that the differences arise primarily because of the heightened degree of correlation and perhaps strength of spin-orbit coupling in the latter.  If that is all in fact true, how does one explain the fact that higher $T_c$s seem to exist in systems without hole pockets?   Note that we refer here not only to the monolayer system, where it has been plausibly argued that substrate-induced phonons can bootstrap the spin fluctuation interaction, but also to the e-doped FeSe intercalates with $T_c$s above 40 K, where~no such effect is obviously present.  No convincing explanation involving realistic materials-specific parameters for this phenomenon has yet been put forward.

The proposals for phonon-assisted $T_c$s in the monolayer systems have stimulated a renewed interest in the role of the lattice in Fe-based superconductors generally.  While early estimates of $T_c$s to be expected from electron--phonon coupling suggested that these physics could be neglected, these~questions need to be revisited.  The obvious question is whether, and under what circumstances, phonons can assist spin fluctuations to enhance $T_c$.  Naively, this can happen only if they are of forward scattering character, since otherwise they drive $s_{++}$ pairing that competes with $s_\pm$ and $d$-wave pair channels.  Nevertheless, it will be interesting to perform materials-specific calculations, including both phonons and spin fluctuations on the same footing.

The role of the lattice is also interesting with regard to normal state physics.  For example, the role of nematic fluctuations near the nematic critical point in promoting superconductivity has been questioned, with the suggestion that these fluctuations do not diverge due to a lattice cutoff. At~present, we have no material-specific theory that can explain even the balance of cooperation vs. competition between nematic order and superconductivity, which appears to be distinctly different in the Fe-chalcogenides compared to the Fe-pnictides.  Going beyond phenomenological theories of pairing due to nematic fluctuations is a current challenge.

$\mu$SR experiments have reported signals of time reversal symmetry breaking in both Fe-pnictides and chalcogenides.  More detailed studies are needed to distinguish between  TRSB states with macroscopic spontaneous currents,  and nonchiral complex admixtures that create only local, impurity induced current. Theoretical studies need to make clear predictions for the signals expected for various states from $\mu$SR and Kerr measurements, including the disorder dependence, and additionally, find ways to distinguish between TRSB arising from spin (e.g., nonunitary order parameter) and from orbital (e.g., complex order parameter orbital combinations like $p_x+ip_y$) degrees of freedom. 

The recent theoretical and experimental studies of topologically nontrivial effects in FeSCs highlight a new exciting direction within this area of research. It is remarkable that band-inversion far off the Fermi-level from DFT predictions, fortuitously gets shifted down to energy scales relevant for superconductivity. While the ``smoking gun'' proof of topological superconductivity might still be argued to be missing, there certainly exists mounting evidence in its favor at present. In particular, the detection of MZMs inside vortices by several different STM groups points to nontrivial band topology and associated self-proximitized topological surface superconductivity. This is remarkable since  intrinsic topological superconductors are considered rare, and the fact that they may inherently exist within a family of correlated materials exhibiting unconventional bulk superconductivity, makes the development all the more noteworthy. Of course there are many unsolved questions and we still lack quantitative analysis of most experimental results in terms of realistic material-specific models.

Some current open topics for FeSCs and nontrivial topological superconductivity refer, e.g., to~the questions of MZM variability in the vortex cores. Why do only some vortices host a MZM on the surface of FeTe$_{1-x}$Se$_x$, and why do apparently no vortices host MZMs in Co-doped LiFeAs. In fact, the latter compound seems particularly elusive regarding its potential topological properties. At this point is seems unclear exactly how bulk and surface states intertwine in the final superconducting condensate. Another open question refers to the robustness of the spin-helical surface or edge states in these systems. Are they topologically protected from basic deterioration? In this regard future experiments able to test, for example, for backscattering blockade would be highly desirable. Furthermore, the~generation of  defect centers seemingly favorable for MZMs is unresolved; why do point-like Fe ions apparently support MZMs, why do some domain walls stabilize $\pi$-shifted regions, and how do strong correlations and local induced magnetism enter the game? While useful theories exist for several of these open points, it is nevertheless also clear that at present we lack quantitative models. These and many more exciting questions may hopefully constitute some of the many research directions pursued in the near future. Thus, even though iron-based superconductors have kept the community busy for more than a decade at present, we have not yet understood all their fascinating electronic properties, and most likely we have not yet unlocked all their secrets.

\section*{Acknowledgments}

We thank M.M. Korshunov for encouraging us to write this paper and acknowledge useful discussions with M. Allan, S. Backes, L. Benfatto, T. Berlijn, S. Bhattacharyya, A. E. B\"ohmer, K. Bj\"ornson. T. Chen, M. H. Christensen, A. V. Chubukov, P. Dai, J.C. Davis, L. Fanfarillo, R. M.  Fernandes, I. Fisher, T. Hanaguri, K.~Flensberg, J. Kang, A. Kostin, P. Kotetes, C. Setty, P. O. Sprau, D. Steffensen, R. Valent\'i, and K. Zantout.

B.M.A. acknowledges support from the Independent Research Fund Denmark, grant number 8021-00047B. P.J.H.  was supported by the U.S. Department of Energy under grant number DE-FG02-05ER46236, and by the National Science Foundation under grant number NSF-DMR-1849751.

%

\end{document}